\documentclass[a4paper,12pt]{article}
\usepackage{epstopdf}
\usepackage[onehalfspacing]{setspace}
\usepackage[top=2.3cm, bottom=2.3cm, left=2.4cm, right=2.4cm]{geometry}
\usepackage[english]{babel}
\usepackage[ansinew]{inputenc}
\usepackage{amsmath, amsthm, amssymb}
\usepackage{yfonts}%\usepackage{eufrak}
\usepackage{dsfont}
\usepackage{color}
\usepackage{graphicx,txfonts}
\usepackage{verbatim}
\usepackage[utf8]{inputenx}
\usepackage{tikz}
\usepackage{pgfplots}
\usetikzlibrary{arrows}
\usepackage{caption}
\usepackage{subcaption}
\usepackage{wrapfig}
\usepackage{multirow}

\allowdisplaybreaks[3]

\newtheorem{dependence}{Distribution-dependent Proposition}

\usepackage[makeroom]{cancel}
\usepackage{setspace}
\usepackage{thmtools,thm-restate}
\usepackage{sgamevar}
\usepackage{bm}
\usepackage{bbm}
\usepackage{float}
\usepackage{enumitem}
\usepackage{appendix}
\usepackage{booktabs}
\usepackage{array}
\usepackage{threeparttable}
\usepackage{adjustbox}
\usepackage{pifont}

\usepackage{mathtools} % for underbracket

\usepackage{pbsi}

\usepackage{url}
\RequirePackage[colorlinks,citecolor=blue,linkcolor=blue,urlcolor=blue]{hyperref}

\usepackage{epigraph}
\usepackage{mathtools}

\DeclareMathOperator*{\argmax}{arg\,max}

\newtheorem{proposition}{Proposition}
\newtheorem{result}{Result}
\newtheorem{remark}{Remark}
\newtheorem{theorem}{Theorem}

\newtheorem{corollary}{Corollary}
\newtheorem{example}{Example}

\newtheorem{definition}{Definition}

\newtheorem{lemma}{Lemma}

\usepackage{pgfplots}

\usepackage{xcolor}
\hypersetup{
	colorlinks   = true,
	citecolor    = olive,
	linkcolor = black,
	menucolor= black,
	urlcolor =blue,
}

\usepackage{natbib}

\begin{document}

\title{Search and Matching without Reciprocity
\footnote{This paper supersedes the 2021 working paper ‘Probabilistic Assortative Matching under Nash Bargaining.' We thank Emmanuelle Auriol, Roland Bénabou, Hector Chade, Yeon-Koo Che, Pierre-André Chiappori, Laura Doval, Christian Hellwig, Thomas Mariotti, Bernard Salanié, François Salanié, Robert Shimer, Lones Smith, and Jean Tirole for valuable feedback, as well as seminar and conference audiences at Stanford SITE, Arizona State University, Bristol, UPenn, Penn State, UT Austin, Rice, SMU, Royal Holloway, York, the Rice--LEMMA Monetary Conference in Paris, Mannheim, Vanderbilt, Bocconi, UCL, PSE, Université Paris-Saclay, CREST, Sciences Po, the MaCCI Summer Institute in Competition Policy, EEA, EAYE, SED and ESEM.}}
\author{Nicolas Bonneton\thanks{Department of Economics, Vanderbilt University. Email: nicolas.bonneton@gmail.com} \and Christopher Sandmann\thanks{Department of Economics, London School of Economics. Email: c.sandmann@lse.ac.uk}}
\date{\today}
\maketitle

\begin{abstract}
We study equilibrium sorting in the canonical search-and-matching model with Nash bargaining and ex-ante and ex-post heterogeneity. Under standard conditions for assortative matching under search, match surplus retains two properties of Becker's frictionless benchmark---single-crossing and single-peakedness---but necessarily loses the third, reciprocity. Instead, search generates consensus: intervals of types share the same surplus-maximizing partner. For such intervals, mismatched pairs are more likely to match, conditional on meeting, than perfectly assortative ones. Two forces drive this: with search, outside options rise less than one-for-one with productivity and fall more for agents with fewer suitable partners. Globally, upper contour sets of the match-probability surface are lattices below an endogenous threshold but almost never above it. This means that MLRP fails for some log-concave survival function.
\end{abstract}

\section{Introduction}

In frictionless matching markets with transferable utility, supermodularity of output generates positive assortative matching: the equilibrium assignment is increasing and traces an increasing ridge of surplus-maximizing partners (\cite{becker1973theory}). Partner rankings along this ridge satisfy \emph{reciprocity}: if one type is another's surplus-maximizing partner, the reverse is also true. In the canonical search-and-matching model under Nash bargaining, by contrast, we show that for every strictly increasing supermodular output function, every equilibrium exhibits \emph{consensus}: intervals of types share the same surplus-maximizing partner, and at most one of them is reciprocated. Geometrically, surplus is centered around two increasing ridges, not one.

We extend the canonical random-search model of \cite{shimer2000assortative} by introducing match-specific heterogeneity, as is common in structural matching models,\footnote{Match-specific heterogeneity enters additively in \cite{Hagedornetal2017} and \cite{LamadonLiseMeghirRobin2024}, and multiplicatively in \cite{gousse2017marriage}. Additive unobserved heterogeneity is also standard in frictionless matching models; see, e.g., \cite{choo2006marries,chiappori2016econometrics}.} so that surplus differences translate into differences in match probabilities rather than only a binary matching decision; Figure~1 illustrates this distinction. Agents have persistent types $x$ and $y$ and, upon meeting, draw an additional i.i.d.\ pair-specific production shock $\xi$. Writing surplus---output net of search values---as $S(x,y)$, a pair matches whenever $S(x,y)+\xi\geq0$, with transfers determined by Nash bargaining. Match probability is therefore increasing in surplus, so the ranking of potential partners by surplus coincides with their ranking by match probability.\footnote{The framework nests three canonical theoretical cases: constant $f$ recovers the homogeneous-search model of \cite{Pissarides2000} (Chap. 6); $\xi=0$ recovers \cite{shimer2000assortative}; and exponentially distributed $\xi$ yields the optimal-transport formulation of \cite{choo2006marries}, a workhorse of empirical matching as shown in \cite{bonnetonsandmann2026limit}.}

The paper makes two contributions. First, we develop an ordinal theory of assortative matching on match-probability surfaces, which we call probabilistic assortative matching (P-PAM). P-PAM decomposes sorting into properties of partner rankings and probability contours, allowing some aspects of sorting to be assortative while others are not. We characterize which combinations of these properties underlie Becker's frictionless benchmark, the lattice criterion of \cite{shimer2000assortative}, and the monotone likelihood ratio property (MLRP) used in empirical work. Second, we use P-PAM to characterize which features of frictionless assortative matching survive random search and which necessarily fail, and identify \emph{productive disadvantage} and \emph{positional disadvantage} as the forces behind these failures.

\paragraph{Contour Geometry.} In \cite{shimer2000assortative} and related random-search models, assortative matching requires the set of pairs with non-negative surplus, $S(x,y)\geq0$---those who would match upon meeting---to form a lattice with convex sections: intuitively, higher types match with sets of higher types (see Figure~1c). We extend this criterion across every probability level. Since match probability is increasing in surplus, this amounts to requiring every surplus upper contour set, $S(x,y)\geq k$, to form a lattice.

At first sight, this all-contour lattice criterion may seem unrelated to a workhorse empirical criterion for assortative matching: MLRP of matching flows \citep{graham2011econometric,siow2015testing,borovivckova2024assortative,chiappori2026assortative}. In fact, for supermodular surplus, lattice geometry is equivalent to robust MLRP: every surplus upper contour set forms a lattice if and only if MLRP holds for every shock distribution with log-concave survival (Theorem~\ref{thm:lattice-mlrp-logconcave}). The quantifier over shock distributions is essential because, as we show, MLRP is sensitive to the distribution of unobserved match heterogeneity: even holding the surplus surface fixed, one shock distribution can induce MLRP, while another induces reverse MLRP over an arbitrarily large region.

The all-contour lattice criterion also has an equivalent characterization in terms of partner rankings. Theorem~\ref{thm:weakPPAM} shows that all upper contour sets are lattices with convex sections if and only if surplus satisfies three properties: weak single-crossing, single-peakedness, and weak reciprocity---meaning that, relative to their surplus-maximizing partners, two agents cannot both gain by moving up the ranking, nor both by moving down.  This characterization gives us a common language for comparing search with Becker's frictionless benchmark. We show that, in the frictionless benchmark, supermodularity delivers more than the increasing assignment itself: surplus satisfies single-crossing, single-peakedness, and strict reciprocity, which goes beyond weak reciprocity by requiring, in particular, that “if you maximize my surplus, I maximize yours” (Result~\ref{rem:becker}). The frictionless benchmark therefore satisfies the all-contour lattice criterion and additionally imposes a stronger restriction on partner rankings.

\paragraph{Equilibrium Sorting.} Which probability contours retain the lattice geometry of the matching set? A single endogenous threshold separates lattice from non-lattice contours. Let $\widehat k\equiv\min_x\max_y S(x,y)$ denote the lowest peak surplus across
types. Theorem~\ref{theorem:lattice} shows that, under the same conditions that
deliver assortative matching in \cite{shimer2000assortative}, every upper contour set at or below $\widehat k$ forms a lattice, whereas almost every non-empty upper contour set above $\widehat k$ does not. This characterization applies to both steady-state and non-stationary equilibria. Lattice failures are confined to high-surplus---equivalently, high-probability---contours. In the no-shock benchmark ($\xi=0$), every type has a non-empty matching set, so $\widehat k\geq0$ and the $k=0$ contour lies in the surviving region. Pair-specific shocks make these lattice failures visible by translating different surplus levels into different match probabilities, thereby separating contours that are pooled together absent shocks.

Two Beckerian properties survive search---single-crossing and single-peakedness---but reciprocity does not. Theorem~\ref{thm:neverPAM} shows what takes its place: \emph{consensus}. Under conditions weaker than strict monotonicity and supermodularity, whenever search is active, a non-degenerate interval of types on each side shares the same unique likeliest partner type. Because consensus occurs on both sides, even weak reciprocity fails. This result holds regardless of how small search frictions are, how types are distributed, or whether equilibrium is in steady-state or non-stationary. For strictly increasing supermodular output, consensus always occurs at the top: the highest type is the likeliest partner of an interval of high types, and greater search frictions expand the range guaranteed to match up. These changes in partner rankings can also alter measured sorting:
Table \ref{table1} illustrates how fluctuations in the meeting rate
alone can move standard sorting correlations from positive to negative
under the standard technology $f(x,y)=xy$, holding all other primitives fixed.\footnote{\cite{chiappori2016econometrics}
highlight the empirical challenge of distinguishing search frictions from
unobserved match heterogeneity; \cite{ciscato2025assessing} provides a
recent attempt to separate the two forces quantitatively.}

Two distinct forces break reciprocity. First, in the frictionless benchmark, a productivity advantage that raises output equally with every partner is offset by equilibrium payoffs and has no effect on partner rankings. Search breaks this offset because matching takes time: discounting causes search values to respond less than one-for-one to the additional output that productive agents bring to a match. This creates what we call \emph{productive disadvantage} (Proposition~\ref{proposition:productive-disadvantage}). 

The mechanism is clearest with increasing additively separable output, where productive disadvantage operates in isolation: every type has the top type as its unique likeliest partner (Proposition~\ref{prop:additiveq}). The two peak-partner correspondences therefore cannot be further apart, and strict reciprocity fails maximally. At the same time, every surplus upper contour satisfies the lattice criterion for negative assortative matching (Corollary~\ref{corr:add:P-nam}), illustrating that lattice-based assortativity and Beckerian partner rankings capture different aspects of sorting.\footnote{This implication may seem surprising because additively separable technologies lie outside the sufficient conditions used by \cite{shimer2000assortative}. Yet we show that the conclusion does not rely on pair-specific shocks: in the shockless case, whenever matching sets are non-empty, $f(x,y)=x+y$ yields NAM and $f(x,y)=x-y$ yields PAM under the lattice criterion (Result~\ref{rem:shimer-smith-3}).}

Second, \emph{positional disadvantage} arises from limited access to suitable partners. For instance, boundary types have suitable partners on only one side, weakening their search options and making them relatively cheap to match with. This force requires no absolute productivity ranking. With horizontally differentiated output $f(x,y)=1-(x-y)^2$, each type produces most with its equal. Nevertheless, in every equilibrium with active search, all types sufficiently close to either boundary have that boundary type as their unique likeliest partner (Proposition~\ref{proposition:positional-disadvantage}).

\paragraph{Related literature.} This paper contributes to the
search-and-matching literature (see \cite{chade2017sorting} for a review), and more specifically to the strand in which match formation is probabilistic conditional on meeting. We do so by introducing a match-specific shock drawn upon meeting, a device common in empirical matching models with unobserved match heterogeneity (\cite{gousse2017marriage}, \cite{Hagedornetal2017},\cite{LamadonLiseMeghirRobin2024}). Closest to us is \cite{borovivckova2024assortative}, who derive MLRP results for meeting-contingent match probabilities under multiplicative Pareto and exponential shocks and use the Pareto case to quantify how selection shapes worker--firm sorting and wages. We instead abstract from any particular shock distribution or single criterion for assortative matching to isolate how random search changes the underlying ordinal structure of sorting.

Other papers generate probabilistic match formation through different mechanisms. \cite{chade2006matching} makes matching decisions probabilistic by letting agents observe noisy signals of one another's types, and identifies an acceptance curse. \cite{antler2026learning} instead lets couples learn about match-specific compatibility while dating, so a meeting leads to marriage only if favorable information arrives before either partner ends the relationship. 

Related ``matching-up'' mechanisms arise in other search environments. Under non-transferable utility, all agents prefer higher-ranked partners but may be rejected by them, so each type's preferred acceptable partner is the highest type willing to accept them \citep{burdett1997marriage,smith2006marriage}. Absolute productivity also matters in directed search: \citet{EeckhoutKircher2010} show that a ``trading-security'' motive can generate negative assortative matching, while sufficiently strong complementarity, root-supermodularity, restores positive assortative matching. In our random-search model, by contrast, matching up in partner rankings persists even under strong complementarity.

Most of the search-and-matching literature studies sorting in steady-state equilibrium. This paper belongs to a smaller body of work that studies sorting outside the steady state (\cite{boldrin1993}, \cite{shimer2001nonstationary}, \cite{LiseRobin2017}, \cite{baley2022mismatch}). We show that the restrictions on output underlying our ordinal results continue to deliver the same qualitative implications outside steady state, even though the extent of matching up and measured sorting can vary over time (Table~\ref{table1}). This allows us to generalize the steady-state results in \cite{shimer2000assortative} and stands in contrast to our parallel results under non-transferable utility (\cite{BonnetonSandmann2025}), where assortative matching can fail in non-stationary environments under conditions for which \cite{morgan1994model} and \cite{smith2006marriage} show that it occurs in steady state.\footnote{Under Nash bargaining, agents price anticipated future fluctuations in matching opportunities into match payoffs.}

\subsection{A stripped-down model} 

\noindent Before turning to the analysis, we present a \emph{simple} stripped-down model that isolates why search breaks reciprocity and reconciles our findings with the positive-assortative-matching result of \cite{shimer2000assortative}. Consider a frictionless matching market as in \cite{becker1973theory}. Agents with types $x\in[0,1]$ and $y\in[0,1]$ generate complementary match output $f(x,y)=xy$. The unique core allocation is assortative matching. Assuming symmetric populations, this means that $x$ matches with an identical partner type $y=x$. Equilibrium utility is $f(x,x)/2$.

Now suppose that the frictionless market takes place in period 1. Prior, in period 0, two agents $x,y \in [0,1]$  meet. At this point agents $x,y$ can either match and exit prematurely, or wait for their respective frictionless match payoffs.\footnote{The two-period structure is analogous to that in \cite{eeckhout2011identifying}, with discounting and a pair-specific shock.} The discounted period 1 payoff is $V(x)=\delta \dfrac{f(x,x)}{2}$. If instead they match in period 0, $x$ and $y$ generate joint output $f(x,y) +  \sigma \xi$ where $\xi$ is an idiosyncratic shock that has a continuous distribution with unbounded support, and $\sigma\ge 0$ scales the shock. Period 0 matching is intratemporally efficient: A pair $(x,y)$ matches if joint output exceeds the sum of discounted period one payoffs:
\[
f(x,y) +  \sigma \xi \geq V(x) + V(y).
\]
We refer to $f(x,y)-V(x)-V(y)$ as surplus. Since shocks $\xi$ are i.i.d. across potential partners, the probability that $(x,y)$ match increases with their surplus. Thus, the type  $y^\star(x)$ that maximizes surplus is also the type with whom $x$ is \emph{likeliest} to match upon meeting:
\begin{align*}
    y^\star(x)=\argmax\limits_{y\in [0,1]} f(x,y) - V(x) - V(y) \quad \text{and} \quad x^\star(y)=\argmax\limits_{x\in [0,1]} f(x,y) - V(x) - V(y).
\end{align*}
Our question is how search frictions shape equilibrium sorting. To that end, we introduce a new property, \emph{reciprocity}. Reciprocity requires that modal matching probabilities be reciprocal: if $x$ is most likely to match with $y^\star(x)$, then $y^\star(x)$ is most likely to match with $x$, that is,
\[
x=x^\star(y^\star(x)).
\]
Reciprocity highlights how the impact of search frictions ($\delta<1$) on equilibrium matching differs from the effect of probabilistic matching in the absence of search frictions ($\sigma>0$ and $\delta=1$). Reciprocity holds without search frictions, but fails when search frictions are present.

\begin{result}
For each $x,y\in[0,1]$, $
y^\star(x)=\min\!\left\{\frac{x}{\delta},\,1\right\}$ and $x^\star(y)=\min\!\left\{\frac{y}{\delta},\,1\right\}$.
\end{result}

\begin{proof}
Surplus is $xy-\delta\frac{x^2+y^2}{2}$. The FOC gives $y=\frac{x}{\delta}$. Imposing $y\in[0,1]$ yields the result.
\end{proof}

This result already exhibits the consensus pattern emphasized in the introduction: all types $x\geq\delta$ share $y=1$ as their surplus-maximizing partner. Search frictions break reciprocity because  $x^\star(y^\star(x))>x$ for every $x\in(0,1)$: individuals are more likely to match with partners who rank more highly than themselves. Moreover, the distortion increases as search frictions increase, i.e., as $\delta$ falls. The intuition is the productive-disadvantage mechanism described in the introduction: search prevents productive agents' outside options from fully reflecting their productivity, creating a tendency to match up.

\begin{figure}[!htbp]
  \centering
  \begin{subfigure}[t]{0.333\textwidth}
    \includegraphics[width=\linewidth,trim=3.8cm 0cm 3.8cm 0cm,clip]
    {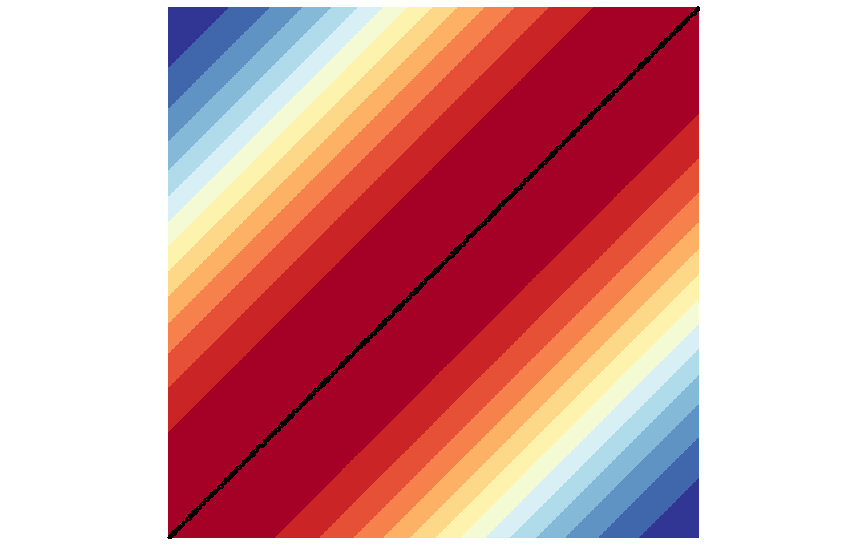}
    \caption{search \ding{55} shocks \ding{51}}
  \end{subfigure}%
  \begin{subfigure}[t]{0.333\textwidth}
    \includegraphics[width=\linewidth,trim=3.8cm 0cm 3.8cm 0cm,clip]
    {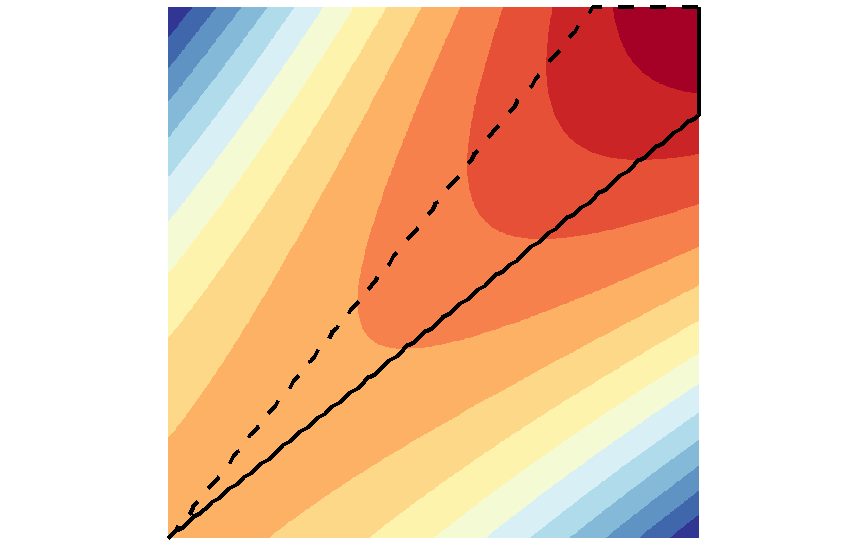}
    \caption{search \ding{51} shocks \ding{51}}
  \end{subfigure}%
  \begin{subfigure}[t]{0.333\textwidth}
    \includegraphics[width=\linewidth,trim=3.8cm 0cm 3.8cm 0cm,clip]
    {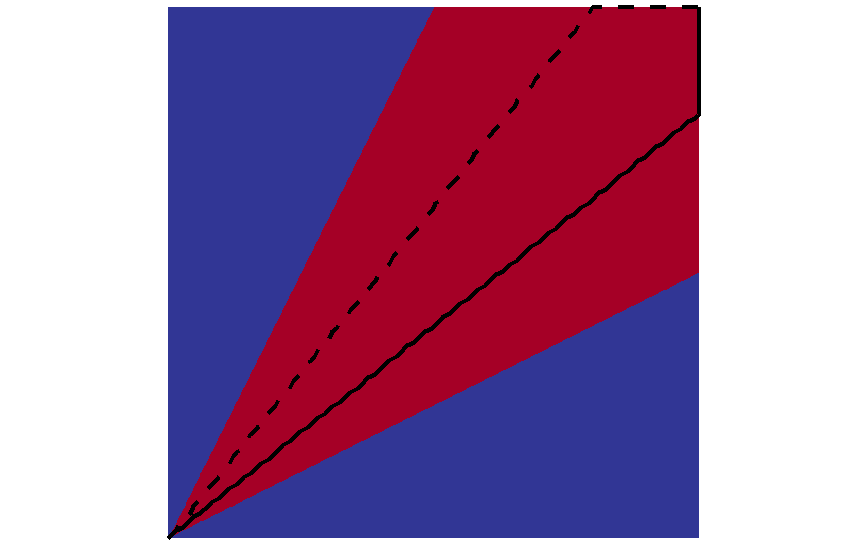}
    \caption{search \ding{51} shocks \ding{55}}
  \end{subfigure}

  \caption{Match probability: the role of search frictions and shocks}
 
  \caption*{\footnotesize
\textit{Notes:} The figure depicts the probability that a pair matches in period 0, conditional on meeting, with $x$ and $y$ on the two axes; warmer colors indicate a higher probability of matching. ``Search'' corresponds to $\delta=0.8$, while ``shocks'' corresponds to pair-specific shocks drawn from a logistic distribution. The dashed line represents $y^\star(x)$ and the solid line $x^\star(y)$.
}
  \label{fig:stripped-down}
\end{figure}

Observe that the identity of likeliest partners does not depend on shocks $\xi$. At the exact zero-shock case ($\sigma=0$) as in \cite{shimer2000assortative}, however, the acceptance rule collapses to a $0$–$1$ cone (Fig.~1c): pairs inside match with probability one and those outside with zero. As a result, low-but-positive-surplus pairs match as often as pairs that maximize surplus\textemdash masking agent's tendency to match up.

The body of the article shows that the failure of reciprocity is not an artifact of this stripped-down model.\footnote{In the stripped-down model, purely horizontally differentiated output, such as $f(x,y)=1-(x-y)^2$, preserves reciprocity. This is no longer true in the general model, where positional disadvantage breaks reciprocity.} Rather, it emerges robustly in general equilibrium search models with heterogeneous agents, a broad class of match-output functions, and non-stationary search dynamics. As we will see, reciprocity proves central to understanding how search frictions affect equilibrium sorting.

\section{The Model}

We study two populations (workers and firms, bachelors and bachelorettes, or buyers and sellers of houses), denoted $X$ and $Y$, whose members seek to form a match with someone from the other population. Each population contains a continuum of agents, and each agent has a type in $[0,1]$. Prior to matching, agents engage in random search: Unmatched agents gradually meet prospective partners over time. Upon meeting, they observe each other’s type and draw a pair-specific shock. If both agree, they match and permanently exit the search pool; otherwise, they continue searching. Each agent maximizes the expected present value of payoffs, discounted at rate $\rho>0$.

We use \emph{Result} for statements outside our main framework and \emph{Distribution-dependent Proposition} for statements that depend on the shock distribution. All other propositions and theorems are distribution-robust. Unless otherwise noted, proofs are in the appendix.

\subsection{Surplus and Match Probabilities}
\label{subsec:surplus-match-probabilities}

Fundamental match output is $f(x,y)$, with $f\in\mathcal C^1$. Upon meeting at date $t$, agents draw a pair-specific production shock $\xi$ from a common cdf $\Xi_t$, independently of their types and of other meetings.  The shock captures unmodeled determinants of match quality. The distribution may vary over time. Total output is $f(x,y)+\xi$.

Match payoffs are determined by Nash bargaining. The values of search $V_t^X(x)$ and $V_t^Y(y)$ serve as threat points, and transfers divide realized gains from trade, $S_t(x,y)+\xi$, where
\begin{equation}
\label{eq:systematic-surplus}
 S_t(x,y):=f(x,y)-V_t^X(x)-V_t^Y(y)
\end{equation}
is systematic surplus, hereafter surplus. Bargaining weights satisfy
$\alpha^X,\alpha^Y\in(0,1)$ and $\alpha^X+\alpha^Y=1$.\footnote{At
$\alpha^X=0$ or $\alpha^Y=0$, the model reduces to one-sided search,
a special case of the NTU model (cf. \cite{smith2006marriage} and \cite{BonnetonSandmann2025} ).}
Agent $x$ receives $V_t^X(x)+\alpha^X[S_t(x,y)+\xi]$, and symmetrically
for $y$.\footnote{Whether shocks are bargained over, as we assume, is not crucial for our ordinal results. In some applications, shocks may represent taste components that are not contractible or observable, and hence not subject to bargaining. The ordinal equivalence between surplus rankings and match-probability rankings is unaffected as long as match probabilities are increasing in surplus.}  Both therefore agree to match if and only if
$S_t(x,y)+\xi\geq0$. The meeting-contingent match probability is
\begin{equation}
\label{eq:m}
 m_t(x,y)=\Pr[S_t(x,y)+\xi\geq0]=1-\Xi_t(-S_t(x,y))=\bar\Xi_t(-S_t(x,y)).
\end{equation}
We focus on additive shocks, as they uniquely ensure that match probabilities are increasing in surplus conditional on meeting.\footnote{For any strictly increasing ex-post output $\phi(f, \xi)$ that is not additively separable there exist $f$ and $\Xi$ for which such co-monotonicity fails. In particular, if ex-post output is multiplicative as in $\phi(f, \xi)=f(x,y)\cdot \xi$, a larger surplus can accompany a smaller acceptance probability. That is a substantive alternative model explored in \cite{borovivckova2024assortative}.}

Except when explicitly considering the shockless benchmark, assume that $\Xi_t$ is continuous, has a finite first moment, and has a strictly positive continuous density on the interior of its support $[\underline\xi_t,\infty)$, allowing $\underline\xi_t=-\infty$.
The lower endpoint satisfies $\underline\xi_t+\max_{x,y}f(x,y)<0$ at every date.\footnote{The additional uniform regularity and positive-tail bounds used for existence are stated in the online appendix.} Since search values are non-negative, every acceptance threshold satisfies $-S_t(x,y)\geq-\max_{x,y}f(x,y)>\underline\xi_t$. Consequently, every pair matches with probability strictly between zero and one, and $m_t$ is a common strictly increasing transformation of $S_t$.

Define the surplus-maximizing partner correspondences by
\begin{align}
\label{eq:likeliest-partners}
 Y_t^\star(x)&:=\arg\max_{y\in[0,1]}S_t(x,y)\qquad \text{and} \qquad
 X_t^\star(y):=\arg\max_{x\in[0,1]}S_t(x,y).
\end{align}
We call these the \emph{likeliest-partner correspondences}, as they equivalently maximize match probability conditional on meeting. Write $\underline y_t(x),\overline y_t(x)$ and $\underline x_t(y),\overline x_t(y)$ for their least and greatest elements. When they are singletons, denote their unique elements by $y_t^\star(x)$ and $x_t^\star(y)$.

Without shocks, as in \cite{shimer2000assortative}, $m_t(x,y)=\mathbf 1\{S_t(x,y)\geq0\}$. A barely acceptable pair then matches upon meeting just as surely as a surplus-maximizing pair. Additive heterogeneity translates these surplus differences into differences in match probabilities. %In the shockless benchmark, $X_t^\star$ and $Y_t^\star$ continue to denote surplus maximizers, which the binary match probabilities need not distinguish from other acceptable partners.

\subsection{Search Pool Dynamics and Value-of-Search} 

An equilibrium is a solution to a coupled system of equations governing the evolution of the pool of unmatched agents and the associated values-of-search. The construction is standard; we describe it for population $X$, with population $Y$ symmetric.

\noindent \emph{Meetings.} The meeting rate is anonymous: any type $x$ meets type $y$ at rate $\lambda_t \,\mu_t^Y(y)$, where $\mu_t^Y(y)$ is the mass of unmatched $y$-types at time $t$. The parameter $\lambda_t$ is a base meeting rate that may depend on time and on the aggregate masses of unmatched agents (accommodating standard linear and quadratic specifications). The resulting flow of match creation between $x$ and $y$ is
\[
q_t(x,y) \equiv \lambda_t \,\mu_t^X(x) \mu_t^Y(y) m_t(x,y).
\]

\noindent \emph{Evolution of the search pool.} Population dynamics are governed by entry and exit. Entry by types $x$ into the search pool is given by some exogenous $\eta_t^X(x)$ that may depend on time $t$ and the state $\mu_t^X,\mu_t^Y$ (as in hot and cold housing markets or seasonal job markets).
\begin{align}
\label{eq:popltn-dynmcs}
{\mu}_{t+h}^X(x)={\mu}_{t}^X(x)+\int\limits_{t}^{t+h}\Bigg\{-\int\limits_{0}^{1}q_{\tau}(x,y)dy+\eta^X_{\tau}(x)\Bigg\}d\tau.
\end{align}

\noindent \emph{Value-of-search.} Agent $x$'s value-of-search is the expected discounted match payoff. We first define the expected payoff from a match. Let $\Pi_t^X(y|x)$ denote $x$'s expected time $t$ match payoff conditional on matching with $y$ (e.g.,  mean wages or sold property prices):
\begin{equation*}
\Pi_t^X(y|x) = V_t^X(x) + \alpha^X \, \mathbb{E}\big[ S_t(x,y) + \xi \mid S_t(x,y) + \xi \geq 0 \big].
\end{equation*}
Match densities are derived from the individual hazard rate $\Lambda_t^X(y|x)=\lambda_t\,\mu_t^Y(y)m_t(x,y)$. Correspondingly,
\[
p_{t,\tau}^X(y|x)
= \Lambda_\tau^X(y|x)\exp\left\{-\int_t^\tau \int_0^1 \Lambda_s^X(y'|x)\,dy'\,ds\right\}
\]
is the match density that type $x$, unmatched at time $t$, matches with type $y$ at time $\tau$. 

Agent $x$'s value-of-search is then given by
\begin{align}\label{eq:value-of-search}
     V_t^X(x) & \equiv\int\limits_{t}^{\infty} e^{-\rho(\tau-t)}\int\limits_{0}^1 \Pi_{\tau}^X(y|x)\,p_{t,\tau}^X(y|x)dy\,\,d\tau.
\end{align}

\bigskip

\noindent \emph{Non-stationary Equilibrium.} Given initial conditions $\mu_0^X, \mu_0^Y$, an equilibrium is a tuple $(\boldsymbol{\mu}, \mathbf{V}) = (\mu_t^X, \mu_t^Y, V_t^X, V_t^Y)_{t \geq 0}$ such that the population masses satisfy \eqref{eq:popltn-dynmcs} and the values-of-search satisfy \eqref{eq:value-of-search} for both populations. Throughout, we restrict attention to equilibria with locally integrable individual meeting rates and values of search that, when discounted to date zero, are uniformly bounded across types and dates. Under the regularity conditions stated in the online appendix, an equilibrium exists for every bounded initial search pool.\footnote{Without pair-specific shocks, existence of a steady-state equilibrium is established by \cite{shimer2000assortative} and extended to more general search technologies by \cite{lauermann2020balance}, while \cite{SandmannBonneton2025} establishes existence of a non-stationary equilibrium. The online appendix adapts the latter proof to accommodate ex-post match heterogeneity.}

\section{The Ordinal Geometry of Probabilistic Matching}
\label{sec:ordinal-geometry}

We develop a conceptual framework for studying sorting on matching-probability surfaces that connects three notions usually studied separately: the lattice structure of the matching set, Beckerian partner rankings, and the monotone likelihood ratio property (MLRP).

\subsection{From One Contour to All Contours}
\label{subsec:all-contours}

Our object of interest is the family of upper contour sets of the matching-probability surface. For each $p\in[0,1]$, define $U_t(p):=\{(x,y):m_t(x,y)\geq p\}$ and its individual sections $U_t^X(p;x):=\{y:m_t(x,y)\geq p\}$ and $U_t^Y(p;y):=\{x:m_t(x,y)\geq p\}$. 

Without pair-specific shocks, $m_t(x,y)=\mathbf 1\{S_t(x,y)\geq0\}$, so all positive probability thresholds select the same matching set: $U_t(p)=U_t(1)=\{(x,y):S_t(x,y)\geq0\}$, $p\in(0,1]$. Following \cite{shimer2000assortative}, and as in subsequent work such as \cite{atakan2006assortative,smith2006marriage, BonnetonSandmann2025}, matching is positively assortative when this set is a lattice:  $U_t(1)$ is a lattice if, for all $x_2>x_1$ and $y_2>y_1$,
\begin{equation}
\label{eq:lattice-contour}
    (x_1,y_2),(x_2,y_1)\in U_t(1)
    \quad\Longrightarrow\quad
    (x_1,y_1),(x_2,y_2)\in U_t(1).
\end{equation}

We ask whether the family of upper contour sets has the same lattice geometry that \cite{shimer2000assortative} requires of the matching set.

\begin{definition}[Weak P-PAM]
\label{def:weak-ppam}
There is \emph{weak probabilistic positive assortative matching} (weak P-PAM) at time $t$ if, for every $p\in[0,1]$, (i) $U_t(p)$ is a lattice and (ii) every individual upper contour set $U_t^X(p;x)$ and $U_t^Y(p;y)$ is convex.
\end{definition}

Convexity of individual sections rules out holes in a type's set of sufficiently likely partners; under continuity of $m_t$, it is equivalent to single-peakedness in partner type.\footnote{Strictly speaking, \cite{shimer2000assortative} define positive assortative matching solely through the lattice property. This is because the convexity of individual sections can be derived from the lattice property in their environment: if $U_t^X(p;x)$ is non-empty for every $x$, then the lattice property implies convexity of $U_t^Y(p;y)$ (cf. Proposition~2 from \cite{shimer2000assortative}). In our probabilistic environment, we impose convexity of both individual upper contour sets explicitly because upper contour sets may be empty for some types at high thresholds.} Visually, at thresholds for which every individual section is non-empty, the lattice condition is equivalent to non-decreasing lower and upper contour boundaries (refer to Proposition 3 in \cite{shimer2000assortative} for a proof). 

Observe that the family $\{U_t(p)\}_{p\in[0,1]}$ coincides, up to a relabeling of thresholds, with the family of surplus upper contour sets. For a fixed surplus function, its geometry is therefore ordinal: changing $\Xi_t$ changes the probability attached to a contour, not the contour itself.

\subsection{From Lattice to Empirical Measures: MLRP}

The lattice geometry of the family of upper contour sets connects to the workhorse empirical notion of probabilistic assortative matching, the monotone likelihood ratio property (MLRP) \citep{graham2011econometric,siow2015testing,borovivckova2024assortative,chiappori2026assortative}.\footnote{MLRP is far-reaching. It implies first-order stochastic dominance in the sense that the conditional CDF $\int_0^y q_t(x,y')\,dy'\big/\int_0^1 q_t(x,y')\,dy'$ is non-increasing in $x$. Furthermore, MLRP implies \emph{non-negative concordance measures}: Kendall's $\tau \geq 0$, Spearman's $\rho_S \geq 0$, and Pearson correlation $\rho \geq 0$ across matched pairs.}\textsuperscript{,}\footnote{\cite{shimer2000assortative}, footnote 8, note that their lattice criterion can be extended to probabilistic acceptance by requiring the probability-of-matching function to be affiliated. In the present two-dimensional setting, affiliation is MLRP.}

\begin{definition}[Monotone likelihood ratio property]
The matching flow $q_t(x,y)$ satisfies MLRP if
\[
q_t(x_2,y_2)\,q_t(x_1,y_1)\;\geq\; q_t(x_2,y_1)\,q_t(x_1,y_2)
\qquad\text{for all } x_2>x_1,\ y_2>y_1.
\]
\end{definition}

Since matching flows satisfy $q_t(x,y)=\lambda_t\mu_t^X(x)\mu_t^Y(y)m_t(x,y)$, the type-specific multiplicative factors cancel from the MLRP cross-product inequality. Hence, on the active type support, $q_t$ satisfies MLRP if and only if $m_t$ does.

The link between MLRP and the lattice geometry is as follows.

\begin{theorem}[All contours and distribution-robust MLRP]
\label{thm:lattice-mlrp-logconcave}
Fix a supermodular surplus $S_t$. The following are equivalent:
\begin{enumerate}[label=(\roman*)]
    \item For every $c\in\mathbb{R}$, the upper contour set
    $\{(x,y):S_t(x,y)\geq c\}$ is a lattice.
\item For every CDF $\Xi_t$ with log-concave survival function,
$m_t=\bar\Xi_t(-S_t)$ satisfies MLRP.\footnote{Log-concave survival includes many standard shock distributions, such as normal, logistic, uniform, and exponential.}
\end{enumerate}
\end{theorem}

Although this result is established in our search-and-matching framework, it is relevant to the related empirical frictionless-matching literature, as both frameworks assume an additive surplus specification $S(x,y)+\xi$ (see \cite{chiappori2016econometrics} for a review). The underlying matching environments differ, but the inference problem---recovering properties of surplus from observed matching probabilities---is the same.

Read in the contrapositive, the theorem says that a failure of the lattice property is exactly a failure of robust MLRP: if some upper contour set is not a lattice, then there exists a log-concave-survival shock distribution under which MLRP fails.

Theorem \ref{thm:lattice-mlrp-logconcave} highlights the key difference between weak P-PAM and MLRP: weak P-PAM depends only on the ordinal information contained in surplus (or match probabilities), whereas MLRP additionally depends on the cardinal transformation from surplus to match probabilities. The next two examples illustrate.

\begin{figure}[!htbp]
  \centering

  \begin{subfigure}[t]{0.49\textwidth}
    \includegraphics[width=\linewidth,trim=4cm 0cm 4cm 0cm,clip]
    {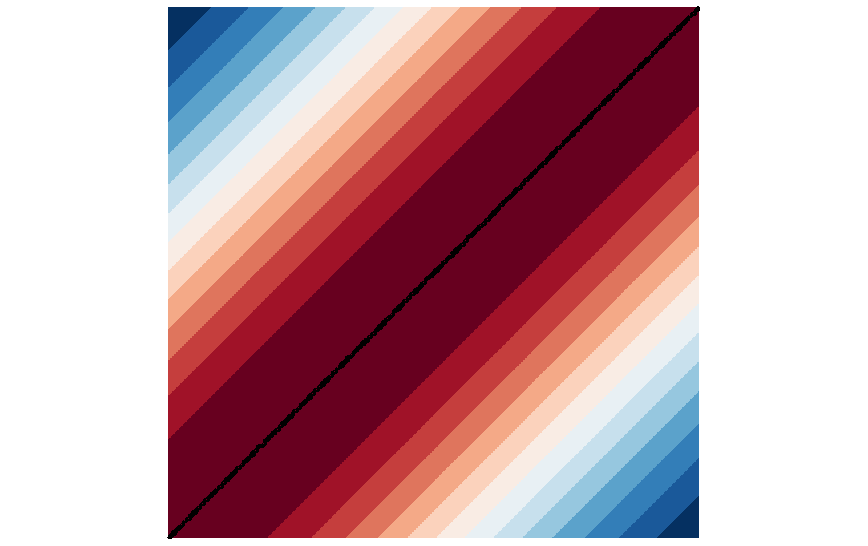}
    \caption{P-PAM \ding{51} \quad MLRP \ding{55}}
    \label{fig:PPAM:noMLRP}
  \end{subfigure}%
  \begin{subfigure}[t]{0.49\textwidth}
    \includegraphics[width=\linewidth,trim=4cm 0cm 4cm 0cm,clip]
    {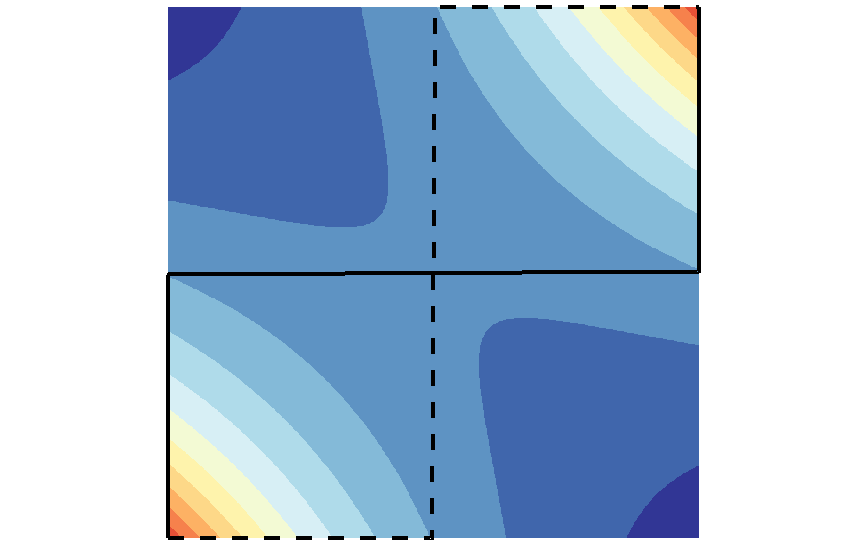}
    \caption{P-PAM \ding{55} \quad MLRP \ding{51}}
    \label{fig:noPPAM:MLRP}
  \end{subfigure}

  \caption{P-PAM and MLRP may disagree on whether matching is positively assortative.}
\caption*{\footnotesize
   \textit{Notes:} The dashed line represents $y_t^\star(x)$ and the solid line $x_t^\star(y)$, the two populations' likeliest partners. Warmer colors indicate higher meeting-contingent match probabilities. The left panel uses $S_t(x,y)=-(x-y)^2/2$ and shifted-Pareto shocks.  The right panel reports the stationary equilibrium with $f(x,y)=(x+y-1)^2$, discount rate $\rho=0.05$, symmetric uniform populations, and shifted-exponential shocks.\footnotemark
}
\label{fig:PPAM_MLRP}
\end{figure}

\begin{example}[P-PAM need not yield MLRP]
For a fixed shock distribution, MLRP may fail in spite of supermodular surplus -- even if weak P-PAM holds. Consider the surplus function from the stripped-down example of the introduction, $S(x,y)=xy-\delta\frac{x^2}{2}-\delta\frac{y^2}{2}$, which is supermodular. P-PAM holds in the frictionless market $\delta=1$ (Figure~\ref{fig:stripped-down}(a)) and fails at any positive level of search friction (Figure~\ref{fig:stripped-down}(b)).

Yet under shifted-Pareto shocks with survival function $\bar\Xi_t(z)=\frac{1}{2}\left(\frac{b}{b+z}\right)^a$, $z\geq0$, MLRP not only fails but is locally reversed whenever $|x-y|>\sqrt{2b}$.  Indeed, by taking $b$ sufficiently small, the local MLRP criterion classifies an arbitrarily large share of the type space as negatively sorted, even at $\delta=1$, where every probability level line is increasing. This example does not contradictTheorem~\ref{thm:lattice-mlrp-logconcave}: the relevant shifted-Pareto tail is log-convex. The left panel of Figure~\ref{fig:PPAM:noMLRP} illustrates.
\end{example}

\begin{example}[MLRP with holes]
The converse failure is equally instructive: MLRP can hold even when the upper contour sets of match probabilities fail to be lattices. Consider the output function $f(x,y)=(x+y-1)^2$. \cite{shimer2000assortative} use this output to show that supermodularity alone does not guarantee assortative matching in a search model: holes in the matching set prevent middle types from matching with one another (their Figure~3). With match-specific shocks there is no matching set\textemdash every pair matches with positive probability\textemdash but the same qualitative feature persists: pairs of equally ranked middle types are the \emph{least} likely to match, so weak P-PAM fails. Under exponential shocks, $\log m_t(x,y)=a_t+\kappa S_t(x,y)$ for some $\kappa>0$. Hence
\[
D_{12}^2\log m_t(x,y)=\kappa f_{12}(x,y)\geq0,
\]
so MLRP holds even though equally ranked middle types are the least likely to match with one another. Thus, more generally, any supermodular output function generates MLRP under exponential shocks. The right panel of Figure~\ref{fig:noPPAM:MLRP} illustrates.
\end{example}
 
\footnotetext{The left panel uses the surplus from the stripped-down model ($\delta=1$) because, as we will show below, equilibrium surplus in the general model cannot satisfy P-PAM.}

Because the comparison between weak P-PAM and MLRP is motivated by empirical measures of sorting, it is useful to note that weak P-PAM itself can be assessed from matching-flow data, even though surplus and meeting-contingent match probabilities are not directly observable. The flow of match creation between types $x$ and $y$ is given by $q_t(x,y)\equiv \lambda_t\,\mu_t^X(x)\,\mu_t^Y(y)\,m_t(x,y)$. 
Hence, an analyst who, as in \cite{choo2006marries}, observes matching flows $q_t$ together with the masses of unmatched agents $\mu_t$ can identify meeting-contingent match probabilities up to the common multiplicative factor $\lambda_t$. Since weak P-PAM is invariant to a common multiplicative normalization of match probabilities, the factor $\lambda_t$ is immaterial.

\subsection{P-PAM and the Frictionless Benchmark}

The lattice geometry of the family of upper contour sets is not an arbitrary strengthening of the binary matching-set requirement. As we show below, it is already satisfied in Becker's frictionless benchmark.

The classic result of \cite{becker1973theory} is that supermodularity, $f(x_2,y_2)+f(x_1,y_1)\geq f(x_1,y_2)+f(x_2,y_1)$ for all $x_2>x_1$ and $y_2>y_1$, gives rise to a monotone assignment. We show that it also imposes three ordinal restrictions on surplus.

\begin{result}[Frictionless Benchmark]
\label{rem:becker}
In \cite{becker1973theory}'s frictionless model with populations of equal size, supermodularity of output ensures that frictionless surplus $S(x,y)$ satisfies the following ordinal structure:%, where
%\[
%Y^\star(x)=\argmax_y S(x,y),\qquad X^\star(y)=\argmax_x S(x,y).
%\]
%and where $\underline y(x),\overline y(x)$ and $\underline x(y),\overline x(y)$ denote the lower and upper endpoints of these sets:
\begin{enumerate}
    \item \textbf{Single-crossing}: $S(x_1,y_2) > S(x_1,y_1) \Rightarrow 
    S(x_2,y_2) > S(x_2,y_1)$ for all $x_2 > x_1$, $y_2 > y_1$, and vice versa for population $Y$.
    \item \textbf{Single-peakedness}: $y \mapsto S(x,y)$ is non-decreasing 
    for $y < \underline{y}^\star(x)$ and non-increasing for $y > \overline y^\star(x)$, and 
    symmetrically for $x \mapsto S(x,y)$.
    \item \textbf{Strict reciprocity}: $
y<\underline y(x) \Longleftrightarrow x>\overline x(y),
\qquad
y>\overline y(x) \Longleftrightarrow x<\underline x(y).
$
\end{enumerate}
\end{result}

In the frictionless model, only zero-surplus pairs are assigned, so the rest of this ordinal structure is behaviorally latent. With pair-specific shocks, matching probabilities are strictly increasing in surplus, making these ordinal comparisons behaviorally relevant.

The connection between Becker's ordinal properties and the lattice geometry of contour sets is not immediate. The next result characterizes this relationship: weak P-PAM corresponds to a weakening of the Beckerian ordinal conditions in Result~\ref{rem:becker}.

\begin{theorem}\label{thm:weakPPAM}
Weak P-PAM holds if and only if $S_t$ satisfies weak single-crossing,\footnote{Weak single-crossing means
$S_t(x_1,y_2)>S_t(x_1,y_1)\Rightarrow
S_t(x_2,y_2)\geq S_t(x_2,y_1)$ for all
$x_2>x_1$, $y_2>y_1$, and symmetrically for population $Y$.
Under single-peakedness and weak reciprocity, weak single-crossing follows automatically.} single-peakedness, and weak reciprocity. Weak reciprocity means
\[y<\underline y_t(x)\Rightarrow x\geq \underline x_t(y),
\quad
y>\overline y_t(x)\Rightarrow x\leq \overline x_t(y),
\]
and symmetrically,
\[
x<\underline x_t(y)\Rightarrow y\geq \underline y_t(x),
\quad
x>\overline x_t(y)\Rightarrow y\leq \overline y_t(x).
\]
\end{theorem}

With unique surplus-maximizing partners, weak reciprocity has a simple interpretation: if $y$ lies below $x$'s surplus-maximizing partner, then $x$ cannot also lie below $y$'s surplus-maximizing partner; and likewise when both lie above. The following remark clarifies the gap between weak and strict reciprocity.

\begin{remark}[Weak and strict reciprocity]
\label{remark-weak-strict-reciprocity}
Suppose that $S_t$ is single-peaked. Then strict reciprocity is equivalent to weak reciprocity together with peak reciprocity: $x\in X^\star(y)\Longleftrightarrow y\in Y^\star(x)$, or in words, $x$ is likeliest to match with $y$ if and only if $y$ is likeliest to match with $x$.
\end{remark}

Visually, the Beckerian ordinal properties organize match probabilities around a peak set shared by both sides of the market. When peak sets are single-valued and interior, this common peak set traces an increasing ridge; in a symmetric model, it coincides with the $45^\circ$ line. Moving away from the ridge lowers match surplus. 

Weak P-PAM does not require this peak set to be shared. Without peak reciprocity, surplus-maximizing, or equivalently, likeliest partners may lie at the boundary, reflecting a tendency to match with extreme types rather than with equally ranked types. (This concern foreshadows Proposition~\ref{proposition:positional-disadvantage}.) This motivates a stronger notion of sorting that retains not only the lattice geometry of all contour sets, but also the full Beckerian ordinal structure.

\begin{definition}[P-PAM]
\label{def:ppam}
There is \emph{probabilistic positive assortative matching} (P-PAM) at time
$t$ if $S_t$ satisfies single-crossing, single-peakedness, and strict
reciprocity.
\end{definition}

We use P-PAM throughout the paper as a diagnostic: we ask which ordinal properties of the frictionless benchmark survive search frictions and which fail.

\section{Equilibrium Matching: Which Contours Are Assortative?}

Section~\ref{sec:ordinal-geometry} developed P-PAM as a diagnostic for the effects of search. We now derive sufficient conditions under which surplus retains two Beckerian ordinal properties---single-crossing and single-peakedness. We then characterize the lattice structure of upper contour sets through an endogenous threshold.

\subsection{Single-Crossing}

We begin our analysis with single-crossing. In a frictionless matching market, the type with whom one matches is also the type with whom one generates the greatest match payoff. Under supermodularity, \cite{becker1973theory} shows that this type is non-decreasing in one’s own type. The next proposition generalizes this result to an environment with search frictions.

\begin{proposition}[Single-crossing]
\label{prop:single-crossing}
If $f$ is supermodular, then $S_t$ and $m_t$ satisfy single-crossing. Moreover, the least and greatest elements of $X_t^\star(y)$ are non-decreasing in $y$, and those of $Y_t^\star(x)$ are non-decreasing in $x$.
\end{proposition}

\begin{proof}
Recall that $m_t(x,y)$ is a strictly increasing transformation of the surplus $f(x,y)-V^X_t(x)-V^Y_t(y)$. Since subtracting separable terms preserves supermodularity, the surplus is supermodular whenever $f$ is. A strictly increasing transformation of a supermodular function satisfies single-crossing. Since $m_t$ and $S_t$ have the same maximizers, monotonicity of the least and greatest elements of $X_t^\star(y)$ follows by Topkis's theorem (Theorem 2.8.3(a)) applied to $S_t$.
\end{proof}

This simple result shows that supermodularity remains a first-order reason for positive sorting in our model.%\footnote{In a related probabilistic matching model, \cite{antler2026learning} identify assortative matching with single-crossing and show that it holds when expected output is supermodular.}

\subsection{Single-Peakedness} \label{sec:singlepeak}
Unlike single-crossing, single-peakedness depends on the curvature of the values of search. To control this curvature away from a steady state, we first establish a key lemma that bounds differences in values of search using an iterative mimicking argument.

\begin{lemma}[Mimicking]
\label{lem:TU-mimicking}
For every date $t$ and type $x'$, define the non-negative measure
\begin{align}
\label{eq:TU-Q-representation}
 Q_t^X(dy\mid x')
 ={}& \alpha^X\int_t^\infty e^{-\rho(\tau-t)}
 \Lambda_\tau^X(y\mid x')\exp\!\left\{-\alpha^X\int_t^\tau\int_0^1
 \Lambda_s^X(z\mid x')\,dz\,ds\right\}d\tau\,dy.
\end{align}
This satisfies $Q_t^X([0,1]\mid x')<1$. Then, for every $x''\in [0,1]$,
\[
 V_t^X(x'')-V_t^X(x')
 \geq
 \int_0^1\bigl[f(x'',y)-f(x',y)\bigr]Q_t^X(dy\mid x').
\]
The symmetric statement holds for population $Y$.
\end{lemma}

This shows that search values rise in proportion---but less than one-to-one---to output. To see the idea, evaluate type $x''$ on the partners and shock realizations accepted by $x'$. Whenever such a realization would generate negative surplus for $x''$, continuing search yields weakly more than the corresponding hypothetical match payoff. This comparison provides a lower bound on $x''$'s value; subtracting $x'$'s value gives
\begin{align*}
    V^X_t(x'')-V^X_t(x')
    \geq
    \int_t^\infty\int_0^1 e^{-\rho(\tau-t)}
    \Big\{
    (1-\alpha^X)\big[V^X_\tau(x'')-V^X_\tau(x')\big]
    +\alpha^X\big[f(x'',y)-f(x',y)\big]
    \Big\}
    p_{t,\tau}^X(y|x')\,dy\,d\tau.
\end{align*}
In a steady state, the continuation-value term can be rearranged directly to obtain the bound in Lemma~\ref{lem:TU-mimicking}. In a non-stationary environment, $V^X_\tau(x'')-V^X_\tau(x')$ may vary with $\tau$, so the argument must instead be iterated forward. The proof formalizes this iteration and shows that its weight vanishes geometrically.\footnote{The steady-state counterpart is Lemma~1 of
\cite{shimer2000assortative}.}

The next corollary gives the derivative version of this bound.
\begin{corollary}[Search-value derivatives]
\label{cor:TU-mimicking}
Suppose that $f\in\mathcal C^1$. Then values of search are continuously differentiable and satisfy
\[
 {V_t^X}'(x)=\int_0^1 f_1(x,y)Q_t^X(dy\mid x),
 \qquad
 {V_t^Y}'(y)=\int_0^1 f_2(x,y)Q_t^Y(dx\mid y).
\]
Therefore $X_t^\star$ and $Y_t^\star$ are non-empty, compact-valued, and upper hemicontinuous.
\end{corollary}

With this bound in hand, proving that surplus is single-peaked largely follows the steady-state argument of \cite{shimer2000assortative}. The same curvature conditions therefore remain sufficient away from steady state. They are satisfied, for example, by $f(x,y)=xy$ and $f(x,y)=1-(x-y)^2$. The extensions to asymmetric populations and pair-specific shocks are immediate.  

\begin{proposition}[Single-peakedness]
\label{mconvex}
Suppose that $f$ is supermodular with $f_{12}\not \equiv 0$ and that $f_1$ and $f_{12}$ are log supermodular. Then, for every $y$, the functions $x\mapsto S_t(x,y)$ and $x\mapsto m_t(x,y)$ are single-peaked. Symmetrically, if $f_2$ and $f_{12}$ are log supermodular, then, for every $x$, the functions $y\mapsto S_t(x,y)$ and $y\mapsto m_t(x,y)$ are single-peaked.\footnote{Throughout, following \cite{shimer2000assortative}, log supermodularity of a possibly signed function $g$ refers to the cross-product condition $g(x_1,y_1)g(x_2,y_2)\geq g(x_1,y_2)g(x_2,y_1)$ for $x_2>x_1$ and $y_2>y_1$.}
\end{proposition}

Taken together, single-crossing and single-peakedness characterize the monotonicity of meeting-contingent match probabilities. Single-crossing implies that the least and greatest elements of $X_t^\star(y)$ and $Y_t^\star(x)$ are non-decreasing, while single-peakedness determines the direction in which match probabilities increase relative to these correspondences.

\begin{figure}\centering
\includegraphics[scale=0.16]{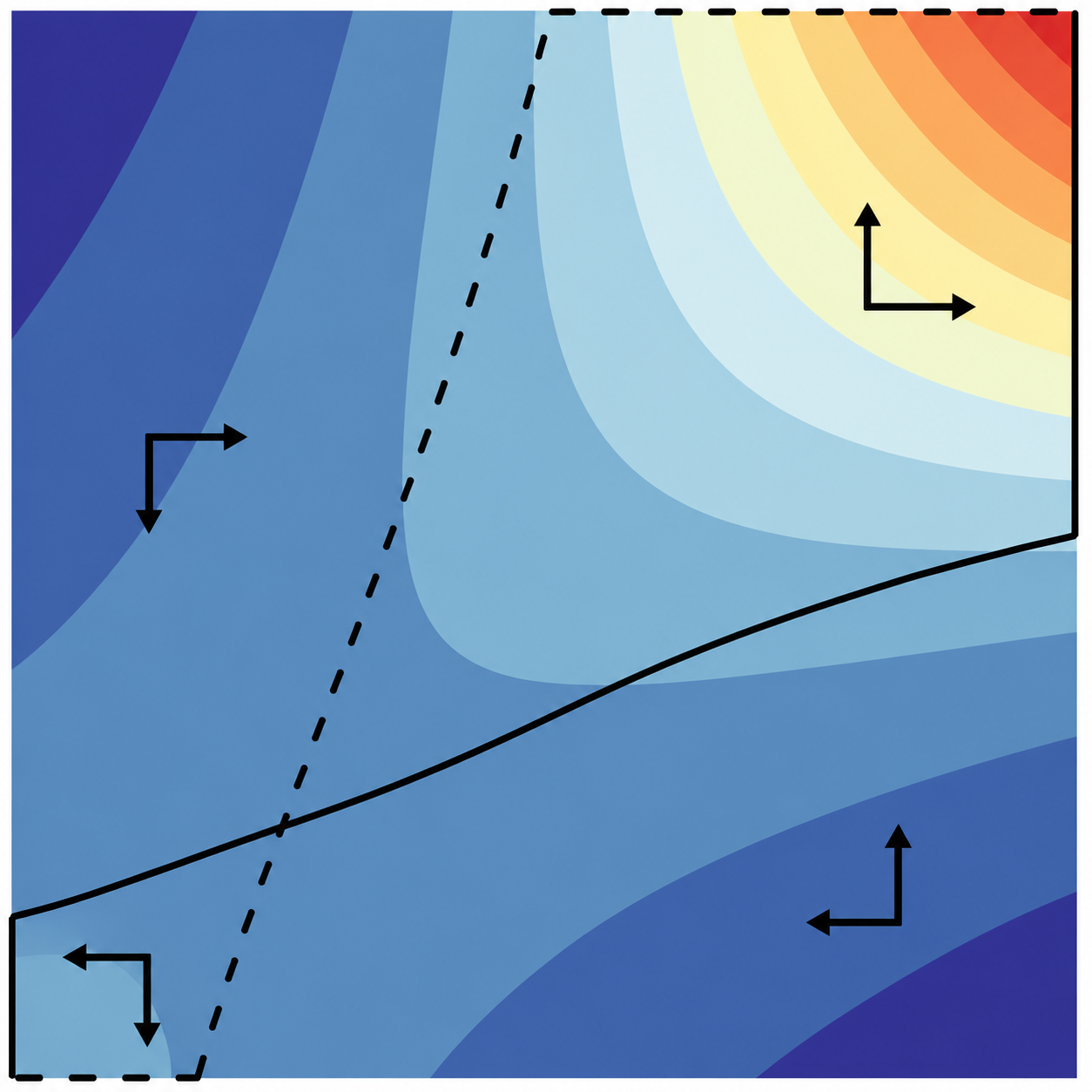}
\caption{ $f(x,y)=xy$ -- Match probabilities satisfying single-crossing and single-peakedness}
\caption*{\footnotesize
\textit{Notes:} The dashed line represents $y^\star(x)$ and the solid line $x^\star(y)$. Arrows indicate the direction of increasing match probabilities. Warmer colors indicate a higher probability of matching conditional on meeting. We numerically solve the stationary equilibrium, discount rate $\rho=0.05$, symmetric and uniform populations, and logistic shocks.  
}
\label{figure:main}
\end{figure}

These results partition the product type space by the two likeliest type correspondences. If a pair lies below both correspondences, match probability rises with both types as depicted by north-east arrows
\begin{tikzpicture}[baseline=-.4ex]
    \draw[->] (0,0) -- (0.3,0);
    \draw[->] (0,0) -- (0,0.3);
\end{tikzpicture}
in Figure~\ref{figure:main}; if it lies above both, match probability falls with both types as depicted by south-west arrows\begin{tikzpicture}[baseline=-1.6ex]
    \draw[->] (0,0) -- (-0.3,0);
    \draw[->] (0,0) -- (0,-0.3);
\end{tikzpicture}\,.

\paragraph{Single-peaked surplus and single-peaked wages.} Single-peaked match probabilities coincide with single-peaked expected payoff conditional on \emph{meeting} (e.g., offered wages). In contrast, expected payoffs conditional on \emph{matching} (e.g., accepted wages) need not be single-peaked. This is because low surplus pairs only match when drawing a high shock. And for sufficiently fat-tailed shock distributions (e.g., $\xi\sim Pareto$), the mean realized shock conditional on matching rises faster than surplus falls. 

\begin{dependence}[Single-peaked and single-dipped payoffs]
\label{rmk:sppayoffs}
Suppose that $S_t(x,\cdot)$ is single-peaked and define $\gamma_t(z):=\int_z^{\bar\xi_t}\bar\Xi_t(u)\,du$.  Then the conditional match payoffs $\Pi_t^X(\cdot\mid x)$ are single-peaked if $\gamma_t$ is log-concave and single-dipped if $\gamma_t$ is log-convex.\footnote{Log-concavity (log-convexity) of $\gamma$ is implied by log-concavity (log-convexity) of the density $\Xi'$, with the log-convex implication requiring $\Xi'(\bar{\xi})=0$ \citep{prekopa1973logarithmic, bagnoli2005log}. Common log-concave densities include the normal, logistic, and uniform; the Pareto density is log-convex.}\
\end{dependence}

This result shows that the theoretical prediction regarding single-peakedness of wages in partner type %---a standard prediction of models without shocks \citep{shimer2000assortative, eeckhout2011identifying, lopesdemelo2018}---
is sensitive to the shock distribution.  Exponential shocks, $\xi\sim \exp(\kappa)$, is the boundary case where the integrated survival function is simultaneously log-concave and log-convex. Expected payoffs conditional on matching are thus exactly constant in partner type, $\Pi_t^X(y|x)=V^X_t(x)+\alpha^X/\kappa$. Thus, exponential shocks capture the case where observed worker wages do not vary with firm productivity type, conditional on worker type. Suggestively, \cite{bonhomme2019distributional} (cf.\ Figure~2) find that firm productivity class has little effect on log wages, conditional on worker type, for all but the most mismatched worker--firm pairs.

\subsection{A Threshold for Assortative Contours}

Which probability contours retain the lattice geometry of the matching set? Under the full set of conditions used by \cite{shimer2000assortative}, a single threshold separates lattice from non-lattice contours. Specifically, we assume: (i) $f$ is supermodular with $f_{12}\not \equiv 0$; (ii) $f_1$ and $f_2$ are log supermodular; (iii) $f_{12}$ is log supermodular; and (iv) output is maximal at like-type boundary pairs, $f(0,0)\geq f(x,0),f(0,y)$ and $f(1,1)\geq f(x,1),f(1,y)$ for all $x,y$. Conditions~(i)--(iii) give single-crossing and single-peakedness. Condition~(iv) anchors the peak correspondences at the two corners.\footnote{The boundary condition is restrictive for vertically differentiated technologies. It is satisfied by standard specifications such as $f(x,y)=xy$ and $f(x,y)=1-(x-y)^2$.} 

\begin{theorem}[A threshold for assortative contours]
\label{theorem:lattice}
Denote $\mathfrak P_t(x):=\max_y m_t(x,y).$ Under conditions~\textup{(i)}--\textup{(iv)},
\begin{enumerate}[label=(\roman*)]
 \item $U_t(p)$ is a lattice for every
 $p\leq\min_x\mathfrak P_t(x)$;
 \item $U_t(p)$ fails to be a lattice for almost every
 $p\in\bigl(\min_x\mathfrak P_t(x),\max_x\mathfrak P_t(x)\bigr)$.
\end{enumerate}
\end{theorem}

Because $m_t(x,y)$ is a strictly increasing transformation of $S_t(x,y)$, Theorem~\ref{theorem:lattice} is equivalently a statement about surplus contours. Let $\widehat k_t:=\min_x\max_y S_t(x,y)$. Then every upper contour set $\{(x,y):S_t(x,y)\geq k\}$ is a lattice for $k\leq\widehat k_t$, whereas for almost every $k\in(\widehat k_t,\max_{x,y}S_t(x,y))$ it is not. This makes the connection to \cite{shimer2000assortative} immediate. Without pair-specific shocks, matching is binary and the only behaviorally relevant contour is $k=0$. If every type has a non-empty matching set, then $\widehat k_t\geq0$, so the $k=0$ contour lies in the lattice region. Pair-specific shocks make the higher-surplus contours behaviorally visible
by mapping different surplus levels into different match probabilities.
For a given surplus function, this transformation preserves the ordering
and geometry of its contours; the equilibrium surplus function itself may
vary with the shock distribution. This conclusion holds whenever matching sets are non-empty, without imposing stationarity or symmetry of the two populations. Theorem~\ref{theorem:lattice} therefore also extends the lattice result of \cite{shimer2000assortative} to non-stationary environments and asymmetric populations.

A useful feature of the threshold characterization is that it has an empirical counterpart. As discussed in Section~\ref{sec:ordinal-geometry}, when matching flows and unmatched type masses are observed by type, meeting-contingent match probabilities are identified up to the common factor $\lambda_t$. This normalization rescales the probability threshold but leaves unchanged which contours lie in the lattice and non-lattice regions. Hence, the threshold can be assessed directly from matching-flow data.

The theorem also identifies a necessary equilibrium condition for P-PAM. If peak match probabilities differ across types, $\mathfrak{P}_t(x)\neq \mathfrak{P}_t(x')$, then weak P-PAM—and hence P-PAM—must fail. Equivalently, P-PAM can arise in equilibrium only if all types attain the same maximal meeting-contingent match probability.\footnote{This is an equilibrium implication of the present TU search model, not a structural implication of the definition of P-PAM.} Can peak match probabilities be common across types—and if not, what feature of search prevents this?

\section{Sorting Without Reciprocity}
\label{sec:reciprocity}

The previous section showed that two Beckerian ordinal properties survive search---single-crossing and single-peakedness. The remaining property is reciprocity. To recover the full Beckerian ordinal structure, and hence P-PAM, we need strict reciprocity; to recover only the lattice geometry, and hence weak P-PAM, we need weak reciprocity. We now show that, for a broad class of output functions, search frictions break both reciprocity requirements, thereby isolating reciprocity as the distinctive departure from Beckerian ordinal sorting over partners. 

The way search breaks reciprocity is through consensus in partner rankings. There exists a pair $(\bar x,\bar y)$ such that a non-degenerate interval of types on side $X$ has $\bar y$ as its unique surplus-maximizing partner, while a non-degenerate interval on side $Y$ has $\bar x$ as its unique surplus-maximizing partner. Consensus is the opposite of reciprocity: many types point to the same surplus-maximizing partner, but that partner points back only to $\bar x$ or $\bar y$. Peak reciprocity therefore fails. And because consensus occurs on both sides, one can choose $x<\bar x$ and $y<\bar y$ so that $y<\underline y_t(x)=\bar y$ and $x<\underline x_t(y)=\bar x$, which violates weak reciprocity as well.

As a preliminary intuition, consider increasing output when meetings are very rare. Search values then become small, so surplus approaches output and many types rank the most productive partner first: consensus replaces reciprocity. Theorem~\ref{thm:neverPAM} shows that this phenomenon is much broader. It arises away from the rare-meeting limit and does not rely on monotone output; even when output rankings are themselves reciprocal—for example, each type produces most with an equally ranked partner—search generates consensus and violates weak reciprocity.

We now state the general result. Call $(\hat x,\hat y)$ a \emph{regular common productivity peak} if $\hat y$ uniquely maximizes $f(x,\cdot)$ for every $x$, $\hat x$ uniquely maximizes $f(\cdot,y)$ for every $y$, and the two peaks are regular. At a boundary, regularity means $(2\hat x-1)f_1(\hat x,\hat y)>0$; at an interior peak it means $f_{11}(\hat x,\hat y)<0$, and symmetrically for $\hat y$.

\begin{theorem}[Sorting without reciprocity]
\label{thm:neverPAM}
Let $f\in\mathcal C^2$. Suppose either of the following holds.
\begin{enumerate}[label=(\roman*),leftmargin=2em,itemsep=.5em]
\item \emph{Productive disadvantage.} There is a regular common
productivity peak
$(\hat x,\hat y)\notin\{(0,1),(1,0)\}$, and $f$ exhibits peak
increasing differences there:
\[
 f(\hat x,\hat y)+f(x,y)
 \geq f(x,\hat y)+f(\hat x,y)
 \qquad\text{for every }x,y.
\]
\item \emph{Positional disadvantage.} There is a boundary type
$b\in\{0,1\}$ such that
$f(b,b)\geq f(x,b),f(b,y)$ for every $x,y$, and
\[
\begin{aligned}
 (2b-1)f_1(x,b)&>(2b-1)f_1(x,z),\\
 (2b-1)f_2(b,y)&>(2b-1)f_2(z,y)
\end{aligned}
\qquad\text{for every }x,y\text{ and }z\neq b.
\]
\end{enumerate}
For every equilibrium and every date $t$ at which search remains active, let $(\bar x,\bar y)=(\hat x,\hat y)$ under \textup{(i)} and $(\bar x,\bar y)=(b,b)$ under \textup{(ii)}. Then there are non-degenerate relative intervals $I_t^X\ni\bar x$ and $I_t^Y\ni\bar y$ such that
\[
 Y_t^\star(x)=\{\bar y\}\quad\text{for every }x\in I_t^X,
 \qquad
 X_t^\star(y)=\{\bar x\}\quad\text{for every }y\in I_t^Y.
\]
Consequently, $S_t$ fails weak reciprocity, and neither weak P-PAM nor P-PAM obtains.
\end{theorem}

The strength of Theorem~\ref{thm:neverPAM} lies in its uniformity across equilibria: for either class of output functions, weak reciprocity fails in every equilibrium whenever search remains active, regardless of how small search frictions are or how types are distributed. The theorem also answers the question posed at the end of the previous section: under either condition, search cannot preserve a common peak match probability across types. Indeed, for any
$x\in I_t^X\setminus\{\bar x\}$,
\[
\max_y m_t(x,y)=m_t(x,\bar y)<m_t(\bar x,\bar y)=\max_y m_t(\bar x,y).
\]
Thus peak match probabilities are non-constant. Whenever the conditions of Theorem~\ref{theorem:lattice} hold, the interval of thresholds over which almost every contour fails to be a lattice is therefore non-empty.

Both theorem cases are satisfied by the standard case of strictly increasing and strictly complementary output, $f_1>0$, $f_2>0$, and $f_{12}>0$. The two conditions nevertheless differ. Condition~(i) allows the common output-maximizing types to be interior, as with $f(x,y)=e^{-(x-1/2)^2-(y-1/2)^2}$, whereas condition~(ii) does not. Condition~(i) can hold under additive separability, whereas condition~(ii) cannot. Conversely, condition~(ii) allows output-maximizing partners to differ across types, as with $f(x,y)=1-(x-y)^2$, whereas condition~(i) does not.

The two conditions capture two distinct mechanisms behind the failure of weak reciprocity: \emph{productive disadvantage} and \emph{positional disadvantage}. The following two subsections develop each mechanism in turn and prove the corresponding part of the theorem.

\subsection{Productive Disadvantage}
\label{subsec:productive-disadvantage}

In a frictionless transferable-utility market, absolute productivity is fully absorbed by equilibrium payoffs, so only complementarity matters for sorting. Random search breaks this offset. A more productive agent raises match output, but delay prevents the agent's value of search---its threat point under Nash bargaining---from rising one-for-one with that contribution. Productive types therefore become relatively cheap partners. We first isolate this force without complementarity and then reintroduce complementarity.

\paragraph{Additive output.} Let $f(x,y)=g(x)+h(y)$, with $g,h\in\mathcal C^1$. Here mimicking via Corollary~\ref{cor:TU-mimicking} pins down the slope of surplus:
\begin{align*}
    D_1 S_t(x,y)
    &=g'(x)-{V'}_t^X(x) \\
    &=g'(x)-\int_0^1 g'(x)Q_t^X(dy\mid x)
    =g'(x)\Big(1-Q_t^X([0,1]\mid x)\Big).
\end{align*}
In the frictionless benchmark, $Q_t^X([0,1]\mid x)=1$, so absolute productivity is fully absorbed by equilibrium values. Under search, $Q_t^X([0,1]\mid x)<1$, so more productive individuals cannot fully claim the greater output they create when matched.\footnote{This role of delay is absent with explicit search cost. In \cite{atakan2006assortative}, equilibrium
satisfies $\int [s(x,y)]_+\,dG(y)=2c$ for every $x$. With additive output, $s(x,y)=a(x)+a(y)$; since the left-hand side is strictly increasing in $a(x)$, $a(x)$, and hence surplus, is constant across types.} Hence $D_1S_t(x,y)$ has the sign of $g'(x)$. This immediately yields the direction of sorting:

\begin{proposition}\label{prop:additiveq}
Without complementarity, $f(x,y)=g(x)+h(y)$, match probabilities are higher for more productive types:
\[
\operatorname{sign}\big(D_1 m_t(x,y)\big)=\operatorname{sign}\big(g'(x)\big) \quad \text{ and } \quad \operatorname{sign}\big(D_2 m_t(x,y)\big)=\operatorname{sign}\big(h'(y)\big).
\]
In particular, if $g$ and $h$ are strictly increasing, then
$X_t^\star(y)=\{1\}$ and $Y_t^\star(x)=\{1\}$ for every $x,y$. 
\end{proposition}

%\begin{proposition}\label{prop:additiveq}
%If $f(x,y)=g(x)+h(y)$, then for all $(x,y)$,
%\[
%\operatorname{sign}\big(D_x m_t(x,y)\big)=\operatorname{sign}\big(g'(x)\big).
%\]
%\end{proposition}%not high, the more productive the partner's type, the higher the proability of matching.

Thus all agents match up: the more productive the partner, the higher the probability of matching. This may seem counterintuitive, since higher types cannot match with everyone. The reason is that higher types match faster. Delaying a match is relatively more costly for them, so they accept more matches. One can think of a housing market in which properties with more desirable features (e.g., cul-de-sac, south-facing garden) sell quickly and are rarely available, leaving a larger share of less desirable but same category properties on the market at any point in time.

Matching up implies that match probabilities increase with both agents’ types, so level lines of $m_t(x,y)$ (pairs of types that yield the same match probability) are decreasing. This is not weak P-PAM but weak P-NAM, \emph{probabilistic negative assortative matching}, in that match probabilities for re-ordered types, $m_t(x,1-y)$, satisfy weak P-PAM.

\begin{corollary}\label{corr:add:P-nam}
Let $g,h$ be nonconstant and set
$f(x,y)=g(x)+h(y)$. Then equilibrium matching satisfies weak P-PAM (but not P-PAM) if $g'(x)h'(y)\leq 0$ $\forall x,y$ and weak P-NAM (but not P-NAM) if $g'(x)h'(y)\geq 0$.\footnote{Corollary~\ref{corr:add:P-nam} provides intuition for why Theorem~\ref{thm:neverPAM}, Item~(i), excludes the opposite-corner productivity peaks $(0,1)$ and $(1,0)$. In the additive case, when productivity is increasing on one side and decreasing on the other, the common productivity peaks lie at one of these opposite corners, and weak P-PAM obtains. Thus, these cases cannot be included in a general result establishing failure of weak reciprocity.}
\end{corollary}

\begin{result}\label{rem:shimer-smith-3}
Corollary~\ref{corr:add:P-nam} applies directly to the shockless case. Whenever matching sets are non-empty, $f(x,y)=x+y$ satisfies the lattice criterion for negative assortative matching, whereas $f(x,y)=x-y$ satisfies the lattice criterion for positive assortative matching. Neither technology satisfies the sufficient conditions for assortative matching in \cite{shimer2000assortative}. We stress that complementarity is not necessary for the binary matching set to have an assortative lattice orientation; hence, the corresponding weak P-PAM/P-NAM result does not rely on pair-specific heterogeneity.
\end{result}

The ordinal ranking is shock-robust; its classification by MLRP is not.

\begin{dependence}\label{Remark:additiveseparable-logsupermodular}
Suppose $f(x,y)=g(x)+h(y)$ with $g$ and $h$ increasing. Then matching
flows satisfy reverse MLRP if the survival function $1-\Xi_t$ is strictly
log-concave, and MLRP if $1-\Xi_t$ is log-convex. Thus, with the same output technology---and the same partner rankings---the MLRP-based classification of sorting can reverse solely with the shock distribution.
\end{dependence}
We note that for a broad class of shock distributions used in empirical work (e.g., logistic in \cite{LamadonLiseMeghirRobin2024}), the survival function is log-concave. As a result, standard measures of assortative matching based on match flows also predict negative assortative matching for all increasing additively separable output functions.

\begin{figure}[!htbp]
  \centering
  \begin{subfigure}[t]{0.49\textwidth}
    \includegraphics[width=\linewidth,trim=4cm 0cm 4cm 0cm,clip]
    {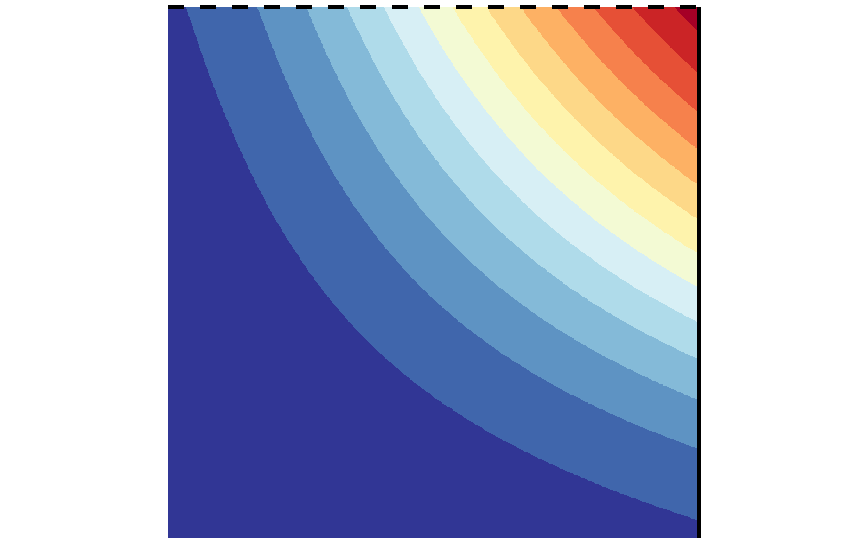}
    \caption{  shocks \ding{51}  }
    \label{fig:addititve:shock}
  \end{subfigure}%
  \begin{subfigure}[t]{0.49\textwidth}
    \includegraphics[width=\linewidth,trim=4cm 0cm 4cm 0cm,clip]
    {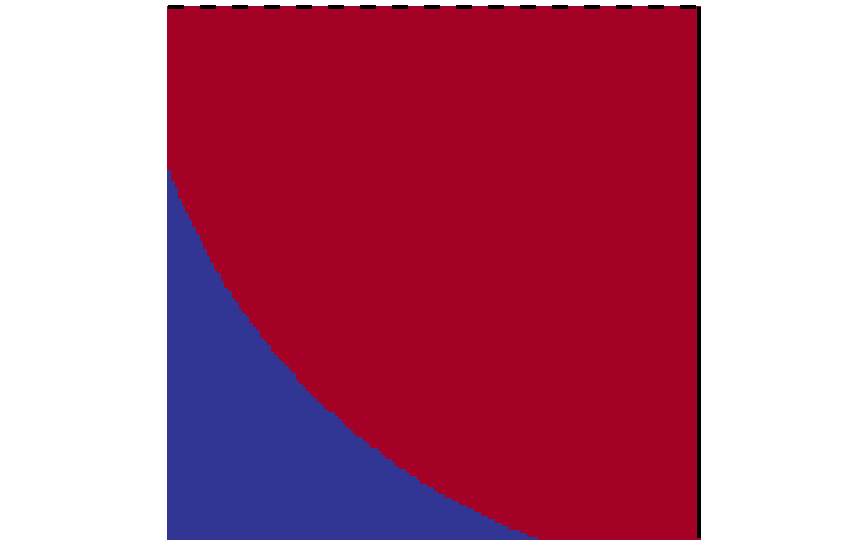}
    \caption{ shocks \ding{55}} 
    \label{fig:addtitive:noshock}
  \end{subfigure}
  \caption{$f(x,y)=x+y$ -- Sorting without complementarity}
\caption*{\footnotesize
\textit{Notes:} The dashed line represents $y^\star(x)$ and the solid line $x^\star(y)$. Warmer colors indicate a higher probability of matching. The left panel shows logistic shocks, and the right panel shows no shocks. We numerically solve for the stationary equilibrium, discount rate $\rho=0.05$, symmetric and uniform population.
}
\label{fig:PPAM_x+y}
\end{figure}

\paragraph{With complementarity.} Complementarity pulls agents toward similarly ranked partners, while productive disadvantage pulls them toward more productive partners. The next proposition gives a sufficient condition under which the latter force dominates locally around a common productivity peak. Applying it symmetrically to both populations yields Theorem~\ref{thm:neverPAM}, Item~(i). We state the theorem under simpler but slightly stronger uniformity conditions for expositional clarity; here, instead, we impose only the conditions strictly needed for the result, making the proof more transparent.

\begin{proposition}[Productive disadvantage]
\label{proposition:productive-disadvantage}
Suppose %that $f\in\mathcal C^1$ and 
that there exists
$(\hat x,\hat y)\in[0,1]^2$ such that:

\begin{enumerate}[label=(\roman*)]
\item $\hat{y}$ is a common peak productive type, strictly so for $\hat{x}$:
\[
\begin{aligned}
f(x,\hat{y})&\geq f(x,y) &&\text{for every }x,y,\\
f(\hat{x},\hat{y})&>f(\hat{x},y) &&\text{for every }y\neq \hat{y}.
\end{aligned}
\]

\item $f$ exhibits peak increasing differences at $(\hat x,\hat y)$:
\[
    f(\hat x,\hat y)+f(x,y)
    \geq
    f(x,\hat y)+f(\hat x,y)
    \qquad
    \text{for every }x,y;
\]
\item the peak-advantage ratio
\[
    \chi^X(x)
    :=
    \inf_{y\in[0,1]\setminus\{\hat y\}}
    \frac{
        f(x,\hat y)-f(x,y)
    }{
        f(\hat x,\hat y)-f(\hat x,y)
    }
\]
is finite in a relatively open neighborhood of $\hat x$ and continuous at $\hat x$.
\end{enumerate}
Then at every date $t$,
\[
    \chi^X(x)> Q_t^Y([0,1]\mid\hat y)
    \quad\Longrightarrow\quad
    Y_t^\star(x)=\{\hat y\}.
\]
In particular, there exists a non-degenerate relatively open interval $I_t^X\subseteq[0,1]$ containing $\hat x$ such that \(Y_t^\star(x)=\{\hat y\}\) for every $x\in I_t^X.$
\end{proposition}
\begin{proof}
Fix a date $t$ and write
\(
    q_t^Y:=Q_t^Y([0,1]\mid\hat y)<1.
\)
By item~(i), the denominator in the definition of $\chi^X$ is strictly positive for every $y\neq\hat y$. Moreover,
\(
    \chi^X(\hat x)=1.
\)

Fix a type $x$ such that $\chi^X(x)> q_t^Y$. Suppose, toward a contradiction, that some $y'\neq\hat y$ weakly improves upon $\hat y$:
\(
    S_t(x,y')\geq S_t(x,\hat y).
\)
Rearranging and applying the mimicking argument with type $y'$ mimicking type $\hat y$ gives
\begin{align*}
f(x,\hat y)-f(x,y')
&\leq 
V_t^Y(\hat y)-V_t^Y(y')\leq
\int_0^1
\bigl[
    f(z,\hat y)-f(z,y')
\bigr]
Q_t^Y(dz\mid\hat y)\leq
q_t^Y
\bigl[
    f(\hat x,\hat y)-f(\hat x,y')
\bigr].
\end{align*}
The final inequality follows from peak increasing differences. On the other hand, (i) and the definition of $\chi^X$ gives
\[
    f(x,\hat y)-f(x,y')\geq 
    \chi^X(x)
    \bigl[
        f(\hat x,\hat y)-f(\hat x,y')
    \bigr]
    >q_t^Y
    \bigl[
        f(\hat x,\hat y)-f(\hat x,y')
    \bigr],
\]
a contradiction. Hence $\{\hat y\}= Y_t^\star(x)$.

Finally, $\chi^X(\hat x)=1>q_t^Y$. Continuity of $\chi^X$ at $\hat x$ therefore yields a non-degenerate relative interval $I_t^X\ni\hat x$ on which $\chi^X(x)>q_t^Y$. The uniqueness conclusion follows on that interval.
\end{proof}

\subsection{Positional Disadvantage}
\label{subsec:positional-disadvantage}

\begin{figure}[!htbp]
\centering
\includegraphics[width=.70\textwidth]{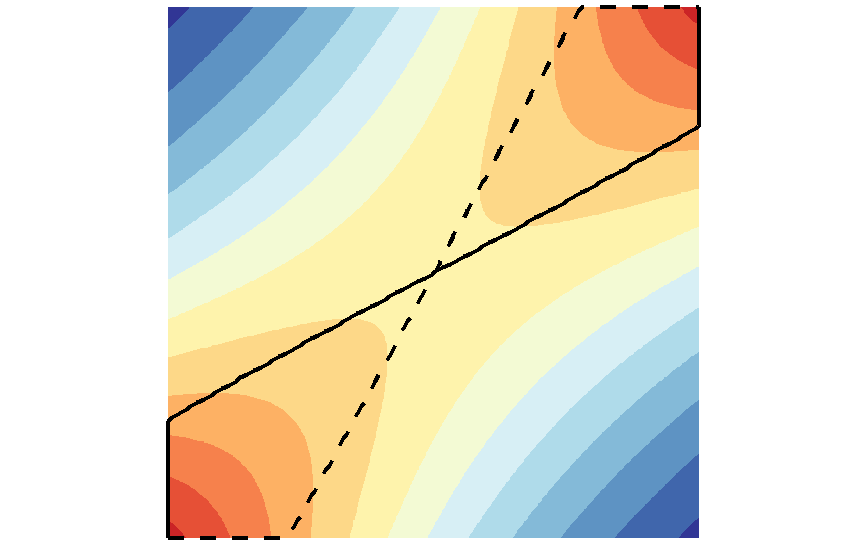}
\caption{Positional disadvantage under horizontal differentiation}
\caption*{\footnotesize
\textit{Notes:} Output is $f(x,y)=1-(x-y)^2$. The dashed line represents $y_t^\star(x)$ and the solid line $x_t^\star(y)$; warmer colors indicate a higher meeting-contingent match probability. The figure reports a stationary equilibrium with discount rate $\rho=0.05$, symmetric uniform populations, and logistic shocks.}
\label{figure:main:horizontal}
\end{figure}

Productive disadvantage cannot explain all failures of reciprocity. For horizontally differentiated output such as $f(x,y)=1-(x-y)^2$, each type produces most with an equally ranked partner. Yet Figure~\ref{figure:main:horizontal} shows that types near a boundary attain their highest surplus with the boundary type rather than with their equal.

We now provide a second, distinct explanation why reciprocity fails: Outside options of agents with few suitable partners, i.e., the most specialized or differentiated agents, deteriorate faster when search frictions rise. A type close to the top has potential partners below but none above; symmetrically for a type near the bottom. As a result, boundary types have fewer outside options and thus are willing to accept less favorable terms to match. Hence, agents slightly away from the boundary, who still have access to partners on both sides, are more likely to match with the boundary type than with an equally ranked partner. Applied symmetrically to both populations, this pattern violates weak reciprocity. We state the result for population $X$ only; applying it symmetrically to population $Y$ yields Theorem~\ref{thm:neverPAM}, Item~(ii).

\begin{proposition}[Positional disadvantage at a boundary]
\label{proposition:positional-disadvantage}
Suppose $f\in\mathcal C^1$ and there is $b\in\{0,1\}$ such that $f(b,b)\geq f(x,b)$ for every $x$ and
\[
 (2b-1)f_1(x,b)>(2b-1)f_1(x,z)
 \qquad\text{for every }x\text{ and }z\neq b.
\]
At every date $t$ at which type $b$ has a positive probability of a future match, there is a non-degenerate relative interval $I_t^Y\ni b$ such that $X_t^\star(y)=\{b\}$ for every $y\in I_t^Y$.
\end{proposition}

The boundary output condition puts $b$ among type $b$'s most productive partners. Boundary complementarity and active search make it the unique surplus-maximizing partner and tilt surplus strictly toward $b$ in a neighborhood of $(b,b)$. Upper hemicontinuity then extends this choice to nearby partner types. When the symmetric conditions hold for population $Y$, the proposition gives Theorem~\ref{thm:neverPAM}\textup{(ii)}.

Positional disadvantage is not intrinsically a boundary phenomenon. Any type with few suitable partners may face it, including an interior type in a thin part of the search pool. Boundary types are useful because the conclusion does not depend on the distribution of types. With vertically differentiated output such as $f(x,y)=xy$, positional disadvantage pulls near-bottom types downward and can offset productive disadvantage there; near the top, the two forces reinforce one another (cf. Figure \ref{figure:main}).

\subsection{Matching up}

With strictly increasing and strictly complementary output, productive and positional disadvantage both operate at the top. Theorem~\ref{thm:neverPAM} therefore implies that a non-degenerate interval of types has the top type as its unique surplus-maximizing partner. To obtain a primitive lower bound on the size of this interval, we use the productive-disadvantage argument, which yields a convenient expression in terms of search frictions, productive advantage, and complementarity.

%The next result, a consequence of the explicit bounds $\chi^X,\chi^Y$ from Proposition \ref{proposition:productive-disadvantage}, quantifies the size of the resulting interval as a function of search frictions, absolute productive advantage, and complementarity.

\begin{corollary}[Quantitative impact of search frictions] 
\label{cor:quantifying}
Suppose that output $f\in\mathcal C^2$ is weakly increasing and strictly complementary, with $f_1(x,y)\geq 0$, $f_2(x,y)\geq 0$, and $f_{12}(x,y)>0$ for all $x,y$. Let $\overline\lambda^X,\overline\lambda^Y>0$ be upper bounds, from date $t$ onward, on individual meeting rates. Then
\begin{align*}
Y_t^\star(x)=\{1\}
\quad\text{whenever}\quad
x&>1-\frac{\min_y f_2(x,y)}{\max_{z,u} f_{12}(z,u)}\frac{\rho}{\overline{\lambda}^Y},\\
X_t^\star(y)=\{1\}
\quad\text{whenever}\quad
y &>1-\frac{\min_x f_1(x,y)}{\max_{z,u} f_{12}(z,u)}\frac{\rho}{\overline{\lambda}^X}.
\end{align*}
\end{corollary}
The bound separates the two forces. A larger ratio $\rho/\overline\lambda$ or a steeper absolute productivity gradient expands the set of types guaranteed to target the top, while stronger complementarity contracts it. These are comparative statics of the \emph{guaranteed} region; under strong distributional assumptions, our companion paper, \cite{bonnetonsandmann2026limit}, shows that the exact likeliest-partner set is monotone in search frictions. The following example illustrates.

\begin{example} \label{example:output}
Suppose that $f(x,y)=g(x)+h(y)+cxy+\mathrm{const}$, with $g',h'\geq 0$ and $c>0$. Corollary~\ref{cor:quantifying} implies that
\[
Y_t^\star(x)=\{1\}
\qquad\text{whenever}\qquad
x>\frac{c\lambda^Y-\rho\min_y h'(y)}{c(\rho+\lambda^Y)}.
\]
Thus stronger absolute productivity expands the region guaranteed to match up, whereas stronger complementarity contracts it. 
\end{example}

\paragraph{Implications for measured sorting.} The preceding bound concerns partner rankings. The gap between the two likeliest-partner correspondences also has implications for measured sorting.  With log-concave survival, wherever the relevant
derivatives exist,
\begin{equation}
\label{eq:MLRP-upset}
D_{12}^2\log m_t =\left. \frac{d^2}{dz^2}\log\bar\Xi_t(z)\right|_{z=-S_t}
D_1S_tD_2S_t-\left. \frac{d}{dz}\log\bar\Xi_t(z) \right|_{z=-S_t} f_{12}.
\end{equation}
The first coefficient is non-positive and the second is positive. Hence local MLRP necessarily holds when $D_1S_tD_2S_t\leq0$. When $D_1S_tD_2S_t>0$, instead, complementarity and log concavity work in opposite directions: complementarity favors MLRP, while log concavity favors its reversal. Under single-peakedness, this is precisely the region between the two likeliest-partner correspondences. A wider gap therefore enlarges the region in which MLRP may fail.

How large can these effects on measured sorting become? To assess their magnitude, we solve the equilibrium numerically for $f(x,y)=xy$, the technology used by \cite{shimer2000assortative}, to illustrate positive assortative matching. The exercise brings together the different notions of sorting studied above. As meetings become scarcer, more types have the top type as their likeliest partner and the two surplus-maximizing ridges move apart. Consistent with \eqref{eq:MLRP-upset}, local MLRP violations become more widespread, while standard sorting correlations become weak or slightly negative. Table~\ref{tab:xy-ns1-normal} shows how partner rankings, MLRP, and sorting correlations evolve over a cycle, holding output, the shock distribution, and the search-pool composition (via ``clone'' entry replacement as in \cite{burdett1997marriage}) fixed. An animation showing how the match-probability contours and likeliest-partner curves evolve over the cycle is available on the authors’ website.\footnote{\url{https://sites.google.com/view/nicolas-bonneton/p-pam-video}} In this exercise, periods with more frequent meetings tend to exhibit more positive measured sorting. At the same current meeting rate, $\lambda_t/\rho=0.141$, all three standard sorting correlations are negative during cooling and positive during heating. Thus, standard measures of sorting can change sign over the business cycle, or across hot and cold phases of housing and labor markets, purely through changes in meeting opportunities over time.

\begin{table}[!htbp] \
\centering
\small
\caption{Sorting around cycles under $f(x,y)=xy$.}
\label{tab:xy-ns1-normal}
\begin{tabular}{lcccccc}
\toprule
 & \multicolumn{6}{c}{Date in the cycle, $t/T$} \\
\cmidrule(lr){2-7}
 & $0$ & $\tfrac{1}{6}$ & $\tfrac{1}{3}$ & $\tfrac{1}{2}$ & $\tfrac{2}{3}$ & $\tfrac{5}{6}$ \\
 & peak & cooling & cooling & trough & heating & heating \\
\midrule
Meeting rate $\lambda_t/\rho$ &   50 & 7.07 & 0.141 & 0.02 & 0.141 & 7.07 \\
Average match probability $\bar m_t$ & 0.20 & 0.53 & 0.92 & 0.91 & 0.79 & 0.42 \\
\midrule
Kendall's $\tau_t$ & $0.508$ & $0.183$ & $-0.020$ & $-0.012$ & $0.063$ & $0.332$ \\
Spearman's $\rho^S_t$ & $0.703$ & $0.274$ & $-0.029$ & $-0.018$ & $0.094$ & $0.484$ \\
Pearson $r_t$ & $0.773$ & $0.348$ & $-0.029$ & $-0.017$ & $0.105$ & $0.550$ \\
\midrule
Types with $y_t^\star(x)=1$ & 0.27 & 0.67 & 0.98 & 0.96 & 0.82 & 0.45 \\
Share of type pairs with $D^2_{12}\log m_t<0$ & 0.03 & 0.32 & 0.52 & 0.51 & 0.44 & 0.15 \\
\bottomrule
\end{tabular}
\caption*{\footnotesize \textit{Notes:} All entries come from a non-stationary equilibrium, read at equally spaced dates around one cycle. One population, symmetric; types uniform on $[0,1]$ with $\mu_t(x)=1$ constant in $t$, with entry offsetting match exits so that the unmatched type distribution remains constant over time; $f(x,y)=xy$; $\alpha^X=\alpha^Y=0.5$; $\rho=0.05$; shocks i.i.d.\ normal, mean $0$, s.d. $\sigma=0.035$. $\lambda_t=\exp\{\bar\ell+A\cos(2\pi t/T)\}$ with $T=200$ ($\rho T=10$), sweeping $\lambda/\rho\in[0.02,50]$. The exercise is illustrative rather than calibrated.}\label{table1}
\end{table}

\section{Discussion}

Throughout the paper, we study sorting through the contours of match surplus---or, equivalently, meeting-contingent match probabilities. The same object also appears naturally in applied work: a number of articles estimate and plot match surplus (e.g., \cite{lise2016matching,gulyas2020firm,lacava2023matching,LamadonLiseMeghirRobin2024,schulz2024labor,chiappori2026assortative}). Looking at surplus directly reveals features of equilibrium sorting---including each type's surplus-maximizing partner and the slopes of level lines---that cannot generally be summarized by a single statistic. Our results provide a language for interpreting this geometry in terms of the economic forces that generate it.

\begin{figure}[!htbp]
\centering
\includegraphics[scale=0.42]{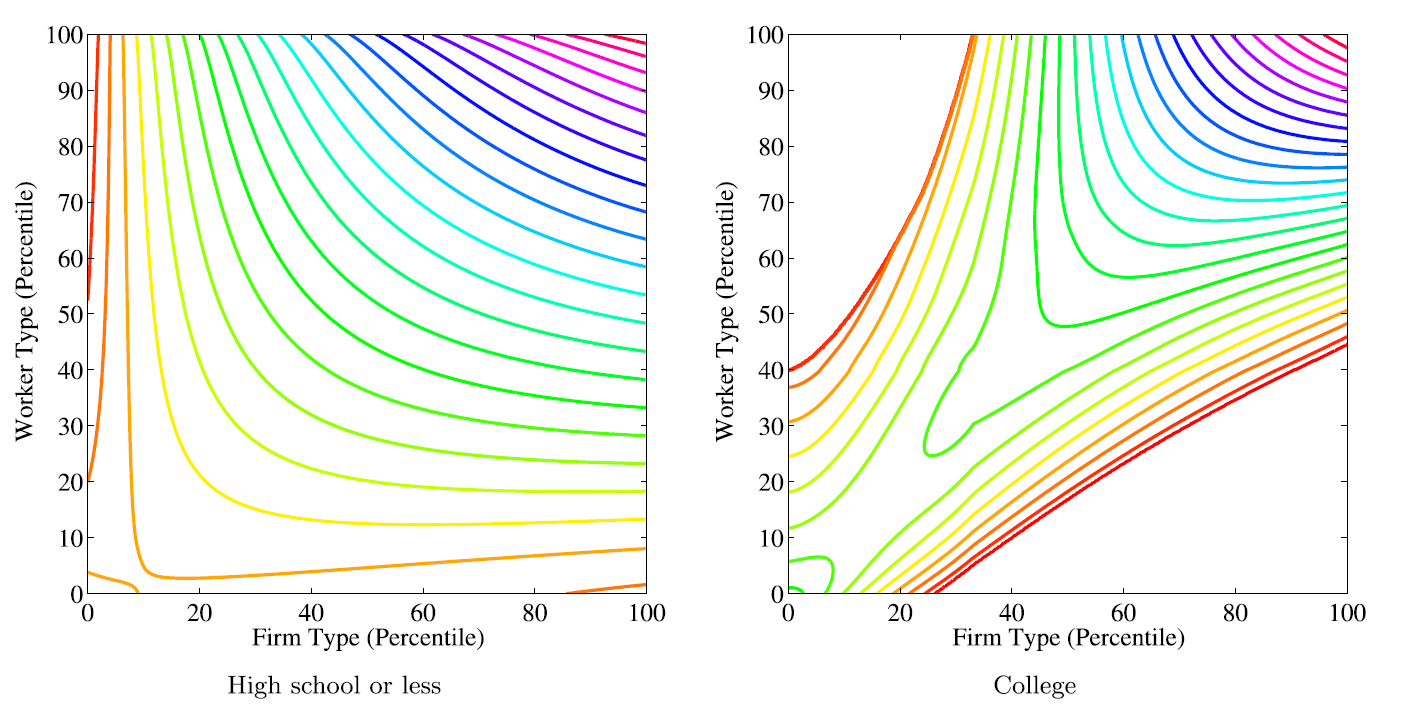}
\caption{\textit{Estimated equilibrium surplus contours in \cite{lise2016matching}.}}\caption*{\footnotesize Notes: The left panel corresponds to workers with a high-school education or less; the right panel corresponds to college graduates. Source: Figure~3 of \cite{lise2016matching}.}
\label{fig}
\end{figure}

To illustrate this claim, consider Figure~\ref{fig}, which reproduces the estimated equilibrium surplus contours of the related search-and-matching model of \cite{lise2016matching}. In both panels, surplus is single-peaked in partner type, consistent with Proposition~\ref{mconvex}. Near the boundaries, highest-surplus partner types are boundary types, consistent with positional disadvantage. Away from the boundaries, highest-surplus partner types are systematically higher than Becker's assortative assignment, consistent with productive disadvantage and the resulting tendency to ``match up.'' Overall, the departure from the frictionless benchmark is much more pronounced in the high-school market, a difference that our results show can arise from weaker complementarity, stronger search frictions, or both. \cite{lise2016matching} support the first channel: they estimate essentially no production complementarity for high-school workers, but strong complementarity for college graduates.\footnote{They also estimate a lower job-contact rate for unemployed high-school workers than for college graduates (17.1 versus 23.1 percent per month), which, on this margin, points in the same direction.}

The figure also illustrates what would be lost by summarizing sorting with a single measure of assortative matching (e.g., MLRP or correlations). The high-school panel is close to our additive benchmark: \cite{lise2016matching} estimates essentially no production complementarity, and almost all surplus-level lines are downward-sloping. Yet, as we have shown, this pattern is compatible with positive, zero, or negative measured sorting.\footnote{With $f(x,y)=h(x)+g(y)$ with both functions increasing, matching flows satisfy MLRP under Pareto shocks, are log-modular under exponential shocks, and are log-submodular under logistic shocks. Recall that MLRP implies nonnegative Pearson, Kendall, and Spearman correlations across matched pairs.} This disconnect is important because measures of assortative matching are often used to learn about complementarities in production (e.g., \citealp{eeckhout2011identifying,bagger2019empirical}). Our analysis therefore delineates which features of measured sorting are driven by the ordinal geometry of surplus and which instead depend on distributional assumptions about unobserved heterogeneity as captured by shocks.

\begingroup
\setstretch{1}\small
\setlength{\bibsep}{0pt plus 0.3ex}
\bibliographystyle{apalike}
\bibliography{a}
\endgroup

\appendix

\section{The Ordinal Geometry of Probabilistic Matching}

\subsection{From Lattice to Empirical Measures: MLRP}

\begin{proof}[\textbf{Proof of Theorem~\ref{thm:lattice-mlrp-logconcave}}]
Fix $x_2>x_1$ and $y_2>y_1$. For this rectangle, the requirement that every upper contour set of $S_t$ be a lattice is equivalent to
\[
\min\bigl\{S_t(x_1,y_1),S_t(x_2,y_2)\bigr\}
\geq \min\bigl\{S_t(x_1,y_2),S_t(x_2,y_1)\bigr\}.
\]
\medskip
\noindent
\emph{We use log-concavity of the survival function to prove
$(i)\Rightarrow(ii)$.}

Suppose that all upper contour sets of $S_t$ are lattices and that $\bar\Xi_t$ is log-concave. We claim that
\begin{align}\label{eq:logcavsurv}
&\log\bar\Xi_t\bigl(-S_t(x_1,y_1)\bigr) +\log\bar\Xi_t\bigl(-S_t(x_2,y_2)\bigr)\geq \log\bar\Xi_t\bigl(-S_t(x_1,y_2)\bigr) +\log\bar\Xi_t\bigl(-S_t(x_2,y_1)\bigr).
\end{align}
\emph{Proof of inequality \eqref{eq:logcavsurv}}: Without loss, suppose that $S_t(x_1,y_2)\leq S_t(x_2,y_1).$ The lattice condition then implies that $S_t(x_1,y_1)\geq S_t(x_1,y_2)$ and $S_t(x_2,y_2)\geq S_t(x_1,y_2).$ If, moreover, $S_t(x_2,y_1)<S_t(x_2,y_2)$, then \eqref{eq:logcavsurv} readily follows. Therefore consider $S_t(x_2,y_1)\geq S_t(x_2,y_2)$. It holds that
\[
\begin{aligned}
&\log\bar\Xi_t\bigl(-S_t(x_1,y_1)\bigr)-\log\bar\Xi_t\bigl(-S_t(x_1,y_2)\bigr)\\
&\quad\geq \log\bar\Xi_t\!\left(-\left[S_t(x_1,y_2)+S_t(x_2,y_1)-S_t(x_2,y_2)\right]\right)-\log\bar\Xi_t\bigl(-S_t(x_1,y_2)\bigr)\\
&\quad\geq \log\bar\Xi_t\!\left( -\left[
S_t(x_2,y_2)+S_t(x_2,y_1)-S_t(x_2,y_2)\right]\right)-\log\bar\Xi_t\bigl(-S_t(x_2,y_2)\bigr)\\
&\quad=\log\bar\Xi_t\bigl(-S_t(x_2,y_1)\bigr)-\log\bar\Xi_t\bigl(-S_t(x_2,y_2)\bigr).
\end{aligned}
\]
The first inequality follows from supermodularity, $S_t(x_1,y_1)\geq S_t(x_1,y_2)+S_t(x_2,y_1)-S_t(x_2,y_2)$, and monotonicity of $s\mapsto \log \bar\Xi_t \bigl(\cdot\bigr)$. For the second inequality, this follows from $S_t(x_2,y_2)\geq S_t(x_1,y_2)$ and concavity of $s\mapsto \log \bar\Xi_t \bigl(\cdot\bigr)$: an increment of equal length, $S_t(x_2,y_1)-S_t(x_2,y_2)\geq 0$, produces a weakly larger increase when it starts from the weakly smaller surplus value. Rearranging the preceding inequalities gives
\eqref{eq:logcavsurv}.

Since $m_t(x,y)=\bar\Xi_t(-S_t(x,y))$, exponentiating yields
\[
m_t(x_1,y_1)m_t(x_2,y_2) \geq m_t(x_1,y_2)m_t(x_2,y_1).
\]
The rectangle was arbitrary, so $m_t$ satisfies MLRP.

\medskip
\noindent
\emph{We prove $(ii)\Rightarrow(i)$ via the contrapositive.}

Suppose that $(i)$ is upset, i.e., some upper contour set of $S_t$ is not a lattice. Then there exist $x_2>x_1$ and $y_2>y_1$ such that $a:=\min\bigl\{S_t(x_1,y_1),S_t(x_2,y_2)\bigr\}<\min\bigl\{S_t(x_1,y_2),S_t(x_2,y_1)\bigr\}=:b.$ For $\kappa>0$, consider the survival function $1-\Xi^\kappa(z)=e^{-e^{\kappa z}}.$ It is a valid survival function because it is non-increasing with range in $[0,1]$. And it is log-concave because $\log\bigl(1-\Xi^\kappa(z)\bigr)=-e^{\kappa z}$ is concave. Under this distribution, $\log m_t(x,y)=-e^{-\kappa S_t(x,y)}.$

Then note the following survival function lower bound 
\begin{align*}
\log m_t(x_1,y_1)+\log m_t(x_2,y_2)  & =-e^{-\kappa S_t(x_1,y_1)}
 -e^{-\kappa S_t(x_2,y_2)} \leq %-e^{-\kappa\min\{S_t(x_1,y_1),S_t(x_2,y_2)\}}=
 -e^{
-\kappa a}
\end{align*}
and the survival function upper bound
\begin{align*}
-2e^{-\kappa b}% =-2e^{-\kappa\min\{S_t(x_1,y_2),S_t(x_2,y_1)\}}
\leq - e^{-\kappa S_t(x_1,y_2)} -e^{-\kappa S_t(x_2,y_1)} & =\log m_t(x_1,y_2)+\log m_t(x_2,y_1).
\end{align*}
To contradict $(ii)$ it hence suffices to show that $-e^{-\kappa a}<-2 e^{-\kappa b}$. Since $b>a$ this holds for $\kappa$ sufficiently large. So MLRP fails under a shock distribution with a log-concave survival function.
\end{proof}

\subsection{P-PAM as a Diagnostic for the Effects of Search}

\begin{proof}[\textbf{Proof of Result \ref{rem:becker}}]
Let $V^X$ and $V^Y$ be stable payoff functions in the frictionless transferable-utility benchmark. Define the net surplus by
\begin{equation}
    S(x,y):= f(x,y)-V^X(x)-V^Y(y).
    \label{eq:Phi}
\end{equation}
Then $S(x,y)\leq 0$ for all $(x,y)$ with equality for matched pairs. Since $f$ is supermodular and search values enter additively into surplus, surplus is equally supermodular and hence satisfies single-crossing.

Next, recall that populations are balanced so that all agents are matched in the frictionless benchmark. Hence, for every $x$ and $y$,
\begin{equation}
    \max_{\tilde y}S(x,\tilde y)=0, \qquad \max_{\tilde x}S(\tilde x,y)=0.
    \label{eq:max-zero}
\end{equation}
We then establish strict reciprocity. Suppose that $x<\underline{x}(y)$ (the case when $x>\overline{x}(y)$ follows a symmetric reasoning). Then $x\notin X^\star(y)$ and therefore $S(x,y)<0$. We claim that every $\tilde y\in Y^\star(x)$ satisfies $\tilde y<y$. Suppose, toward a contradiction, that there exists $\tilde y\in Y^\star(x)$ with $\tilde y\geq y$. If $\tilde y=y$, then $S(x,y)=0$, contradicting $x\notin X^\star(y)$. Hence $\tilde y>y$.

Choose any $\tilde x\in X^\star(y)$. By the premise that $x<\underline{x}(y)$, $\tilde x>x$. We therefore have $x<\tilde x,$ $y<\tilde y,$ $S(x,\tilde y)=0,$ $S(\tilde x,y)=0.$ And supermodularity gives
\begin{equation}
    S(x,y)+S(\tilde{x},\tilde{y})\geq S(x,\tilde{y})+S(\tilde{x},y) =0.
    \label{eq:uncrossing-supermod}
\end{equation}
This yields $S(x,y)=0$, a contradiction to $x<\underline{x}(y)$. Therefore $\tilde y<y$ for every $\tilde y\in Y^\star(x)$, and hence $y>\overline{y}(x)$.

It remains to verify single-peakedness. Fix $x$. Take $y_1<y_2\leq \underline{y}(x)$. (A symmetric reasoning applies for $\overline{y}(x)\leq y_1<y_2$.) If $y_2\in Y^\star(x)$, then $S(x,y_2)=0\geq S(x,y_1)$, so the desired inequality is immediate. Suppose instead that $y_2<\underline{y}(x)$. By strict reciprocity, $x>\overline{x}(y_2)$. Choose $x'\in X^\star(y_2)$. Then $x'<x$ and $S(x',y_2)=0$. Supermodularity applied to $x'<x$ and $y_1<y_2$ gives
\begin{equation}
    S(x,y_2)-S(x,y_1) \geq S(x',y_2)-S(x',y_1)=-S(x',y_1)\geq 0,
\end{equation}
where the last inequality uses $S\leq 0$. Hence $S(x,y_1)\leq S(x,y_2)$.
\end{proof}

\begin{proof}[\textbf{Proof of Theorem~\ref{thm:weakPPAM}}]
Suppose first that weak P-PAM holds. Convexity of individual upper contour sets gives single-peakedness. If weak reciprocity failed with $y<\underline y_t(x)$ and $x<\underline x_t(y)$, both $(x,\underline y_t(x))$ and $(\underline x_t(y),y)$ would have strictly higher match probability than $(x,y)$. At the lower of their two probabilities, the contour would contain both crossed pairs but not their meet. The case $y>\overline y_t(x)$ and $x>\overline x_t(y)$ similarly produces a missing join. Thus weak reciprocity holds.

Conversely, suppose single-peakedness and weak reciprocity hold. Individual upper contour sets are convex. Take $x_1<x_2$ and $y_1<y_2$ with $(x_1,y_2),(x_2,y_1)\in U_t(p)$. If their meet were excluded, then
\[
 m_t(x_1,y_1)<\min\{m_t(x_1,y_2),m_t(x_2,y_1)\}.
\]
Single-peakedness would imply $y_1<\underline y_t(x_1)$ and $x_1<\underline x_t(y_1)$, contrary to weak reciprocity. If their join  were excluded, the analogous inequalities would imply $y_2>\overline y_t(x_2)$ and $x_2>\overline x_t(y_2)$, again a contradiction. Hence every upper contour set is a lattice.

Finally, a strict reversal $m_t(x_1,y_2)>m_t(x_1,y_1)$ and $m_t(x_2,y_1)>m_t(x_2,y_2)$ would make the smaller diagonal probability strictly smaller than both crossed probabilities, contradicting the lattice property. This proves weak single-crossing; the argument with the populations interchanged is identical.
\end{proof}

\begin{proof}[\textbf{Proof of Remark~\ref{remark-weak-strict-reciprocity}}]
Suppose first that strict reciprocity holds. It immediately implies weak reciprocity. To establish peak reciprocity, suppose that $x\in X_t^\star(y)$ but $y\notin Y_t^\star(x)$. Single-peakedness makes $Y_t^\star(x)$ an interval, so either $y<\underline y_t(x)$ or $y>\overline y_t(x)$. Strict reciprocity then implies, respectively, $x>\overline x_t(y)$ or $x<\underline x_t(y)$, contradicting $x\in X_t^\star(y)$. The reverse implication follows symmetrically.

Conversely, suppose that weak reciprocity and peak reciprocity hold. If $y<\underline y_t(x)$, weak reciprocity gives $x\geq\underline x_t(y)$. Were $x\leq\overline x_t(y)$, we would have $x\in X_t^\star(y)$ and hence, by peak reciprocity, $y\in Y_t^\star(x)$, a contradiction. Thus $x>\overline x_t(y)$. Conversely, if $x>\overline x_t(y)$, weak reciprocity gives $y\leq\overline y_t(x)$. Were $y\geq\underline y_t(x)$, peak reciprocity would again imply $x\in X_t^\star(y)$, a contradiction. Hence $y<\underline y_t(x)$. This proves the first equivalence in the definition of strict reciprocity. The second follows by the symmetric argument: $y>\overline y_t(x)$ if and only if $x<\underline x_t(y)$.
\end{proof}

\section{Equilibrium Matching: Which Contours Are Assortative?}

\subsection{Mimicking}

\begin{proof} [ \textbf{Proof of Lemma \ref{lem:TU-mimicking}} ]
We prove the result for population $X$; the argument for population $Y$ is symmetric.

\noindent \emph{Bound on differences in search values:} For any $x',x''$, define
\begin{align*}
    \Pi_\tau^X(y\mid x'';x'):= V_\tau^X(x'')+\alpha^X \mathbb E^\xi\left[S_\tau(x'',y)+\xi \,\middle|\, S_\tau(x',y)+\xi\geq0 \right].
\end{align*}
Thus, $\Pi_\tau^X(y\mid x'';x')$ is type $x''$'s expected payoff when it accepts the same realizations of $\xi$ as type $x'$. Since $\Pi_\tau^X(y\mid x'';x')$ and $\Pi_\tau^X(y\mid x';x')$ condition on the same shock realizations, the conditional shock term cancels when taking the difference:
\begin{align}
\label{eq:mimicked-payoff-difference}
    &\Pi_\tau^X(y\mid x'';x')-\Pi_\tau^X(y\mid x';x') =(1-\alpha^X)\bigl[V_\tau^X(x'')-V_\tau^X(x')\bigr]+\alpha^X \bigl[f(x'',y)-f(x',y)\bigr].
\end{align}
Then consider that some agent type $x{''}$ chose to accept a match with some $y$ for the exact same realizations of $\xi$ that $x{'}$ accepts the match. Because meetings are anonymous, the mimicked rule generates type $x'$'s distribution of match times and partner types. It follows that
\begin{align}
\label{eq:TU-intratemporal-mimicking}
    V_t^X(x'') \geq \int_t^\infty e^{-\rho(\tau-t)} \int_0^1 \Pi_\tau^X(y\mid x'';x') p_{t,\tau}^X(y\mid x')\,dy\,d\tau.
\end{align}
On the other hand, the definition of the value-of-search gives
\begin{align}
\label{eq:value-mimicked-type}
    V_t^X(x') =\int_t^\infty e^{-\rho(\tau-t)}
    \int_0^1 \Pi_\tau^X(y\mid x';x') p_{t,\tau}^X(y\mid x')\,dy\,d\tau.
\end{align}
Subtracting \eqref{eq:value-mimicked-type} from \eqref{eq:TU-intratemporal-mimicking} and using \eqref{eq:mimicked-payoff-difference} gives
\begin{align}
\label{eq:TU-one-step-mimicking}
    V_t^X(x'')-V_t^X(x') \geq & \int_t^\infty e^{-\rho(\tau-t)}\int_0^1 \Big\{ \alpha^X\bigl[f(x'',y)-f(x',y)\bigr] \\
    &\qquad\qquad+ (1-\alpha^X) \bigl[V_\tau^X(x'')-V_\tau^X(x')\bigr] \Big\} p_{t,\tau}^X(y\mid x')\,dy\,d\tau. \notag
\end{align}

We now iterate \eqref{eq:TU-one-step-mimicking}, following the proof in \cite{SandmannBonneton2025}. Write $\overline p_{t_0,t_1}^X(x)
    :=\int_0^1p_{t_0,t_1}^X(y\mid x)\,dy.$ For $k=1,2,\ldots$, set $\tau_0=t$ and define
\begin{align*}
    M_t^{X,[k]}(y\mid x') :=&
    \int_t^\infty \int_{\tau_1}^\infty \cdots \int_{\tau_{k-1}}^\infty e^{-\rho(\tau_k-t)} \alpha^X p_{\tau_{k-1},\tau_k}^X(y\mid x') \\
    &\quad\times (1-\alpha^X)^{k-1} \prod_{\ell=1}^{k-1} \overline p_{\tau_{\ell-1},\tau_\ell}^X(x') \,d\tau_k\cdots d\tau_1, \notag\\[1ex]
    R_t^{X,[k]}(x',x''):=
    & \int_t^\infty \int_{\tau_1}^\infty\cdots \int_{\tau_{k-1}}^\infty e^{-\rho(\tau_k-t)}
    \bigl[
        V_{\tau_k}^X(x'')-V_{\tau_k}^X(x')
    \bigr] \\
    &\quad\times (1-\alpha^X)^k  \prod_{\ell=1}^{k}\overline p_{\tau_{\ell-1},\tau_\ell}^X(x') \,d\tau_k\cdots d\tau_1. \notag
\end{align*}

When $k=1$, the multiple integrals are understood as single integrals over $\tau_1\in[t,\infty)$, and an empty product equals one.

We prove by induction that, for every $k\geq1$,
\begin{align}
\label{eq:TU-mimicking-induction}
    V_t^X(x'')-V_t^X(x') \geq{}&\int_0^1\bigl[f(x'',y)-f(x',y)\bigr] \sum_{\ell=1}^k M_t^{X,[\ell]}(y\mid x')\,dy+R_t^{X,[k]}(x',x'').
\end{align}
For $k=1$, this is precisely \eqref{eq:TU-one-step-mimicking}. Suppose that \eqref{eq:TU-mimicking-induction} holds for $k-1$. Applying \eqref{eq:TU-one-step-mimicking} at time $\tau_{k-1}$ to the value difference appearing in $R_t^{X,[k-1]}(x',x'')$ gives
\begin{align}
\label{eq:R-iteration}
    R_t^{X,[k-1]}(x',x'')\geq\int_0^1 \bigl[f(x'',y)-f(x',y)\bigr] M_t^{X,[k]}(y\mid x')\,dy+R_t^{X,[k]}(x',x'').
\end{align}
Substituting \eqref{eq:R-iteration} into the induction hypothesis proves \eqref{eq:TU-mimicking-induction} for $k$.

It remains to pass to the infinite iteration. Since every future-match density is a subprobability density,
\begin{align*}
    \int_0^1M_t^{X,[k]}(y\mid x')\,dy \leq \alpha^X(1-\alpha^X)^{k-1}.
\end{align*}
Consequently, the non-negative series
\begin{align}
\label{eq:def-Q-infinite}
    Q_t^X(dy\mid x'):=\sum_{k=1}^\infty M_t^{X,[k]}(y\mid x')\,dy
\end{align}
defines a finite measure. Moreover, boundedness of discounted values-of-search and the subprobability property of the future-match densities imply
\begin{align}
\label{eq:R-convergence}
\left|R_t^{X,[k]}(x',x'')\right|
&\leq 2(1-\alpha^X)^k \sup_{\tau\geq t,\,x\in[0,1]} \left\{ e^{-\rho(\tau-t)} \left|V_\tau^X(x)\right| \right\} \longrightarrow0.
\end{align}
Letting $k\to\infty$ in \eqref{eq:TU-mimicking-induction}, using \eqref{eq:R-convergence} and the absolute convergence of the series of measures, gives
\begin{align*}
    V_t^X(x'')-V_t^X(x')\geq \int_0^1\bigl[f(x'',y)-f(x',y)\bigr]Q_t^X(dy\mid x').
\end{align*}

\noindent \emph{Proof of closed-form~\eqref{eq:TU-Q-representation}:} We use the particular measures constructed in the proof of
Lemma~\ref{lem:TU-mimicking}. Recall that
\[
Q_t^X(dy\mid x)=\sum_{k=1}^{\infty}M_t^{X,[k]}(y\mid x)\,dy,
\qquad
h_s^X(x):=\int_0^1\Lambda_s^X(z\mid x)\,dz,
\qquad
H_{t,\tau}^X(x):=\int_t^\tau h_s^X(x)\,ds.
\]
Since
\[
p_{u,v}^X(y\mid x)=\Lambda_v^X(y\mid x)\exp\left\{-\int_u^v h_s^X(x)\,ds\right\}
\]
and
\[
\overline p_{u,v}^X(x)=h_v^X(x)\exp\left\{-\int_u^v h_s^X(x)\,ds\right\},
\]
the survival terms in \(M_t^{X,[k]}\) telescope. Moreover,
\[
\int_{t<\tau_1<\cdots<\tau_{k-1}<\tau}\prod_{\ell=1}^{k-1}h_{\tau_\ell}^X(x)\,d\tau_1\cdots d\tau_{k-1}=\frac{H_{t,\tau}^X(x)^{k-1}}{(k-1)!}.
\]
Consequently,
\begin{align}
\label{eq:TU-M-representation}
M_t^{X,[k]}(y\mid x)={}&
\alpha^X(1-\alpha^X)^{k-1}\int_t^\infty e^{-\rho(\tau-t)}\Lambda_\tau^X(y\mid x)e^{-H_{t,\tau}^X(x)}\frac{H_{t,\tau}^X(x)^{k-1}}{(k-1)!}\,d\tau .
\end{align}
All terms are non-negative, so Tonelli's theorem and the exponential series give \eqref{eq:TU-Q-representation}:
\begin{align*}
\sum_{k=1}^{\infty}M_t^{X,[k]}(y\mid x)
&=\alpha^X\int_t^\infty e^{-\rho(\tau-t)}\Lambda_\tau^X(y\mid x)e^{-H_{t,\tau}^X(x)}\exp\bigl\{(1-\alpha^X)H_{t,\tau}^X(x)\bigr\}\,d\tau\\
&=\alpha^X\int_t^\infty e^{-\rho(\tau-t)} \Lambda_\tau^X(y\mid x)\exp\bigl\{-\alpha^XH_{t,\tau}^X(x)\bigr\}\,d\tau.
\end{align*}

\noindent \emph{Subprobability bound:}
By \eqref{eq:TU-Q-representation}, $Q_t^X([0,1]\mid x')$ is the discounted first-arrival mass of the auxiliary inhomogeneous Poisson process with intensity $\alpha^X\int_0^1\Lambda_\tau^X(y\mid x')\,dy$. Its locally integrable intensity rules out an arrival at $t$, so $\rho>0$ makes this mass strictly less than one. Equivalently, integration by parts gives
\begin{align}
\label{eq:for-corollary-3}
 Q_t^X([0,1]\mid x')=1-\rho\int_t^\infty e^{-\rho(\tau-t)}\exp\!\left\{-\alpha^X\int_t^\tau\int_0^1\Lambda_s^X(z\mid x')\,dz\,ds\right\}d\tau<1,
\end{align}
where the strict inequality follows because the integrand is positive at every finite $\tau$.
\end{proof}

\begin{proof}[\textbf{Proof of Corollary~\ref{cor:TU-mimicking}}]
We prove the result for population $X$; the argument for population $Y$ is symmetric. Applying Lemma~\ref{lem:TU-mimicking} in both directions gives, whenever $x,x+\epsilon\in[0,1]$,
\begin{equation}
\label{eq:TU-mimicking-sandwich}
\begin{aligned}
 \int_0^1 [f(x+\epsilon,y)-f(x,y)]Q_t^X(dy\mid x)&\leq V_t^X(x+\epsilon)-V_t^X(x)\\
 &\leq\int_0^1 [f(x+\epsilon,y)-f(x,y)]Q_t^X(dy\mid x+\epsilon).
\end{aligned}
\end{equation}
Since both measures have mass below one,
\[
 |V_t^X(x+\epsilon)-V_t^X(x)|\leq \|f_1\|_\infty|\epsilon|.
\]
Thus values of search are Lipschitz in own type at every date.

Let $x_n\to x$. Continuity of $f$, $\Xi_\tau$, and $V_\tau^X$ implies $m_\tau(x_n,y)\to m_\tau(x,y)$. Since
\[
 \Lambda_\tau^X(y\mid x_n)=\lambda_\tau\mu_\tau^Y(y)m_\tau(x_n,y)\leq\lambda_\tau\mu_\tau^Y(y),
\]
local finiteness of meetings and dominated convergence give, for every $R>t$,
\[
 \int_t^R\int_0^1|\Lambda_\tau^X(y\mid x_n)-\Lambda_\tau^X(y\mid x)|\,dy\,d\tau\longrightarrow0.
\]
The cumulative hazards therefore converge uniformly on $[t,R]$. By \eqref{eq:TU-Q-representation}, the densities contributed to $Q_t^X(\cdot\mid x_n)$ by dates in $[t,R]$ consequently converge in $L^1$ to those for $x$. The mass contributed by later dates is, by the same representation,
\[
 e^{-\rho(R-t)}\exp\!\left\{-\alpha^X\int_t^R\int_0^1\Lambda_s^X(z\mid x)\,dz\,ds\right\}Q_R^X([0,1]\mid x)\leq e^{-\rho(R-t)},
\]
uniformly in $x$. Taking first $n\to\infty$ and then $R\to\infty$ therefore yields
\begin{equation}
\label{eq:TU-Q-continuity}
 \bigl\|Q_t^X(\cdot\mid x_n)-Q_t^X(\cdot\mid x)\bigr\|_{\mathrm{TV}}\longrightarrow0.
\end{equation}

Since $f\in\mathcal C^1$, the expansion $f(x+\epsilon,y)-f(x,y)=\epsilon f_1(x,y)+o(|\epsilon|)$ is uniform in $y$. By \eqref{eq:TU-Q-continuity}, both bounds in \eqref{eq:TU-mimicking-sandwich} equal
\[
 \epsilon\int_0^1 f_1(x,y)Q_t^X(dy\mid x)+o(|\epsilon|).
\]
Hence
\begin{equation}
\label{eq:TU-value-derivative}
 {V_t^X}'(x)=\int_0^1 f_1(x,y)Q_t^X(dy\mid x),
\end{equation}
with one-sided derivatives at the endpoints. Continuity of $f_1$ and \eqref{eq:TU-Q-continuity} make this derivative continuous. The symmetric argument gives $V_t^Y\in\mathcal C^1$ and its stated derivative formula. Thus $S_t$ is jointly continuous, and  Berge's maximum theorem implies that $X_t^\star$ and $Y_t^\star$ are non-empty, compact-valued, and upper hemicontinuous.
\end{proof}

\subsection{Single-Peakedness}
\begin{proof}[\textbf{Proof of Proposition~\ref{mconvex}}]
If $x \mapsto m_t(x,y)$ is not single-peaked, there exist $x_1<x_2<x_3$ such that
\[
m_t(x_1,y)>m_t(x_2,y)
\quad\text{and}\quad
m_t(x_3,y)>m_t(x_2,y).
\]
Since $m_t$ is a strictly increasing transformation of surplus, this implies
\begin{align*}
V_t^X(x_2)-V_t^X(x_1) > f(x_2,y)-f(x_1,y),
\qquad
V_t^X(x_3)-V_t^X(x_2) < f(x_3,y)-f(x_2,y).
\end{align*}
Twofold application of the mimicking argument (Lemma~\ref{lem:TU-mimicking}), with $x_2$ the mimicked type, gives $Q_t(\cdot\mid x_2)$, non-decreasing, non-negative and continuous with $Q_t(1\mid x_2)<1$, satisfying
\begin{align*}
\int_0^1 [f(x_2,z)-f(x_1,z)]Q_t(dz\mid x_2)
&>f(x_2,y)-f(x_1,y),\\
\int_0^1 [f(x_3,z)-f(x_2,z)]Q_t(dz\mid x_2)
&<f(x_3,y)-f(x_2,y).
\end{align*}
Using the mean value theorem,\footnote{We apply the mean value theorem to $x\mapsto\int_0^1 f(x,z)Q_t(dz\mid x_2)-f(x,y)$.} there exist $x_1<\tilde x_1<x_2<\tilde x_2<x_3$ such that
\begin{equation} \label{th:singlepmain}
\int_0^1 f_1(\tilde{x}_1,z)\,Q_t(dz| x_2)\;>\; f_1(\tilde{x}_1,y)
\quad \text{and} \quad
\; \int_0^1 f_1(\tilde{x}_2,z)\,Q_t(dz| x_2)  \;< \; f_1(\tilde{x}_2,y)
\end{equation}
Write $u_i(z):=f_1(\tilde x_i,z)$, $i=1,2$, and $q:=Q_t(1\mid x_2)$. Both $u_i$ are non-decreasing by supermodularity. If $q=0$, \eqref{th:singlepmain} gives $u_1(y)<0<u_2(y)$, which is covered by Case~3 below.

\medskip
\noindent
\emph{Claim (Arrow--Pratt comparison).}
If $q>0$, then, for every $c\in[0,1]$,
\[
\int_0^1 u_1(z)\frac{Q_t(dz\mid x_2)}q>u_1(c)
\quad\Longrightarrow\quad
\int_0^1 u_2(z)\frac{Q_t(dz\mid x_2)}q\geq u_2(c).
\]

\noindent
\emph{Proof of the claim.}
For $z<c<w$, integrating log-supermodularity of $f_{12}$ over $[z,c]\times[c,w]$ gives
\[
[u_1(c)-u_1(z)][u_2(w)-u_2(c)]
\geq
[u_2(c)-u_2(z)][u_1(w)-u_1(c)].
\]
Let $L_i:=\int_0^c[u_i(c)-u_i(z)]Q_t(dz\mid x_2)$ and $G_i:=\int_c^1[u_i(z)-u_i(c)]Q_t(dz\mid x_2)$. Integrating the displayed inequality gives $L_1G_2\geq L_2G_1$. A reversal would require $G_1>L_1$ and $G_2<L_2$, hence $L_1G_2<L_2G_1$, a contradiction.
\hfill$\diamond$

\medskip
\noindent
\emph{Case 1: $u_1(y)>0$.}
By \eqref{th:singlepmain} and monotonicity,
\(
q\,u_1(1)>u_1(y)>q\,u_1(y).
\)
Thus $q>0$, and continuity gives $y'\in(y,1)$ with $q\,u_1(y')=u_1(y)$. Log-supermodularity of $f_1$ gives
\[
u_1(y)u_2(y')\geq u_1(y')u_2(y)
\quad\Longrightarrow\quad
q\,u_2(y')\geq u_2(y),
\]
where we use $u_1(y')>0$. Consequently,
\eqref{th:singlepmain} implies
\[
\int_0^1 u_1(z)\frac{Q_t(dz\mid x_2)}q>u_1(y'),
\qquad
\int_0^1 u_2(z)\frac{Q_t(dz\mid x_2)}q <\frac{u_2(y)}q\leq u_2(y'),
\]
contradicting the claim.

\medskip
\noindent
\emph{Case 2: $u_2(y)<0$.}
Now \eqref{th:singlepmain} and monotonicity give
\(
q\,u_2(0)<u_2(y)<q\,u_2(y).
\)
Thus $q>0$, and continuity gives $y'\in(0,y)$ with
$q\,u_2(y')=u_2(y)$. Log-supermodularity of $f_1$ gives
\[
u_1(y')u_2(y)\geq u_1(y)u_2(y')
\quad\Longrightarrow\quad
q\,u_1(y')\leq u_1(y),
\]
where division by $u_2(y')<0$ reverses the inequality. Hence
\[
\int_0^1 u_1(z)\frac{Q_t(dz\mid x_2)}q>\frac{u_1(y)}q\geq u_1(y'),
\qquad
\int_0^1 u_2(z)\frac{Q_t(dz\mid x_2)}q<u_2(y'),
\]
again contradicting the claim.

\medskip
\noindent
\emph{Case 3:
$f_1(\tilde x_1,y)\leq0\leq f_1(\tilde x_2,y)$.}
If both marginals vanish at $y$, then \eqref{th:singlepmain} contradicts the Arrow-Pratt comparison with $c=y$. Otherwise, for $z>y$, log-supermodularity and supermodularity give
\[
\begin{aligned}
0\leq{}&f_1(\tilde x_1,y)[f_1(\tilde x_2,z)-f_1(\tilde x_2,y)]-f_1(\tilde x_2,y)[f_1(\tilde x_1,z)-f_1(\tilde x_1,y)]\leq0.
\end{aligned}
\]
For $z<y$, the analogous inequalities hold with the increments reversed. If exactly one marginal vanishes at $y$, these inequalities force it to vanish at every partner type, contradicting \eqref{th:singlepmain}. Therefore, $f_1(\tilde x_1,\cdot)$ and
$f_1(\tilde x_2,\cdot)$ are constant, with strictly negative and strictly positive values, respectively. So both increments must vanish, i.e., $f_1(\tilde{x}_2,\cdot)$ and $f_1(\tilde{x}_1,\cdot)$ are constant.

For any $x>\tilde x_1$, the cross-product inequality comparing $x$ with the negative constant row $f_1(\tilde x_1,\cdot)$ makes $f_1(x,\cdot)$ non-increasing. For any $x<\tilde x_2$, comparison with the positive constant row $f_1(\tilde x_2,\cdot)$ gives the same conclusion. Since $\tilde x_1<\tilde x_2$, these comparisons cover every $x$. Supermodularity therefore makes $f_1(x,\cdot)$ constant for every $x$, implying $f_{12}\equiv 0$, a contradiction.

All cases contradict \eqref{th:singlepmain}. Thus $m_t(\cdot,y)$, and equivalently $S_t(\cdot,y)$, is single-peaked. Interchanging the populations proves the symmetric statement.
\end{proof}

\begin{proof}[\textbf{Proof of Distribution-dependent Proposition~\ref{rmk:sppayoffs}}]
Write $\bar\Xi_t:=1-\Xi_t$ and $\gamma_t(z):=\int_z^{\bar\xi_t}\bar\Xi_t(u)\,du$. Integration by
parts gives
\[
 \Pi_t^X(y\mid x)=V_t^X(x)+\alpha^X\frac{\gamma_t(-S_t(x,y))}{\bar\Xi_t(-S_t(x,y))}.
\]
Let $H_t(S):=\gamma_t(-S)/\bar\Xi_t(-S)$. Since $\gamma_t'=-\bar\Xi_t$,
\[
 H_t'(S)=
 \frac{\gamma_t'(-S)^2-\gamma_t(-S)\gamma_t''(-S)}{\gamma_t'(-S)^2}.
\]
Thus $H_t$ is non-decreasing when $\gamma_t$ is log-concave and non-increasing when $\gamma_t$ is log-convex. Composing $H_t$ with the single-peaked function $S_t(x,\cdot)$ yields, respectively, a single-peaked or single-dipped conditional match payoff.
\end{proof}

\subsection{A Threshold for Assortative Contours}

\begin{proof}[\textbf{Proof of Theorem~\ref{theorem:lattice}}]
Throughout, $\underline{x}_t(y)$ and $\overline{x}_t(y)$ denote the least or greatest elements of $x_t(y) = \argmax_x m_t(x,y)$, and similarly for $\underline{y}_t(x)$, $\overline{y}_t(x)$. Define the peak match probability function $\mathfrak{P}_t(x) := \max_y m_t(x,y)$.

\medskip
\noindent\emph{Step 0 (boundary peaks).}
Conditions~\textup{(i)} and~\textup{(iv)} imply $ 0\in X_t^\star(0)$, $0\in Y_t^\star(0)$, $1\in X_t^\star(1)$, $1\in Y_t^\star(1).$

We prove $1\in Y_t^\star(1)$; the other statements are symmetric. Suppose by contradiction that $S_t(1,y')>S_t(1,1)$ for some $y'<1$. Rearranging and applying Lemma~\ref{lem:TU-mimicking} with type $y'$ mimicking type $1$ gives
\begin{align*}
0
&\leq f(1,1)-f(1,y')<V_t^Y(1)-V_t^Y(y')\leq \int_0^1\bigl[f(x',1)-f(x',y')\bigr]Q_t^Y(dx'\mid1).
\end{align*}
By supermodularity, $
f(x',1)-f(x',y')\leq f(1,1)-f(1,y')$, and $Q_t^Y([0,1]\mid1)<1$. Combining these inequalities with the preceding display yields a contradiction.

\medskip
\noindent\emph{Step 1.} We show that for any pair $(x',y')$ satisfying either $x' \leq \underline{x}_t(y')$ and $y' \leq \underline{y}_t(x')$, or $x' \geq \overline{x}_t(y')$ and $y' \geq \overline{y}_t(x')$, it holds that $m_t(x',y') \geq \min_x \mathfrak{P}_t(x)$.

We prove the first case; the second is symmetric. If $y'=\underline{y}_t(x')$, then $y'\in Y_t^\star(x')$ and hence $m_t(x',y')=\mathfrak P_t(x')\geq\min_x\mathfrak P_t(x).$ Suppose therefore that $y'<\underline y_t(x').$ Define
\[
A(y'):=\{x:\underline y_t(x)\leq y'\},
\qquad
\hat x:=\sup A(y').
\]
By Step~0, $\underline y_t(0)=0\leq y'$, so $0\in A(y')$, and $A(y')$ is a  nonempty interval. It is also closed. Indeed, if $x_n\in A(y')$ and $x_n\to x$, due to sequential compactness of $[0,y']$ there exists a subsequence such that $\underline y_t(x_n)\to\tilde y\leq y'$. Upper hemicontinuity of $Y_t^\star$ implies $\tilde y\in Y_t^\star(x)$, and therefore $\underline y_t(x)\leq\tilde y\leq y'$. Thus
\[
A(y')=[0,\hat x]
\quad\text{and}\quad
\underline y_t(\hat x)\leq y'.
\]

Since $y'<\underline y_t(x')$, we have $x'\notin A(y')$ and hence $\hat x<x'$. Choose $x_n\downarrow\hat{x}$ with $\hat x<x_n<x'$. Then $x_n\notin A(y')$, so
\(
\underline y_t(x_n)>y'.
\)
By monotonicity, $\underline y_t(x_n)$ converges to some $\bar y\geq y'$. Upper hemicontinuity gives $\bar y\in Y_t^\star(\hat x)$. Since $Y_t^\star(\hat x)$ is an interval and
\(
\underline y_t(\hat x)\leq y'\leq\bar y,
\)
we conclude that
\(
y'\in Y_t^\star(\hat x).
\)

Finally, $\hat x<x'\leq\underline x_t(y')$, so single-peakedness of
$m_t(\cdot,y')$ yields
\[
m_t(x',y')\geq m_t(\hat x,y') = \mathfrak P_t(\hat x) \geq \min_x\mathfrak P_t(x).
\]

\medskip
\noindent\emph{Step 2.} We show that $U_t(p)$ is a lattice for $p \leq \min_x \mathfrak{P}_t(x)$.

Consider $x_1 < x_2$ and $y_1 < y_2$ with $m_t(x_1, y_2), m_t(x_2, y_1) \geq p$. We show $m_t(x_1, y_1) \geq p$; the argument for $m_t(x_2, y_2)$ is symmetric.

\textit{Case 1:} $y_1 \leq \underline{y}_t(x_1)$. If also $x_1 \leq \underline{x}_t(y_1)$, conclude via Step~1. If $x_1 > \underline{x}_t(y_1)$, then $x_2 > x_1 > \underline{x}_t(y_1)$, so single-peakedness of $m_t(\cdot, y_1)$ gives $m_t(x_1, y_1) \geq m_t(x_2, y_1) \geq p$.

\textit{Case 2:} $y_1 \geq \overline{y}_t(x_1)$. Symmetric to Case~1.

\textit{Case 3:} $y_1 \in (\underline{y}_t(x_1), \overline{y}_t(x_1))$. Then, by single-peakedness (using (i)-(iii) and Proposition~\ref{mconvex}) it holds that $m_t(x_1, y_1) = \mathfrak{P}_t(x_1) \geq \min_x \mathfrak{P}_t(x) \geq p$.

\medskip
\noindent\emph{Step 3.}
We show that $U_t(p)$ fails to be a lattice for almost every $p\in(\min_x\mathfrak P_t(x),\max_x\mathfrak P_t(x))$.

\medskip
\noindent\textit{Claim.}
The peak-value functions
$\mathfrak P_t(x):=\max_y m_t(x,y)$ and $\mathfrak P_t^Y(y):=\max_xm_t(x,y)$ are Lipschitz continuous.

\medskip
\noindent\textit{Proof of Claim.}
Corollary~\ref{cor:TU-mimicking} and the continuous differentiability of $\Xi_t$ imply that $m_t\in\mathcal C^1([0,1]^2)$. Hence $L_1:=\sup_{x,y}|D_1m_t(x,y)|<\infty$, and, for any $x_1,x_2$,
\[
\bigl|\mathfrak P_t(x_2)-\mathfrak P_t(x_1)\bigr| \leq \max_y\bigl|m_t(x_2,y)-m_t(x_1,y)\bigr| \leq L_1|x_2-x_1|.
\]
Thus $\mathfrak P_t$ is Lipschitz. The symmetric argument establishes the claim for $\mathfrak P_t^Y$. \hfill$\lozenge$

Consider the set of types belonging to a reciprocal pair of likeliest partners, $
\mathcal X := \{x:x\in X_t^\star(\hat y) \text{ for some }\hat y\in Y_t^\star(x)\}.$

\noindent\textit{Claim.} The set $\mathfrak P_t(\mathcal{X})$ has zero measure, $\operatorname{Leb}(\mathfrak P_t(\mathcal X))=0$.

\medskip
\noindent\textit{Proof of Claim.} Since $\mathfrak P_t$ is Lipschitz, it is differentiable outside a null set $N$. Take any $x\in\mathcal X\cap(0,1)\setminus N$ and choose $\hat y\in Y_t^\star(x)$ such that $x\in X_t^\star(\hat y)$. Because $x$ is an interior maximizer of $m_t(\cdot,\hat y)$, $D_1m_t(x,\hat y)=0$. The envelope theorem (cf. \cite{MilgromSegal2002} Theorem 1) therefore gives $\mathfrak P_t'(x)=D_1m_t(x,\hat y)=0$.

Let $C:=\{x\in(0,1)\setminus N:\mathfrak P_t'(x)=0\}$. Then $\mathcal X\subseteq C\cup N\cup\{0,1\}$. By the one-dimensional area formula,
\[
\operatorname{Leb}\bigl(\mathfrak P_t(C)\bigr) \leq \int_C|\mathfrak P_t'(x)|\,dx = 0.
\]
Moreover, Lipschitz functions map null sets into null sets, so $\operatorname{Leb}(\mathfrak P_t(N))=0$. The images of the two boundary points are finite. Consequently, $\operatorname{Leb}(\mathfrak P_t(\mathcal X))=0$. \hfill$\lozenge$

Now fix
\[
p\in\bigl(\min_x\mathfrak P_t(x),\max_x\mathfrak P_t(x)\bigr)\setminus\mathfrak P_t(\mathcal X).
\]
Almost every $p$ in the displayed interval is of this form. By continuity of $\mathfrak P_t$, there exists $x'\in(0,1)$ such that $\mathfrak P_t(x')=p$. Let $y_L:=\underline y_t(x')$ and $y_U:=\overline y_t(x')$. Since $p\notin\mathfrak P_t(\mathcal X)$, we have $x'\notin\mathcal X$, and therefore $x'\notin X_t^\star(y)$ for every $y\in [y_L,y_U]$. Hence either $x'<\underline x_t(y)$ or $x'>\overline x_t(y)$. These alternatives cannot both occur on $[y_L,y_U]$. Otherwise, monotonicity would give $y'<y''$ such that $x'>\overline x_t(y')$ and $x'<\underline x_t(y'')$. At a transition point $\hat y\in[y',y'']$, upper hemicontinuity supplies a limit point of $X_t^\star(y)$ weakly below $x'$ from the left and another weakly above $x'$ from the right. Since $X_t^\star(\hat y)$ is an interval, it would contain $x'$, a contradiction. Therefore exactly one of the following holds:
\[
\begin{array}{ll}
\text{\rm(a)}&
x'<\underline x_t(y)
\quad\text{for every }y\in[y_L,y_U],\\[0.3ex]
\text{\rm(b)}&
x'>\overline x_t(y)
\quad\text{for every }y\in[y_L,y_U].
\end{array}
\]

We prove a failure of the lattice property in case~\emph{(a)}; case~\emph{(b)} is symmetric.

\medskip \noindent\textit{Claim.} There exists a failure of the lattice property in case~\emph{(a)}.

\medskip
\noindent\textit{Proof of Claim.} The boundary condition implies $y_L>0$; otherwise $\underline x_t(y_L)=\underline x_t(0)=0$ by Step~0, contradicting~(a). Since $x'$ lies strictly below the peak set of $m_t(\cdot,y_L)$,
\[
m_t(\underline x_t(y_L),y_L)>p=m_t(x',y_L)=\mathfrak P_t(x').
\]
Monotonicity and upper hemicontinuity of $X_t^\star(\cdot)$ imply that its lower endpoint $\underline x_t(\cdot)$ is left-continuous. And due to the preceding claim $\mathfrak P_t^Y$ is continuous. Choose $y_1<y_L$ sufficiently close to $y_L$ that $\underline x_t(y_1)>x'$ and $\mathfrak P_t^Y(y_1)>p$, and let $x_2:=\underline x_t(y_1)$. Then $m_t(x_2,y_1)=\mathfrak P_t^Y(y_1)>p=m_t(x',y_L)$, so $(x',y_L),(x_2,y_1)\in U_t(p)$. Since $y_1<y_L=\underline y_t(x')$, single-peakedness implies $m_t(x',y_1)<p$. Hence the meet $(x',y_1)\notin U_t(p)$, so $U_t(p)$ is not a lattice.
\end{proof}

\section{Sorting Without Reciprocity}

The proof of Theorem~\ref{thm:neverPAM} builds on Propositions~\ref{proposition:productive-disadvantage} and \ref{proposition:positional-disadvantage}. For Item~\textup{(i)}, we first show that regularity of the common productivity peak implies the uniformity condition needed for Proposition~\ref{proposition:productive-disadvantage}. Applying that proposition to both populations yields the result. Item~\textup{(ii)} follows directly from applying Proposition~\ref{proposition:positional-disadvantage} to both populations.

\begin{proof}[\textbf{Proof of Theorem~\ref{thm:neverPAM}, Item~\textup{(i)}}]
Fix arbitrary $q\in[0,1)$. We establish that there exists a non-degenerate interval $I_t^X\ni\hat x$ such that
\begin{equation}
\label{eq:regularity-implies-uniformity}
\frac{f(x,\hat y)-f(x,y)}{f(\hat x,\hat y)-f(\hat x,y)}>q
\end{equation}
for every $x\in I_t^X$ and every $y\neq\hat y$.

\medskip
\noindent
\emph{Case (a): Boundary peak.}
Suppose first that $\hat y\in\{0,1\}$. Define
\[
\mathcal A^X(x,y):=
\begin{cases}
\displaystyle
\frac{f(x,\hat y)-f(x,y)}{|y-\hat y|},& y\neq\hat y,\\[1.2ex]
\displaystyle
(2\hat y-1)f_2(x,\hat y),& y=\hat y.
\end{cases}
\]
By the fundamental theorem of calculus, $\mathcal A^X(x,y)=(2\hat y-1)\int_0^1 f_2\bigl(x,\hat y+s(y-\hat y)\bigr)\,ds$, so $\mathcal A^X$ is continuous on $[0,1]^2$. Because $\hat y$ uniquely maximizes $f(\hat x,\cdot)$, $\mathcal A^X(\hat x,y)>0$ for every $y\neq\hat y$; at $y=\hat y$, regularity gives $\mathcal A^X(\hat x,\hat y) =(2\hat y-1)f_2(\hat x,\hat y)>0$. Hence, by continuity and compactness, $c:=\min_{y\in[0,1]}\mathcal A^X(\hat x,y)>0$.

By uniform continuity, there exists a relatively open interval $I_t^X\ni\hat x$ such that $|\mathcal A^X(x,y)-\mathcal A^X(\hat x,y)|<(1-q)c$ for every $x\in I_t^X$ and $y\in[0,1]$. Therefore $\mathcal A^X(x,y)> \mathcal A^X(\hat x,y)-(1-q)c \geq q\mathcal A^X(\hat x,y)$. Multiplying by $|y-\hat y|$ proves \eqref{eq:regularity-implies-uniformity} for every $y\neq\hat y$.

\medskip
\noindent
\emph{Case (b): Interior peak.}
Suppose now that $\hat y\in(0,1)$. Since $\hat y$ maximizes $f(x,\cdot)$ for every $x$ and is interior, $f_2(x,\hat y)=0$ for every $x$. Define
\[
\mathcal A^X(x,y):=
\begin{cases}
\displaystyle
\frac{f(x,\hat y)-f(x,y)}{(y-\hat y)^2},& y\neq\hat y,\\[1.2ex]
\displaystyle
-\frac12 f_{22}(x,\hat y),& y=\hat y.
\end{cases}
\]
Taylor's formula with integral remainder gives $\mathcal A^X(x,y)=-\int_0^1(1-s)f_{22}\bigl(x,\hat y+s(y-\hat y)\bigr)\,ds$, so $\mathcal A^X$ is continuous on $[0,1]^2$. For $y\neq\hat y$, strict common peakedness gives $\mathcal A^X(\hat x,y)>0$, while regularity gives $\mathcal A^X(\hat x,\hat y)=-\frac12 f_{22}(\hat x,\hat y)>0$. Thus $c:=\min_{y\in[0,1]}\mathcal A^X(\hat x,y)>0$.

Uniform continuity again gives a relatively open interval $I_t^X\ni\hat x$ such that $|\mathcal A^X(x,y)-\mathcal A^X(\hat x,y)|<(1-q)c$ for every $x\in I_t^X$ and $y\in[0,1]$. Hence $\mathcal A^X(x,y)>q\mathcal A^X(\hat x,y)$, and multiplying by $(y-\hat y)^2$ establishes \eqref{eq:regularity-implies-uniformity}.

\medskip
\noindent
\emph{Conclusion.}
The common-peak assumption supplies premise~(i) of Proposition~\ref{proposition:productive-disadvantage}, peak increasing differences supplies premise~(ii), and the argument above supplies premise~(iii). Taking $q:=Q_t^Y([0,1]\mid\hat y)<1$, that proposition gives a relative interval $I_t^X\ni\hat x$ such that $Y_t^\star(x)=\{\hat y\}$ for every $x\in I_t^X$. By the symmetric argument, there is a relative interval $I_t^Y\ni\hat y$ such that $X_t^\star(y)=\{\hat x\}$ for every $y\in I_t^Y$.

Because $(\hat x,\hat y)\notin\{(0,1),(1,0)\}$, the two intervals contain types on a common side of $(\hat x,\hat y)$. If they contain $x<\hat x$ and $y<\hat y$, then $y<\underline y_t(x)=\hat y$ and $x<\underline x_t(y)=\hat x$, violating weak reciprocity. If instead they contain $x>\hat x$ and $y>\hat y$, then $y>\overline y_t(x)=\hat y$ and $x>\overline x_t(y)=\hat x$, violating the corresponding upper-bound implication. Hence weak reciprocity fails.
\end{proof}

\begin{proof}[\textbf{Proof of Theorem~\ref{thm:neverPAM}, Item~\textup{(ii)}}]
The boundary-peak and boundary-complementarity conditions imply Proposition~\ref{proposition:positional-disadvantage} for population $X$ and its symmetric counterpart for population $Y$. Hence there exist relative intervals $I_t^X,I_t^Y\ni b$ such that $Y_t^\star(x)=\{b\}$ for every $x\in I_t^X$ and $X_t^\star(y)=\{b\}$ for every $y\in I_t^Y$.

If $b=1$, choose $x<1$ and $y<1$ in these intervals. Then $y<\underline y_t(x)=1$ and $x<\underline x_t(y)=1$, violating weak reciprocity. If $b=0$, choose $x>0$ and $y>0$ and apply the symmetric upper-bound implication.
\end{proof}

\subsection{Positional Disadvantage}

\begin{proof}[\textbf{Proof of Proposition~\ref{proposition:positional-disadvantage}}]
We prove the result for $b=1$; the case $b=0$ is symmetric. Write $Q_t^X(\cdot\mid1)$ for the measure in Lemma~\ref{lem:TU-mimicking}. By active search it has positive mass, and its total mass is strictly below one.

First, $X_t^\star(1)=\{1\}$. To see this, fix $x<1$ and write $\Delta_x(z):=f(1,z)-f(x,z)$. The boundary-peak condition gives $\Delta_x(1)\geq0$, while boundary complementarity gives
\[
\Delta_x(z)=\int_x^1 f_1(u,z)\,du<\int_x^1 f_1(u,1)\,du=\Delta_x(1)\qquad\text{for every }z<1.
\]
If $S_t(x,1)\geq S_t(1,1)$, the mimicking inequality would imply
\[
\Delta_x(1)\leq V_t^X(1)-V_t^X(x) \leq \int_0^1 \Delta_x(z)Q_t^X(dz\mid1)< \Delta_x(1),
\]
a contradiction.

Next, Corollary~\ref{cor:TU-mimicking} gives
\begin{align*}
D_1S_t(1,1)&=
f_1(1,1)\bigl[1-Q_t^X([0,1]\mid1)\bigr]+\int_0^1\bigl[f_1(1,1)-f_1(1,z)\bigr]Q_t^X(dz\mid1)>0.
\end{align*}
Indeed, the boundary-peak condition implies $f_1(1,1)\geq0$, while boundary complementarity makes the integrand strictly positive for $z<1$.

Continuity now gives $\delta>0$ such that $D_1S_t(x,y)>0$ whenever $x,y\in[1-\delta,1]$. Since $X_t^\star$ is upper hemicontinuous and $X_t^\star(1)=\{1\}$, there is $\varepsilon_t>0$ such that
\[
X_t^\star(y)\subseteq[1-\delta,1]
\qquad
\text{for every }y\in[1-\varepsilon_t,1].
\]
For every such $y$, surplus is strictly increasing on an interval containing all of its maximizers. Hence $X_t^\star(y)=\{1\}$ for every such $y$. Taking $I_t^Y=[1-\varepsilon_t,1]$ proves the claim.
\end{proof}

\subsection{Matching Up\textemdash Quantitative Impact of Search Frictions}

\begin{proof}[\textbf{Proof of Corollary~\ref{cor:quantifying}}]
For every $y<1$,
\begin{align*}
&\bigl[f(1,1)-f(1,y)\bigr] - \bigl[f(x,1)-f(x,y)\bigr]= \int_x^1\int_y^1 f_{12}(z,u)\,du\,dz \leq (1-x)(1-y)\max_{z,u}f_{12}(z,u),
\end{align*}
whereas
\[
f(x,1)-f(x,y) = \int_y^1f_2(x,u)\,du \geq (1-y)\min_u f_2(x,u).
\]
Consequently, the peak-advantage ratio in Proposition~\ref{proposition:productive-disadvantage} satisfies
\[
\chi^X(x) \geq \frac{\min_u f_2(x,u)}{\min_u f_2(x,u)+(1-x)\max_{z,u}f_{12}(z,u)}.
\]

Integrating the representation in Lemma~\ref{lem:TU-mimicking} over
partner type and then by parts gives
\begin{align*}
Q_t^Y([0,1]\mid y) =& 1-\rho\int_t^\infty e^{-\rho(\tau-t)}
\exp\left\{-\alpha^Y\int_t^\tau\int_0^1\Lambda_s^Y(z\mid y)\,dz\,ds \right\}d\tau.
\end{align*}
Since
$\alpha^Y\int_0^1\Lambda_s^Y(z\mid y)\,dz\leq\overline\lambda^Y$, the exponential in the integral is bounded below by $e^{-\overline\lambda^Y(\tau-t)}$. Therefore
\[
Q_t^Y([0,1]\mid y) \leq 1-\frac{\rho}{\rho+\overline\lambda^Y} = \frac{\overline\lambda^Y}{\rho+\overline\lambda^Y}.
\]
Thus $\chi^X(x)>\overline\lambda^Y/(\rho+\overline\lambda^Y)$ is sufficient for $Y_t^\star(x)=\{1\}$. Substituting the lower bound for $\chi^X(x)$ and rearranging gives
\[
x>1-\frac{\min_u f_2(x,u)}{\max_{z,u}f_{12}(z,u)}\frac{\rho}{\overline\lambda^Y}.
\]
The population-$Y$ statement is symmetric.
\end{proof}

\end{document}

\begin{center}
    \Huge Online Appendix
\end{center}

This appendix extends the non-stationary equilibrium-existence result of \citet{SandmannBonneton2025} to transferable-utility economies with state-dependent, pair-specific production shocks of unbounded support. The extension has two opposing effects. Lipschitz shock CDFs smooth aggregate matching decisions and simplify continuity of the fixed-point operator; unbounded shocks and meeting opportunities eliminate the time-uniform value bound used in the model without shocks. We replace that bound by a deterministic, date-dependent envelope and normalize the value operator accordingly.

\section{Assumptions and Main Result}
\label{sec:environment-existence}

We study a continuous-time search-and-matching economy with two populations, denoted by $X$ and $Y$. Agent types are $x,y\in[0,1]$. At date $t$, the distributions of unmatched agents are represented by the non-negative measurable densities $\mu_t^X,\mu_t^Y$. We write $\mu_t=(\mu_t^X,\mu_t^Y)$ and use the seminorm
\[
N(\mu_t,\widetilde\mu_t):=\max\left\{\int_0^1|\mu_t^X(x)-\widetilde\mu_t^X(x)|\,dx,\,\int_0^1|\mu_t^Y(y)-\widetilde\mu_t^Y(y)|\,dy \right\}.
\]

An unmatched type $x$ meets type $y$ at rate $\lambda^X(t,\mu_t)(y\mid x)$; symmetrically, $\lambda^Y(t,\mu_t)(x\mid y)$ is type $y$'s meeting rate with type $x$. For every $t,\mu,x,y$, coherence requires
\[
\lambda^X(t,\mu)(y\mid x)\mu^X(x)=\lambda^Y(t,\mu)(x\mid y)\mu^Y(y).
\]
Entry rates are denoted by $\eta^X(t,\mu_t)(x)$ and $\eta^Y(t,\mu_t)(y)$.

\begin{assumption}[Search and entry]
\label{ass:search-entry}
The meeting and entry technologies are jointly measurable in their
arguments and satisfy the following conditions.

\begin{enumerate}[label=(\roman*)]
\item \emph{Hierarchical search.} The meeting rate $\lambda^X(t,\mu)(y\mid x)$ is non-decreasing in $x$, and $\lambda^Y(t,\mu)(x\mid y)$ is non-decreasing in $y$.

\item \emph{Linear boundedness of meetings.}
There exists $L^\lambda<\infty$ such that
\[
\lambda^X(t,\mu)(y\mid x) \leq L^\lambda\bigl(1+\mu^Y(y)\bigr),
\qquad
\lambda^Y(t,\mu)(x\mid y)\leq L^\lambda\bigl(1+\mu^X(x)\bigr).
\]

\item \emph{State continuity of meetings.}
For every $x,y$,
\begin{align*}
\int_0^1 \bigl|\lambda^X(t,\mu)(z\mid x)-\lambda^X(t,\widetilde\mu)(z\mid x)\bigr|\,dz &\leq L^\lambda N(\mu,\widetilde\mu),\\
\int_0^1 \bigl| \lambda^Y(t,\mu)(z\mid y) - \lambda^Y(t,\widetilde\mu)(z\mid y) \bigr|\,dz &\leq L^\lambda N(\mu,\widetilde\mu).
\end{align*}

\item \emph{Regularity of entry.}
There exists $L^\eta<\infty$ such that
\[
0\leq\eta^X(t,\mu)(x),\eta^Y(t,\mu)(y)\leq L^\eta,
\]
and
\begin{align*}
\int_0^1 \bigl| \eta^X(t,\mu)(x)-\eta^X(t,\widetilde\mu)(x) \bigr|\,dx &\leq L^\eta N(\mu,\widetilde\mu),\\
\int_0^1 \bigl| \eta^Y(t,\mu)(y)-\eta^Y(t,\widetilde\mu)(y) \bigr|\,dy &\leq L^\eta N(\mu,\widetilde\mu).
\end{align*}
\end{enumerate}
\end{assumption}

We assume that the initial population is bounded:
\[
\overline{\overline\mu}_0 := \max\left\{ \mathop{\mathrm{ess\,sup}}_x\mu_0^X(x),\, \mathop{\mathrm{ess\,sup}}_y\mu_0^Y(y) \right\}<\infty.
\]
Since exit is non-negative and entry is bounded, every feasible population path satisfies
\begin{align}
0\leq\mu_t^X(x),\mu_t^Y(y) \leq \overline{\overline\mu}_t := \overline{\overline\mu}_0+L^\eta t.
\label{eq:population-envelope}
\end{align}

Fundamental match output is $f(x,y)$. We do not impose smoothness or
continuity in types.

\begin{assumption}[Output]
\label{ass:output-bv}
The function $f:[0,1]^2\to[0,\overline f]$ is jointly measurable. Its sections admit a uniform bound on total variation:
\[
\sup_{y\in[0,1]} \mathcal V_0^1\bigl(f(\cdot,y)\bigr) \leq L^f,
\qquad
\sup_{x\in[0,1]} \mathcal V_0^1\bigl(f(x,\cdot)\bigr) \leq L^f,
\]
where $\mathcal V_0^1$ denotes total variation on $[0,1]$.
\end{assumption}

Upon meeting at date $t$ in state $\mu$, a pair draws a match-specific production shock with CDF $\Xi(t,\mu)(\cdot)$. Shocks are conditionally independent across meetings. We write
\[
\overline\Xi(t,\mu)(a) := 1-\Xi(t,\mu)(a)
\]
and define the associated stop-loss transform
\[
G(t,\mu,a) := \int_a^\infty \overline\Xi(t,\mu)(z)\,dz = \int_{\mathbb R}(\xi-a)_+\,\Xi(t,\mu)(d\xi).
\]
The stop-loss transform is the natural primitive for the existence proof. It combines the probability of acceptance and the expected gain conditional on acceptance into a single finite object, avoiding conditional expectations that may be undefined when the matching probability is zero. Unbounded shock support is therefore harmless provided the expected positive gain at the lowest reachable threshold is uniformly bounded.

\begin{assumption}[Regularity of shocks]
\label{ass:shocks}
For every $(t,\mu)$, $\Xi(t,\mu)$ is a CDF. Moreover:

\begin{enumerate}[label=(\roman*)]
\item $(t,\mu,a)\mapsto\Xi(t,\mu)(a)$ is jointly measurable;

\item there exists $L^\Xi<\infty$ such that, for all $t,\mu,\widetilde\mu,a,\widetilde a$,
\[
\bigl| \Xi(t,\mu)(a)-\Xi(t,\widetilde\mu)(\widetilde a) \bigr| \leq L^\Xi\bigl( |a-\widetilde a|+N(\mu,\widetilde\mu) \bigr);
\]

\item there exists $L^G<\infty$ such that, for every
$a\geq-\overline f$,
\[
\bigl| G(t,\mu,a)-G(t,\widetilde\mu,a) \bigr| \leq L^G N(\mu,\widetilde\mu);
\]

\item \emph{Uniform positive-tail bound.}
The expected positive gain at the lowest reachable threshold is
uniformly bounded:
\[
\overline G := \sup_{t,\mu}G(t,\mu,-\overline f)<\infty.
\]
\end{enumerate}
\end{assumption}

The stop-loss transform is automatically one-Lipschitz in its final
argument. Indeed, for $a<\widetilde a$,
\begin{align}
0\leq G(t,\mu,a)-G(t,\mu,\widetilde a) = \int_a^{\widetilde a} \overline\Xi(t,\mu)(z)\,dz \leq \widetilde a-a.
\label{eq:G-one-Lipschitz}
\end{align}
Thus no separate Lipschitz assumption on $a\mapsto G(t,\mu,a)$ is needed.

Assumption~\ref{ass:shocks} imposes no lower or upper support bound. It permits, for example, Gaussian and exponential shocks, and Pareto shocks with finite mean. Its Lipschitz-CDF condition excludes atoms. In particular, the degenerate no-shock model is covered by \citet{SandmannBonneton2025}, rather than by the present theorem.

\subsection{Equilibrium}
Let $\rho>0$, and let the Nash bargaining weights satisfy $\alpha^X,\alpha^Y\in(0,1)$ and $\alpha^X+\alpha^Y=1$. Given values-of-search $V_t^X(x)$ and $V_t^Y(y)$, a pair matches if and only if
\[
f(x,y)+\xi \geq V_t^X(x)+V_t^Y(y).
\]
The meeting-contingent match probability is therefore
\begin{align}
m_t(x,y) = \overline\Xi(t,\mu_t)\bigl(V_t^X(x)+V_t^Y(y)-f(x,y)\bigr).
\label{eq:equilibrium-match-probability}
\end{align}

Population dynamics are given by
\begin{align}
\mu_{t+h}^X(x)=&\mu_t^X(x)+\int_t^{t+h}\Bigg[-\mu_\tau^X(x)\int_0^1\lambda^X(\tau,\mu_\tau)(y\mid x)m_\tau(x,y)\,dy+\eta^X(\tau,\mu_\tau)(x)\Bigg]d\tau,
\label{eq:population-dynamics-X}
\end{align}
with a symmetric equation for population $Y$.

Define the equilibrium match hazard
\[
\Lambda_t^X(y\mid x):=\lambda^X(t,\mu_t)(y\mid x)m_t(x,y).
\]
The density that type $x$, unmatched at date $t$, first matches with type $y$ at date $\tau$ is
\[
p_{t,\tau}^X(y\mid x)=\Lambda_\tau^X(y\mid x)\exp\left\{-\int_t^\tau\int_0^1\Lambda_r^X(z\mid x)\,dz\,dr\right\}.
\]

Let $\Pi_t^X(y\mid x)$ denote type $x$'s expected payoff conditional on matching with type $y$. Rather than dividing by a possibly zero match probability, we define the relevant product by
\begin{align}
m_t(x,y)\Pi_t^X(y\mid x) ={}& m_t(x,y)V_t^X(x) \nonumber\\
&+\alpha^X G\left(t,\mu_t,V_t^X(x)+V_t^Y(y)-f(x,y)\right).
\label{eq:match-probability-times-payoff}
\end{align}
Whenever $m_t(x,y)>0$, this is equivalent to
\[
\Pi_t^X(y\mid x)=V_t^X(x)+\alpha^X \mathbb E\left[f(x,y)+\xi-V_t^X(x)-V_t^Y(y)\,\middle|\,f(x,y)+\xi\geq V_t^X(x)+V_t^Y(y)\right].
\]
Whenever the match probability is zero, set $\Pi_t^X(y\mid x)=0$. This convention is immaterial, since conditional payoffs enter all subsequent expressions only through the product $m_t(x,y)\Pi_t^X(y\mid x)$.

The value-of-search is the primitive discounted expected future match payoff:
\begin{align}
V_t^X(x) = \int_t^\infty e^{-\rho(\tau-t)} \int_0^1 \Pi_\tau^X(y\mid x) p_{t,\tau}^X(y\mid x)\,dy\,d\tau,
\label{eq:equilibrium-value}
\end{align}
with a symmetric equation for population $Y$.

\subsection{Main Result}

\begin{definition}[Equilibrium]
\label{def:equilibrium}
An equilibrium is a tuple $(\boldsymbol\mu,\mathbf V,\mathbf m)$ satisfying
\eqref{eq:equilibrium-match-probability},
\eqref{eq:population-dynamics-X}, and
\eqref{eq:equilibrium-value}, and their population-$Y$ counterparts.
\end{definition}

\begin{theorem}[Existence]
\label{thm:existence}
Suppose that Assumptions~\ref{ass:search-entry}, \ref{ass:output-bv}, and~\ref{ass:shocks} hold. Then, for every bounded initial population, every discount rate $\rho>0$, and every pair of bargaining weights $\alpha^X,\alpha^Y\in(0,1)$ satisfying $\alpha^X+\alpha^Y=1$, the search-and-matching economy admits a non-stationary equilibrium.
\end{theorem}

The proof constructs an out-of-equilibrium value operator whose fixed points are equilibria. The next section establishes that this operator is well defined. Subsequent sections establish its continuity and locally uniform bounded variation and then apply the fixed-point theorem in \citet{SandmannBonneton2025}.

\section{The Fixed-Point Problem}
\label{sec:fixed-point-problem}

Let $v=(v^X,v^Y)$ be a finite, non-negative, jointly measurable value-of-search profile. Out of equilibrium, agents take the profile $v$ as describing the values and matching decisions of other agents. The aggregate population path generated by this conjecture must first be determined.

\subsection{Population Dynamics under a Conjectured Value Profile}

For an arbitrary candidate state $\mu$, define the aggregate meeting-contingent match probability under $v$ by
\[
m_t[v,\mu](x,y):=\overline\Xi(t,\mu)\bigl(v_t^X(x)+v_t^Y(y)-f(x,y)\bigr).
\]
The state and aggregate matching probabilities are jointly determined: the state affects meetings and the shock distribution, while matching determines exit from the search pool.

\begin{proposition}[Population dynamics]
\label{prop:population-under-v}
For every non-negative, finite-valued, jointly measurable conjecture $v$, the population dynamics~\eqref{eq:population-dynamics-X} and their population-$Y$ counterpart, evaluated at
\[
m_t(x,y)=m_t[v,\mu_t](x,y),
\]
admit a unique global solution, denoted by $\boldsymbol\mu[v]$. The solution $t\mapsto\boldsymbol\mu_t[v]$ is continuous as an \(L^1\)-valued path. Moreover,
\[
0\leq \boldsymbol\mu_t^X[v](x), \boldsymbol\mu_t^Y[v](y) \leq \overline{\overline\mu}_t
\]
for almost every $x,y$ and every $t\geq0$.
\end{proposition}

We henceforth write
\begin{align}
m_t[v](x,y):=m_t\bigl[v,\boldsymbol\mu_t[v]\bigr](x,y).
\label{eq:aggregate-m-v}
\end{align}

\subsection{The Primitive Out-of-Equilibrium Value}

An individual type $x$ takes the state $\boldsymbol\mu[v]$ and the conjectured partner values $v^Y$ as given. Unlike other type-$x$ agents, however, the individual evaluates a match against her actual out-of-equilibrium value $\boldsymbol V_t^X[v](x)$.

Her subjective probability of matching with $y$ upon meeting is
\begin{align}
\text{\normalfont\bsifamily m}_t^X[v](x,y)=\overline\Xi\bigl(t,\boldsymbol\mu_t[v]\bigr)\left(\boldsymbol V_t^X[v](x)+v_t^Y(y)-f(x,y)\right).
\label{eq:subjective-match-probability}
\end{align}

Let $\boldsymbol\pi_t^X[v](y\mid x)$ denote the associated expected match payoff. We define the product of the match probability and the conditional payoff by
\begin{align}
&\text{\normalfont\bsifamily m}_t^X[v](x,y) \boldsymbol\pi_t^X[v](y\mid x)= \text{\normalfont\bsifamily m}_t^X[v](x,y) \boldsymbol V_t^X[v](x) +\alpha^X G\left(t,\boldsymbol\mu_t[v],\boldsymbol V_t^X[v](x)+v_t^Y(y)-f(x,y)\right).
\label{eq:subjective-payoff-product}
\end{align}
When the subjective match probability is positive, $\boldsymbol\pi_t^X[v](y\mid x)$ is the expected Nash-bargained payoff conditional on matching.

The subjective density of a future match is
\begin{align}
\text{\normalfont\bsifamily p}_{t,\tau}^X[v](y\mid x)={}&\lambda^X\bigl(\tau,\boldsymbol\mu_\tau[v]\bigr)(y\mid x)\text{\normalfont\bsifamily m}_\tau^X[v](x,y) \nonumber\\
&\times \exp\left\{ -\int_t^\tau\int_0^1 \lambda^X\bigl(r,\boldsymbol\mu_r[v]\bigr)(z\mid x) \text{\normalfont\bsifamily m}_r^X[v](x,z) \,dz\,dr\right\}.
\label{eq:subjective-match-density}
\end{align}

\begin{definition}[Out-of-equilibrium value-of-search]
\label{def:out-of-equilibrium-value}
Given $v$, the out-of-equilibrium value-of-search is a non-negative profile $\boldsymbol V[v]=(\boldsymbol V^X[v],\boldsymbol V^Y[v])$ satisfying
\begin{align}
\boldsymbol V_t^X[v](x) = \int_t^\infty e^{-\rho(\tau-t)} \int_0^1 \boldsymbol\pi_\tau^X[v](y\mid x) \text{\normalfont\bsifamily p}_{t,\tau}^X[v](y\mid x) \,dy\,d\tau,
\label{eq:primitive-ooe-value}
\end{align}
and the symmetric population-$Y$ equation, where \eqref{eq:subjective-match-probability}--\eqref{eq:subjective-match-density} are evaluated at $\boldsymbol V[v]$.
\end{definition}

Definition~\ref{def:out-of-equilibrium-value} is recursive: $\boldsymbol V^X[v]$ determines both the payoff from matching and the set of shock realizations for which the individual matches. The next lemma establishes that the definition nevertheless has a unique solution.

Define the deterministic meeting-rate envelope
\[
\overline h(t):=L^\lambda \bigl(1+\overline{\overline\mu}_t\bigr)=L^\lambda \bigl(1+\overline{\overline\mu}_0+L^\eta t\bigr)
\quad \text{ and } \quad
B^Z(t):=\alpha^Z\overline G\int_t^\infty e^{-\rho(\tau-t)}\overline h(\tau)\,d\tau \quad \text{ for } Z\in\{X,Y\}.
\]
Under the linear population bound,
\[
B^Z(t)=\alpha^Z\overline G L^\lambda \left[ \frac{1+\overline{\overline\mu}_0+L^\eta t}{\rho} + \frac{L^\eta}{\rho^2} \right].
\]

\begin{lemma}[Well-posedness of the out-of-equilibrium value]
\label{lem:well-posed-value}
For every non-negative, finite-valued, jointly measurable conjecture $v$, Definition~\ref{def:out-of-equilibrium-value} admits a unique non-negative, jointly measurable solution
\(
\boldsymbol V[v]=\bigl(\boldsymbol V^X[v],\boldsymbol V^Y[v]\bigr).
\)
This solution satisfies
\begin{equation}
\label{eq:value-envelope}
0\leq\boldsymbol V_t^Z[v](z)\leq B^Z(t)\qquad \text{for every }t,z \text{ and }Z\in\{X,Y\}.
\end{equation}
\end{lemma}

\begin{proposition}[Fixed points and equilibria]
\label{prop:fixed-point-equilibrium}
A non-negative profile $V$ is a fixed point of the operator $\boldsymbol V$ if and only if $\bigl(\boldsymbol\mu[V],V,m[V]\bigr)$ is an equilibrium in the sense of Definition~\ref{def:equilibrium}.
\end{proposition}

\begin{proof}
At a fixed point $V=\boldsymbol V[V]$, the aggregate matching probability in \eqref{eq:aggregate-m-v} and the individual's subjective matching probability in \eqref{eq:subjective-match-probability} coincide. Hence the induced population dynamics, match payoffs, and primitive values-of-search satisfy the equilibrium equations. Conversely, every equilibrium satisfies Definition~\ref{def:out-of-equilibrium-value} when the conjecture is set equal to its equilibrium value profile and is therefore a fixed point of $\boldsymbol V$.
\end{proof}

\section{Existence of a Fixed-Point}
\label{sec:fixed-point-theorem}

We apply the fixed-point theorem developed in \citet{SandmannBonneton2025}. Because values-of-search need not be bounded uniformly over the infinite horizon, we first normalize the value operator by the deterministic envelope from Lemma~\ref{lem:well-posed-value}.

\subsection{The Fixed-Point Theorem}

Let $\mathcal F$ denote the space of jointly measurable mappings
\[
F:[0,1]\times\mathbb R_+\longrightarrow[0,1],
\]
and let $\mathcal F^2$ denote its two-population product. For $F=(F^X,F^Y)\in\mathcal F^2$, define the discounted seminorm
\begin{align}
\boldsymbol{\big|\big|}F\boldsymbol{\big|\big|}:=\max_{Z\in\{X,Y\}}\int_0^\infty\int_0^1 e^{-t}|F_t^Z(z)|\,dz\,dt.
\label{eq:fixed-point-seminorm}
\end{align}

For a scalar function $F\in\mathcal F$ and $T<\infty$, define
\[
T\mathcal V \bigl(F,[0,1]\times[0,T]\bigr):=\mathcal V_0^1\bigl(F_0(\cdot)\bigr) + \mathcal V_0^T\bigl(F_\cdot(0)\bigr) +\mathcal V_2\bigl(F,[0,1]\times[0,T]\bigr),
\]
where
\begin{align*}
\mathcal V_2 \bigl(F,[0,1]\times[0,T]\bigr) := \sup_{\mathcal P}
\sum_{i=1}^m \Big|
&F_{t_i}(z_i)-F_{t_{i-1}}(z_i)-F_{t_i}(z_{i-1})+F_{t_{i-1}}(z_{i-1})
\Big|,
\end{align*}
and the supremum is taken over all discrete paths
\[
0=z_0<z_1<\cdots<z_m=1,
\qquad
0=t_0<t_1<\cdots<t_m=T.
\]

\begin{theorem}[Fixed-point theorem;
\citealp{SandmannBonneton2025}]
\label{thm:abstract-fixed-point}
Let $H:\mathcal F^2\to\mathcal F^2$. Suppose that:

\begin{enumerate}[label=(\roman*)]
\item $H$ is continuous under the seminorm \eqref{eq:fixed-point-seminorm}: for every $\overline F\in\mathcal F^2$ and every $\varepsilon>0$, there exists $\delta>0$ such that
\[
\boldsymbol{\big|\big|} H[F]-H[\overline F] \boldsymbol{\big|\big|}
<\varepsilon \qquad \text{whenever $\boldsymbol{\big|\big|}F-\overline F\boldsymbol{\big|\big|}<\delta$;}
\]

\item for every $T<\infty$, there exists $C_T<\infty$ such that
\[
T\mathcal V \bigl(H^Z[F],[0,1]\times[0,T]\bigr) \leq C_T \qquad \text{for every $F\in\mathcal F^2$ and $Z\in\{X,Y\}$.}
\]
\end{enumerate}

Then $H$ admits a fixed point.
\end{theorem}

\subsection{Envelope Normalization}

Although values need not admit a bound uniform over calendar time, they remain finite at every date. The reason is that the population and hence meeting opportunities grow at most linearly, whereas future gains are discounted exponentially.

Let
\(
\overline\alpha:=\max\{\alpha^X,\alpha^Y\}
\)
and define the common value envelope
\(
B(t)
:=
1+
\overline\alpha\,\overline G
\int_t^\infty e^{-\rho(\tau-t)}
\overline h(\tau)\,d\tau.
\) 
By construction, $B$ is positive and affine in calendar time, and
\(B(t)\geq1+B^Z(t)\) for \(Z\in\{X,Y\}.\)

Define the rescaling operator
$\mathcal J:\mathcal F^2\to\mathcal F_B^2$ by
\[
(\mathcal JF)_t^Z(z):=B(t)F_t^Z(z) \qquad \text{where}\qquad \mathcal F_B^2
:=
\left\{
v:
0\leq v_t^Z(z)\leq B(t)
\right\}.
\]
The normalized value operator is
\[
\widehat{\boldsymbol V}[F]
:=
\mathcal J^{-1}\boldsymbol V[\mathcal JF],
\qquad
\widehat{\boldsymbol V}_t^Z[F](z)=\frac{\boldsymbol V_t^Z[\mathcal JF](z)}{B(t)}.
\label{eq:normalized-value-operator}
\]
Lemma~\ref{lem:well-posed-value} implies that
\(
\widehat{\boldsymbol V}:\mathcal F^2\longrightarrow\mathcal F^2.
\)
Moreover,
\begin{align}
F=\widehat{\boldsymbol V}[F]\quad\Longleftrightarrow\quad\mathcal JF=\boldsymbol V[\mathcal JF].
\label{eq:normalized-fixed-point-equivalence}
\end{align}

\subsection{Continuity}

This is the point at which pair-specific shocks simplify the existence proof. Without shocks, the aggregate matching rule is an indicator, so an arbitrarily small change in conjectured values can switch a positive mass of nearly indifferent pairs from matching to not matching. The no-shock model therefore requires an anti-concentration condition on output. 
Smooth shocks replace the output-based anti-concentration condition needed in the model without shocks. Indeed, for any conjectures
$v,\widetilde v$ and states $\mu,\widetilde\mu$,
\begin{align}
&\bigl| m_t[v,\mu](x,y)-m_t[\widetilde v,\widetilde\mu](x,y)\bigr|\leq L^\Xi \Big(|v_t^X(x)-\widetilde v_t^X(x)|+|v_t^Y(y)-\widetilde v_t^Y(y)|+N(\mu,\widetilde\mu)\Big).
\label{eq:aggregate-match-continuity}
\end{align}
No continuity or differentiated-marginal-output condition on $f$ is used in this inequality.

\begin{proposition}[Continuity]
\label{prop:normalized-continuity}
Under Assumptions~\ref{ass:search-entry}, \ref{ass:output-bv}, and~\ref{ass:shocks}, the normalized value operator
\[
\widehat{\boldsymbol V}:\mathcal F^2\longrightarrow\mathcal F^2
\]
is continuous under the seminorm \eqref{eq:fixed-point-seminorm}.
\end{proposition}

The proof proceeds in three steps. First, \eqref{eq:aggregate-match-continuity} and Gronwall's inequality give local continuity of the state mapping $v\mapsto\boldsymbol\mu[v]$. Second, stability of the finite-horizon contraction gives local continuity of $v\mapsto\boldsymbol V[v]$. Finally, the value envelope controls the infinite-horizon tail uniformly.

\subsection{Locally Uniform Bounded Variation}

The type-variation argument is inherited from the TU mimicking argument in \cite{BonnetonSandmann2025}. State-dependent shocks do not alter it: conditional on calendar time and the aggregate state, two types that use the same acceptance rule face the same shock distribution, so the shock component cancels from their payoff difference.
\begin{lemma}[Mimicking]
\label{lem:TU-mimicking-hierarchical}
Fix a conjecture $v$. For every $x_2>x_1$ and every $t$, there exists a non-negative measure $Q_t^X[v](\cdot\mid x_1)$ with
\(
Q_t^X[v]([0,1]\mid x_1)<1
\)
such that
\begin{align}
\boldsymbol V_t^X[v](x_2)-\boldsymbol V_t^X[v](x_1)\geq
\int_0^1\bigl[f(x_2,y)-f(x_1,y)\bigr]Q_t^X[v](dy\mid x_1).
\label{eq:ooe-mimicking}
\end{align}
The symmetric statement holds for population $Y$.
\end{lemma}
Lemma~\ref{lem:TU-mimicking-hierarchical} slightly generalizes the mimicking lemma in the main text which assumes anonymous search and therefore holds in both directions for both $x_2>x_1$ and $x_2<x_1$.
\begin{proposition}[Locally uniform bounded variation]
\label{prop:normalized-bv}
Under Assumptions~\ref{ass:search-entry}, \ref{ass:output-bv}, and~\ref{ass:shocks}, the following hold.
\begin{enumerate}[label=(\roman*)]
\item For every $T<\infty$, there exists $K_T<\infty$ such that
\[
\left|\widehat{\boldsymbol V}_{t_2}^Z[F](z)-\widehat{\boldsymbol V}_{t_1}^Z[F](z)\right|\leq K_T|t_2-t_1|
\]
for all $0\leq t_1<t_2\leq T$, every $z$, every $F\in\mathcal F^2$, and $Z\in\{X,Y\}$.

\item If $\overline G>0$, then, for every $t\geq0$,
\[
\mathcal V_0^1 \left(\widehat{\boldsymbol V}_t^Z[F](\cdot)\right)\leq 1+\frac{2L^f}{\overline\alpha\,\overline G}
\]
for every $F\in\mathcal F^2$ and $Z\in\{X,Y\}$. If $\overline G=0$, the normalized value operator is identically zero.

\item Consequently, if $\overline G>0$, then, for every $T<\infty$,
\[
T\mathcal V \left(\widehat{\boldsymbol V}^Z[F],[0,1]\times[0,T]\right)\leq 1+\frac{2L^f}{\overline\alpha\,\overline G} +
3K_TT
\]
for every $F\in\mathcal F^2$ and $Z\in\{X,Y\}$. If $\overline G=0$, uniformly bounded variation holds trivially.
\end{enumerate}
\end{proposition}

\subsection{Equilibrium Existence}

Proposition~\ref{prop:normalized-continuity} verifies condition~(i) of Theorem~\ref{thm:abstract-fixed-point}, while Proposition~\ref{prop:normalized-bv} verifies condition~(ii). The normalized operator therefore admits a fixed point. Through \eqref{eq:normalized-fixed-point-equivalence}, this fixed point induces a fixed point of the original value operator and hence an equilibrium.

\newpage 

\appendix

\section{Proofs for the Fixed-Point Construction}
\label{app:fixed-point-construction}

\subsection{Proof of Proposition~\ref{prop:population-under-v}}
\label{app:proof-population-under-v}

\begin{proof}[Proof of Proposition~\ref{prop:population-under-v}]
The argument follows the proof of the population-dynamics result in \citet{SandmannBonneton2025}. The only difference is that the meeting-contingent match probability now depends on the state through the shock distribution. Assumption~\ref{ass:shocks}(ii) gives
\[
\bigl|m_t[v,\mu](x,y)-m_t[v,\widetilde\mu](x,y)\bigr|\leq L^\Xi N(\mu,\widetilde\mu).
\]
On every finite interval $[0,T]$, the population envelope and the linear bound on meetings therefore imply that the additional term in the population drift is bounded by
\[
\overline{\overline\mu}_T L^\lambda \bigl(1+\overline{\overline\mu}_T\bigr) L^\Xi N(\mu,\widetilde\mu).
\]
Thus the drift remains locally Lipschitz in the state. The successive-interval contraction argument of \citet{SandmannBonneton2025} applies without further change, and the population envelope rules out finite-time explosion.
\end{proof}

\subsection{Proof of Lemma~\ref{lem:well-posed-value}}
\label{app:proof-well-posed-value}

\begin{proof}
We prove the result for population $X$. The proof for population $Y$ is symmetric. Fix a conjecture $v$ and the induced population path $\boldsymbol\mu[v]$.

For notational convenience, write
\[
\overline\lambda_t^X[v](x):= \int_0^1 \lambda^X\bigl(t,\boldsymbol\mu_t[v]\bigr)(y\mid x)\,dy.
\]
By Assumption~\ref{ass:search-entry} and \eqref{eq:population-envelope},
\begin{align}
\overline\lambda_t^X[v](x) \leq \overline h(t) = L^\lambda \bigl(1+\overline{\overline\mu}_0+L^\eta t\bigr).
\label{eq:total-meeting-rate-bound}
\end{align}

\medskip
\noindent
\emph{Step 1: Finite-horizon contraction.}

Fix $T<\infty$ and a bounded non-negative terminal value $\varphi^X:[0,1]\to\mathbb R_+$. Let $\mathcal B_T^X$ be the complete space of bounded non-negative measurable mappings $\widetilde V^X:[0,T]\times[0,1]\to\mathbb R_+$.

Conditioning on the first meeting after date $t$ gives the operator $\mathcal R_{T,v}^{X,\varphi}$:
\begin{align}
\bigl(\mathcal R_{T,v}^{X,\varphi}\widetilde V^X\bigr)_t(x)=& e^{-\rho(T-t)} \exp\left\{-\int_t^T\overline\lambda_r^X[v](x)\,dr \right\} \varphi^X(x) \nonumber\\
&+ \int_t^T e^{-\rho(\tau-t)} \exp\left\{-\int_t^\tau\overline\lambda_r^X[v](x)\,dr \right\} \int_0^1 \lambda^X\bigl(\tau,\boldsymbol\mu_\tau[v]\bigr)(y\mid x) \nonumber\\
&\qquad\times \Bigg[
\widetilde V_\tau^X(x) +
\alpha^X G\left( \tau,\boldsymbol\mu_\tau[v],\widetilde V_\tau^X(x)+v_\tau^Y(y)-f(x,y) \right)
\Bigg] \,dy\,d\tau.
\label{eq:first-meeting-operator}
\end{align}
The expression in square brackets is the expected value immediately after a meeting: following rejection the agent continues with value $\widetilde V_\tau^X(x)$, whereas acceptance yields that value plus the Nash share of the positive match surplus.

For fixed $(\tau,x,y)$, the mapping
\[
a\longmapsto a+\alpha^X G\left( \tau,\boldsymbol\mu_\tau[v], a+v_\tau^Y(y)-f(x,y) \right)
\]
is non-decreasing and one-Lipschitz. Indeed, if $a_2>a_1$, then
\begin{align*}
0 &\leq G\left(\tau,\boldsymbol\mu_\tau[v],a_1+v_\tau^Y(y)-f(x,y)\right)-
G\left(\tau,\boldsymbol\mu_\tau[v],a_2+v_\tau^Y(y)-f(x,y)\right) \leq a_2-a_1
\end{align*}
by \eqref{eq:G-one-Lipschitz}. Consequently,
\begin{align}
(1-\alpha^X)(a_2-a_1) \leq
\Bigl[a_2+\alpha^XG(a_2+\cdots)\Bigr] - \Bigl[a_1+\alpha^XG(a_1+\cdots)\Bigr] \leq a_2-a_1
\label{eq:meeting-value-Lipschitz}
\end{align}

Equip $\mathcal B_T^X$ with the Bielecki norm
\[
\|\widetilde V^X\|_{\beta,T}:= \sup_{t\in[0,T],\,x\in[0,1]} e^{-\beta(T-t)} |\widetilde V_t^X(x)|.
\]
Let
\(
H_T := L^\lambda \bigl(1+\overline{\overline\mu}_T\bigr).
\)
Using \eqref{eq:meeting-value-Lipschitz} and \eqref{eq:total-meeting-rate-bound},
\begin{align*}
\bigl\|
\mathcal R_{T,v}^{X,\varphi}\widetilde V^X - \mathcal R_{T,v}^{X,\varphi}\widehat V^X \bigr\|_{\beta,T}
&\leq \frac{H_T}{\rho+\beta} \|\widetilde V^X-\widehat V^X\|_{\beta,T}.
\end{align*}
Choose $\beta>0$ sufficiently large that $H_T/(\rho+\beta)<1$. The operator is then a contraction. It therefore admits a unique fixed point, denoted $\boldsymbol V^{X,T,\varphi}[v]$. For zero terminal value, write $\boldsymbol V^{X,T}[v]$.

The operator in \eqref{eq:first-meeting-operator} is monotone in both the candidate value and the terminal value. Therefore,
\begin{align}
\varphi^X\leq\psi^X
\quad\Longrightarrow\quad
\boldsymbol V^{X,T,\varphi}[v] \leq \boldsymbol V^{X,T,\psi}[v].
\label{eq:terminal-monotonicity}
\end{align}

\medskip
\noindent
\emph{Step 2: An equivalent gain equation and the value envelope.}

Consider the finite-horizon gain equation
\begin{align}
\widetilde V_t^X(x) = e^{-\rho(T-t)}\varphi^X(x) &+ \alpha^X \int_t^T e^{-\rho(\tau-t)} \int_0^1 \lambda^X\bigl(\tau, \boldsymbol\mu_\tau[v]\bigr)(y\mid x) \nonumber\\
&\qquad\times G\left(\tau,\boldsymbol\mu_\tau[v], \widetilde V_\tau^X(x)+v_\tau^Y(y)-f(x,y)\right) \,dy\,d\tau.
\label{eq:finite-horizon-gain-equation}
\end{align}
The operator on the right-hand side is also a contraction in the
Bielecki norm, with modulus at most
$\alpha^XH_T/(\rho+\beta)$.

Its fixed point is locally absolutely continuous in time.
Differentiating \eqref{eq:finite-horizon-gain-equation} almost
everywhere and integrating by parts against the first-meeting survival
probability shows that it satisfies
\eqref{eq:first-meeting-operator}. By uniqueness of the fixed point of
the latter operator, the two finite-horizon formulations coincide.
Thus $\boldsymbol V^{X,T,\varphi}[v]$ satisfies
\eqref{eq:finite-horizon-gain-equation}.

For the zero-terminal-value problem,
the acceptance threshold is bounded below by $-\overline f$ because
$\boldsymbol V^{X,T}[v]$ and $v^Y$ are non-negative. Hence
\[
G\left(
\tau,\boldsymbol\mu_\tau[v],
\boldsymbol V_\tau^{X,T}[v](x)+v_\tau^Y(y)-f(x,y)
\right)
\leq
\overline G.
\]
Equation~\eqref{eq:finite-horizon-gain-equation} therefore yields
\begin{align}
0
\leq
\boldsymbol V_t^{X,T}[v](x)
&\leq
\alpha^X\overline G
\int_t^T e^{-\rho(\tau-t)}
\overline h(\tau)\,d\tau \leq
B^X(t).
\label{eq:value-search-bound-B}
\end{align}
This proves the value envelope uniformly in $T$, $x$, and $v$.

\medskip
\noindent
\emph{Step 3: Infinite-horizon limit and uniqueness.}

Let $T_2>T_1$. Restricted to $[0,T_1]$,
$\boldsymbol V^{X,T_2}[v]$ solves the first-meeting problem with
terminal value
$\boldsymbol V_{T_1}^{X,T_2}[v]\geq0$. By
\eqref{eq:terminal-monotonicity},
\[
\boldsymbol V_t^{X,T_1}[v](x)
\leq
\boldsymbol V_t^{X,T_2}[v](x)
\qquad
(t\leq T_1).
\]
Thus the finite-horizon values increase with the horizon. Since they
are bounded above by $B^X(t)$, the pointwise limit
\[
\boldsymbol V_t^X[v](x)
:=
\lim_{T\to\infty}
\boldsymbol V_t^{X,T}[v](x)
\]
exists and satisfies \eqref{eq:value-envelope}.

Dominated convergence in
\eqref{eq:finite-horizon-gain-equation} gives
\begin{align}
\boldsymbol V_t^X[v](x)
={}&
\alpha^X
\int_t^\infty e^{-\rho(\tau-t)}
\int_0^1
\lambda^X\bigl(\tau,\boldsymbol\mu_\tau[v]\bigr)(y\mid x)
\nonumber\\
&\qquad\times
G\left(
\tau,\boldsymbol\mu_\tau[v],
\boldsymbol V_\tau^X[v](x)+v_\tau^Y(y)-f(x,y)
\right)
\,dy\,d\tau.
\label{eq:ooe-gain-equation}
\end{align}
Notice that
\begin{align}
e^{-\rho(T-t)}B^X(T)\longrightarrow0
\qquad
\text{as }T\to\infty,
\label{eq:value-transversality}
\end{align}
because $B^X$ is affine in calendar time.

We next prove uniqueness. Suppose without loss that $\varphi\geq\psi$. Monotonicity gives
$\boldsymbol V^{X,T,\varphi}[v]\geq
\boldsymbol V^{X,T,\psi}[v]$, while subtracting their gain equations
and using that $G$ is non-increasing in its threshold yields
\begin{align}
\label{eq:terminal-stability}
0\leq
\boldsymbol V_t^{X,T,\varphi}[v](x)
-
\boldsymbol V_t^{X,T,\psi}[v](x)
\leq
e^{-\rho(T-t)}
\|\varphi-\psi\|_\infty.
\end{align}
Indeed, a terminal perturbation of size $c$ at $T$ can alter the value
at any earlier date by at most $e^{-\rho(T-t)}c$; at every intervening
meeting the expected continuation-payoff mapping is one-Lipschitz. The general case follows by applying the argument to the upper and
lower envelopes of the terminal conditions.

Let $\widetilde V^X$ and $\widehat V^X$ be two non-negative
infinite-horizon solutions. According to \eqref{eq:value-search-bound-B} these are bounded by $B^X$. Restricting both to $[0,T]$ and treating their date-$T$ values as terminal conditions,
\eqref{eq:terminal-stability} gives
\[
\sup_x
\left|
\widetilde V_t^X(x)-\widehat V_t^X(x)
\right|
\leq
2e^{-\rho(T-t)}B^X(T).
\]
Letting $T\to\infty$ and using
\eqref{eq:value-transversality} proves uniqueness.

\medskip
\noindent
\emph{Step 4: Equivalence with the primitive first-match payoff.}

Evaluate the subjective matching probability
\eqref{eq:subjective-match-probability} at the unique solution
$\boldsymbol V^X[v]$, and define
\[
\Lambda_\tau^X[v](y\mid x):=\lambda^X\bigl(\tau,\boldsymbol\mu_\tau[v]\bigr)(y\mid x) \text{\normalfont\bsifamily m}_\tau^X[v](x,y).
\]
Then
\[
\text{\normalfont\bsifamily p}_{t,\tau}^X[v](y\mid x)
=\Lambda_\tau^X[v](y\mid x)\exp\left\{-\int_t^\tau\int_0^1\Lambda_r^X[v](z\mid x)\,dz\,dr\right\}.
\]

By \eqref{eq:subjective-payoff-product},
\begin{align}
&\int_0^1
\lambda^X\bigl(\tau,\boldsymbol\mu_\tau[v]\bigr)(y\mid x)\text{\normalfont\bsifamily m}_\tau^X[v](x,y)\boldsymbol\pi_\tau^X[v](y\mid x)\,dy
\nonumber\\
&\qquad=\boldsymbol V_\tau^X[v](x)\int_0^1\Lambda_\tau^X[v](y\mid x)\,dy
\nonumber\\
&\qquad\quad+\alpha^X\int_0^1\lambda^X\bigl(\tau,\boldsymbol\mu_\tau[v]\bigr)(y\mid x)G\left(\tau,\boldsymbol\mu_\tau[v],\boldsymbol V_\tau^X[v](x)+v_\tau^Y(y)-f(x,y)\right)dy.
\label{eq:hazard-payoff-identity}
\end{align}

Equation~\eqref{eq:ooe-gain-equation} implies that $t\mapsto\boldsymbol V_t^X[v](x)$ is locally absolutely continuous and, almost everywhere,
\[
\frac{d}{dt}\boldsymbol V_t^X[v](x) =\rho\boldsymbol V_t^X[v](x)-\alpha^X \int_0^1 \lambda^X\bigl(t,\boldsymbol\mu_t[v]\bigr)(y\mid x)G\left(t,\boldsymbol\mu_t[v],\boldsymbol V_t^X[v](x)+v_t^Y(y)-f(x,y)\right)dy.
\]
Consequently,
\begin{align*}
\frac{d}{d\tau}\Bigg[&e^{-\rho(\tau-t)}\exp\left\{-\int_t^\tau\int_0^1 \Lambda_r^X[v](z\mid x)\,dz\,dr \right\} \boldsymbol V_\tau^X[v](x) \Bigg]\\
={}& - e^{-\rho(\tau-t)}\exp\left\{-\int_t^\tau\int_0^1\Lambda_r^X[v](z\mid x)\,dz\,dr \right\}\\
&\times \int_0^1 \lambda^X\bigl(\tau,\boldsymbol\mu_\tau[v]\bigr)(y\mid x) \text{\normalfont\bsifamily m}_\tau^X[v](x,y) \boldsymbol\pi_\tau^X[v](y\mid x)\,dy,
\end{align*}
where the last equality uses
\eqref{eq:hazard-payoff-identity}.

Integrating from $t$ to $T$ and letting $T\to\infty$ gives
\[
\boldsymbol V_t^X[v](x) = \int_t^\infty e^{-\rho(\tau-t)} \int_0^1 \boldsymbol\pi_\tau^X[v](y\mid x) \text{\normalfont\bsifamily p}_{t,\tau}^X[v](y\mid x) \,dy\,d\tau.
\]
The terminal term vanishes by \eqref{eq:value-envelope}--\eqref{eq:value-transversality}. Thus the solution of the gain equation satisfies the primitive first-match payoff formula in Definition~\ref{def:out-of-equilibrium-value}.

Conversely, the same integration-by-parts calculation on finite horizons shows that any locally bounded solution of the primitive first-match formula satisfies the gain equation. Its uniqueness has already been established. This proves that the primitive out-of-equilibrium value is well posed.
\end{proof}

\section{Proofs for the Fixed-Point Theorem}

\subsection{Proof of Proposition~\ref{prop:normalized-continuity}}
\label{app:proof-normalized-continuity}

\begin{proof}
Fix $F,\widetilde F\in\mathcal F^2$, and set
\(
v:=\mathcal JF,
\qquad
\widetilde v:=\mathcal J\widetilde F.
\)
For each $t$, write
\[
d_t(v,\widetilde v) := \int_0^1|v_t^X(x)-\widetilde v_t^X(x)|\,dx +\int_0^1|v_t^Y(y)-\widetilde v_t^Y(y)|\,dy.
\]
For every $R<\infty$,
\begin{align}
\int_0^R d_t(v,\widetilde v)\,dt \leq 2B(R)e^R \boldsymbol{\big|\big|} F-\widetilde F \boldsymbol{\big|\big|}.
\label{eq:local-input-control}
\end{align}

\medskip
\noindent
\emph{Step 1: Continuity of the population path.}

Let
\[
N_t:=N\bigl(\boldsymbol\mu_t[v],\boldsymbol\mu_t[\widetilde v]\bigr).
\]
Equation~\eqref{eq:aggregate-match-continuity}, the population envelope, and the meeting and entry assumptions imply that, for every $R<\infty$, there exists $C_R<\infty$ such that
\[
N_t \leq C_R \int_0^t \bigl[ N_\tau+d_\tau(v,\widetilde v) \bigr]\,d\tau \qquad (t\leq R).
\]
Gronwall's inequality therefore gives
\begin{align}
\sup_{t\leq R}N_t \leq C_R \int_0^R d_\tau(v,\widetilde v)\,d\tau.
\label{eq:state-continuity-local}
\end{align}
Here and below, $C_R$ may change from line to line.

\medskip
\noindent
\emph{Step 2: Continuity of finite-horizon values.}

Let $\boldsymbol V^{X,R}[v]$ and $\boldsymbol V^{X,R}[\widetilde v]$ denote the zero-terminal-value finite-horizon solutions constructed in the proof of Lemma~\ref{lem:well-posed-value}, and define
\[
D_t^{X,R} := \int_0^1 \left|\boldsymbol V_t^{X,R}[v](x)-\boldsymbol V_t^{X,R}[\widetilde v](x)\right|dx.
\]
The finite-horizon gain representation, the one-Lipschitz property of $G$ in its threshold, and Assumption~\ref{ass:shocks}(iii) imply
\begin{align}
D_t^{X,R}\leq{}&\alpha^X\int_t^R e^{-\rho(\tau-t)} \Big[H_RD_\tau^{X,R}+
H_R \int_0^1 |v_\tau^Y(y)-\widetilde v_\tau^Y(y)|\,dy + \bigl( L^\lambda\overline G+H_RL^G \bigr)N_\tau \Big]d\tau,
\label{eq:finite-horizon-value-stability}
\end{align}
where
\[
H_R := L^\lambda \bigl(1+\overline{\overline\mu}_R\bigr).
\]
The term $L^\lambda\overline G N_\tau$ controls the change in the
meeting rate, while $H_RL^G N_\tau$ controls the change in the
stop-loss transform.

Backward Gronwall and
\eqref{eq:state-continuity-local} yield
\begin{align}
\int_0^R D_t^{X,R}\,dt \leq C_R \int_0^R d_t(v,\widetilde v)\,dt.
\label{eq:finite-horizon-value-continuity}
\end{align}
The same estimate holds for population $Y$. Combining \eqref{eq:local-input-control} and \eqref{eq:finite-horizon-value-continuity} shows that the finite-horizon values converge locally in $L^1$ whenever $F\to\widetilde F$ in the fixed-point seminorm.

\medskip
\noindent
\emph{Step 3: Infinite-horizon values and normalization.}

The terminal-value stability estimate established in the proof of Lemma~\ref{lem:well-posed-value} implies that, for $t\leq R$,
\begin{align}
0 \leq \boldsymbol V_t^Z[v](z) - \boldsymbol V_t^{Z,R}[v](z) \leq e^{-\rho(R-t)}B^Z(R) \leq e^{-\rho(R-t)}B(R).
\label{eq:finite-horizon-tail}
\end{align}

Fix $\varepsilon>0$. First choose $T$ sufficiently large that $e^{-T}<\varepsilon/3$. Since $B(R)$ grows at most linearly, we can then choose $R>T$ sufficiently large that
\[
2T e^{-\rho(R-T)}B(R)<\frac{\varepsilon}{3}.
\]
By \eqref{eq:local-input-control}--\eqref{eq:finite-horizon-value-continuity}, there exists $\delta>0$ such that
\[
\boldsymbol{\big|\big|} F-\widetilde F \boldsymbol{\big|\big|}<\delta
\]
implies
\[
\max_{Z\in\{X,Y\}} \int_0^T\int_0^1 \frac{\left| \boldsymbol V_t^{Z,R}[v](z) - \boldsymbol V_t^{Z,R}[\widetilde v](z) \right|}{B(t)} \,dz\,dt < \frac{\varepsilon}{3}.
\]
Equation~\eqref{eq:finite-horizon-tail} then controls the two truncation errors on $[0,T]$.

Finally, since $0\leq\widehat{\boldsymbol V}^Z[F]\leq1$,
\[
\int_T^\infty\int_0^1 e^{-t} \left| \widehat{\boldsymbol V}_t^Z[F](z)-\widehat{\boldsymbol V}_t^Z[\widetilde F](z) \right|\,dz\,dt \leq e^{-T} < \frac{\varepsilon}{3}.
\]
Combining the three estimates proves
\[
\boldsymbol{\big|\big|} \widehat{\boldsymbol V}[F] - \widehat{\boldsymbol V}[\widetilde F] \boldsymbol{\big|\big|} <\varepsilon.
\]
\end{proof}

\subsection{Proof of Lemma~\ref{lem:TU-mimicking-hierarchical}}
\label{app:proof-ooe-mimicking}

\begin{proof}
The proof of the mimicking argument in the main text applies for types $x_2>x_1$: type $x_2$ can disregard any additional meetings generated by hierarchical search and mimic type $x_1$'s meeting and acceptance  strategy. 
\end{proof}

\subsection{Proof of Proposition~\ref{prop:normalized-bv}}
\label{app:proof-normalized-bv}

\begin{proof}
Fix $F\in\mathcal F^2$ and write $v=\mathcal JF$.

\medskip
\noindent
\emph{Step 1: Time regularity.}

The gain representation derived in the proof of Lemma~\ref{lem:well-posed-value} can be written as
\[
\boldsymbol V_t^X[v](x) = \int_t^\infty e^{-\rho(\tau-t)} \Phi_\tau^X[v](x)\,d\tau,
\]
where
\(
\Phi_t^X[v](x):=\alpha^X \int_0^1 \lambda^X\bigl(t,\boldsymbol\mu_t[v]\bigr)(y\mid x) G\left( t,\boldsymbol\mu_t[v],\boldsymbol V_t^X[v](x)+v_t^Y(y)-f(x,y) \right)dy.
\)
For $t\leq T$,
\[
0\leq \Phi_t^X[v](x) \leq \alpha^X L^\lambda \bigl(1+\overline{\overline\mu}_T\bigr) \overline G.
\]
It follows that $\boldsymbol V^X[v](x)$ is locally absolutely continuous and, almost everywhere,
\[
\frac{d}{dt}\boldsymbol V_t^X[v](x) = \rho\boldsymbol V_t^X[v](x) - \Phi_t^X[v](x).
\]
The value envelope therefore gives a bound on this derivative that is uniform in $x$ and $F$ on every interval $[0,T]$.

Because $B$ is positive and affine,
\[
\widehat{\boldsymbol V}_t^X[F](x) = \frac{\boldsymbol V_t^X[v](x)}{B(t)}
\]
is also uniformly Lipschitz on $[0,T]$. Thus there exists $K_T<\infty$ such that
\[
\left| \widehat{\boldsymbol V}_{t_2}^X[F](x) - \widehat{\boldsymbol V}_{t_1}^X[F](x) \right| \leq K_T|t_2-t_1|.
\]
The same argument applies to population $Y$.

\medskip
\noindent
\emph{Step 2: Variation in type.}

We use the particular out-of-equilibrium mimicking measures generated by the iterative proof of Lemma~\ref{lem:TU-mimicking-hierarchical}. These measures satisfy the following uniform domination property: for every measurable $A\subseteq[0,1]$,
\begin{equation}
\label{eq:Q-domination}
Q_t^X[v](A\mid x) \leq |A| \int_t^\infty e^{-\rho(\tau-t)} \overline h(\tau)\,d\tau,
\end{equation}
where $|A|$ denotes Lebesgue measure.

Indeed, the subjective first-match density satisfies
\[
\text{\normalfont\bsifamily p}_{s,\tau}^X[v](y\mid x) \leq \lambda^X\bigl(\tau,\boldsymbol\mu_\tau[v]\bigr)(y\mid x) \leq \overline h(\tau),
\]
and its total mass over all future match dates and partner types is at most one. The $k$th term in the iterative construction of $Q_t^X[v](\cdot\mid x)$ is therefore bounded on $A$ by
\[
\alpha^X(1-\alpha^X)^{k-1}|A| \int_t^\infty e^{-\rho(\tau-t)} \overline h(\tau)\,d\tau.
\]
Summing over $k$ proves~\eqref{eq:Q-domination}.

Now fix an arbitrary partition $0=x_0<x_1<\cdots<x_m=1$. Since
\(
|a-b| = (a-b)+2\max\{b-a,0\},
\)
we have
\begin{align*}
&\sum_{i=1}^m \left| \boldsymbol V_t^X[v](x_i) - \boldsymbol V_t^X[v](x_{i-1}) \right|= \boldsymbol V_t^X - \boldsymbol V_t^X+ 2\sum_{i=1}^m \max\left\{
\boldsymbol V_t^X[v](x_{i-1}) - \boldsymbol V_t^X[v](x_i), 0
\right\}.
\end{align*}
The value envelope gives
\[
\boldsymbol V_t^X - \boldsymbol V_t^X \leq B^X(t).
\]

For every adjacent pair, the mimicking inequality gives
\[
\begin{aligned}
\boldsymbol V_t^X[v](x_i) - \boldsymbol V_t^X[v](x_{i-1}) \geq \int_0^1 \bigl[ f(x_i,y)-f(x_{i-1},y) \bigr] Q_t^X[v](dy\mid x_{i-1}).
\end{aligned}
\]
Consequently,
\begin{align*}
&\max\left\{ \boldsymbol V_t^X[v](x_{i-1}) - \boldsymbol V_t^X[v](x_i), 0
\right\}\leq \int_0^1 \left| f(x_i,y)-f(x_{i-1},y) \right| Q_t^X[v] dy\mid x_{i-1}).
\end{align*}
Combining these inequalities and applying \eqref{eq:Q-domination} yields
\begin{align*}
&\sum_{i=1}^m \left| \boldsymbol V_t^X[v](x_i) - \boldsymbol V_t^X[v](x_{i-1}) \right|\\ 
&\quad\leq B^X(t) + 2 \left( \int_t^\infty e^{-\rho(\tau-t)} \overline h(\tau)\,d\tau \right) \int_0^1 \sum_{i=1}^m \left| f(x_i,y)-f(x_{i-1},y)
\right|dy\\
&\quad\leq B^X(t) + 2L^f \int_t^\infty e^{-\rho(\tau-t)} \overline h(\tau)\,d\tau.
\end{align*}
The final inequality follows from Assumption~\ref{ass:output-bv}. Taking the supremum over all partitions gives
\begin{equation}
\label{eq:unnormalized-type-variation}
\mathcal V_0^1 \left( \boldsymbol V_t^X[v](\cdot) \right) \leq B^X(t) +2L^f \int_t^\infty e^{-\rho(\tau-t)} \overline h(\tau)\,d\tau.
\end{equation}

If $\overline G=0$, the gain equation implies $\boldsymbol V[v]\equiv0$, so the claimed bounded-variation property is immediate. Suppose therefore that $\overline G>0$. Recalling the definitions
\[
B^X(t) = \alpha^X\overline G \int_t^\infty e^{-\rho(\tau-t)} \overline h(\tau)\,d\tau
\]
and
\[
B(t) = 1+\overline\alpha\,\overline G \int_t^\infty e^{-\rho(\tau-t)} \overline h(\tau)\,d\tau,
\]
we obtain
\begin{align*}
\mathcal V_0^1 \left(\widehat{\boldsymbol V}_t^X[F](\cdot)\right)
&=\frac{1}{B(t)}\mathcal V_0^1\left(\boldsymbol V_t^X[v](\cdot)\right)\leq \frac{B^X(t)}{B(t)}+2L^f\frac{\int_t^\infty e^{-\rho(\tau-t)}\overline h(\tau)\,d\tau}{B(t)}\leq 1+ \frac{2L^f}{\overline\alpha\,\overline G}.
\end{align*}
The population-$Y$ bound is symmetric.

\medskip
\noindent
\emph{Step 3: Two-dimensional variation.}

Take any discrete path
\(
0=z_0<\cdots<z_m=1,
\qquad
0=t_0<\cdots<t_m=T.
\)
By the time-Lipschitz estimate,
\begin{align*}
\sum_{i=1}^m
\Big|
&\widehat{\boldsymbol V}_{t_i}^Z[F](z_i)
- \widehat{\boldsymbol V}_{t_{i-1}}^Z[F](z_i)
- \widehat{\boldsymbol V}_{t_i}^Z[F](z_{i-1})
+ \widehat{\boldsymbol V}_{t_{i-1}}^Z[F](z_{i-1})
\Big|
\leq 2K_TT.
\end{align*}
Moreover,
\(
    \mathcal V_0^T \left(t\mapsto \widehat{\boldsymbol V}_t^Z\right)\leq K_T T.
\)

while the type variation at date zero is bounded by \(2L^f+1\).
Therefore,
\[
T\mathcal V \left( \widehat{\boldsymbol V}^Z[F],[0,1]\times[0,T]\right) \leq 1+ \frac{2L^f}{\overline\alpha\,\overline G}+3K_T T.
\]
The bound is uniform over $F\in\mathcal F^2$ and both populations.
\end{proof}

\subsection{Proof of Theorem~\ref{thm:existence}}
\label{app:proof-existence}

\begin{proof}
Lemma~\ref{lem:well-posed-value} implies that the normalized operator $\widehat{\boldsymbol V}$ maps $\mathcal F^2$ into itself. Proposition~\ref{prop:normalized-continuity} verifies condition~(i) of Theorem~\ref{thm:abstract-fixed-point}, and Proposition~\ref{prop:normalized-bv} verifies condition~(ii). Therefore, there exists $F^\star\in\mathcal F^2$ such that
\(
F^\star = \widehat{\boldsymbol V}[F^\star].
\)
Let
\(
V^\star:=\mathcal JF^\star.
\)
By \eqref{eq:normalized-fixed-point-equivalence},
\(
V^\star = \boldsymbol V[V^\star].
\)
Proposition~\ref{prop:fixed-point-equilibrium} therefore implies that
\(
\bigl(\boldsymbol\mu[V^\star],V^\star,m[V^\star]\bigr)
\)
is an equilibrium.
\end{proof}

\begingroup
\setstretch{1}\small
\setlength{\bibsep}{0pt plus 0.3ex}
\bibliographystyle{apalike}
\bibliography{a}
\endgroup